# A Taxonomy and Future Directions for Sustainable Cloud Computing: 360 Degree View

SUKHPAL SINGH GILL and RAJKUMAR BUYYA, The University of Melbourne, Australia

The cloud computing paradigm offers on-demand services over the Internet and supports a wide variety of applications. With the recent growth of Internet of Things (IoT) based applications the usage of cloud services is increasing exponentially. The next generation of cloud computing must be energy-efficient and sustainable to fulfil the end-user requirements which are changing dynamically. Presently, cloud providers are facing challenges to ensure the energy efficiency and sustainability of their services. The usage of large number of cloud datacenters increases cost as well as carbon footprints, which further effects the sustainability of cloud services. In this paper, we propose a comprehensive taxonomy of sustainable cloud computing. The taxonomy is used to investigate the existing techniques for sustainability that need careful attention and investigation as proposed by several academic and industry groups. Further, the current research on sustainable cloud computing is organized into several categories: application design, sustainability metrics, capacity planning, energy management, virtualization, thermal-aware scheduling, cooling management, renewable energy and waste heat utilization. The existing techniques have been compared and categorized based on the common characteristics and properties. A conceptual model for sustainable cloud computing has been proposed along with discussion on future research directions.

## 1. INTRODUCTION

Cloud computing offers a flexible and powerful computing environment to provide on-demand, subscription-based online services over the Internet to host applications on a pay-as-you-go basis. The various cloud providers such as Microsoft, Google and Amazon make extensive use of Cloud Data Centers (CDCs) to fulfill the requirements (memory, data, compute or network) of the digital world. To reduce the service delay and maintain the Service Level Agreement (SLA), fault tolerance should be provided through replicating the compute abilities redundantly [1]. To ensure the availability and reliability of services, the components of CDCs such as network devices, storage devices and servers should be run 24/7 [2]. Large amounts of data are created by digital activities such as data streaming, file sharing, searching and social networking websites, e-commerce, sensor networks and that data can be stored as well as processed efficiently using CDCs [3] [4]. The energy cost is added by creating, processing and storing each bit of data, which increases carbon footprints that further impacts on the sustainability of cloud services. Due to the large consumption of electricity by CDCs, the research community is addressing challenge of designing sustainable CDCs [5].

With the continuous growth of Internet of Things (IoT) based applications, the usage of cloud services is increasing exponentially, which further increases the electricity consumptions of CDCs by 20-25% every year [6]. Existing studies claimed that 78.7 million metric tons of $CO_2$ are produced by datacenters, which is equal to the two percent of global emissions [7]. CDCs in USA consumed 100 billion kilowatt hours (kWh) in 2015, which is sufficient for Washington City [11]. The consumption of electricity will reach 150 billion kWh by 2022 i.e. increase by 50% [12]. Energy consumption in CDCs can be increased to 8000 terawatt hours (TWh) in 2030 if controlled mechanisms are not identified [122]. Due to underloading and overloading of resources in infrastructure (cooling, computing, storage, networking etc.), the energy consumption in cloud datacenters is not efficient and mostly the energy is consumed while some of the resources are in idle state, which increases the cost of cloud services [11]. Figure 1

---

Author's address: S. S. Gill and R. Buyya, Cloud Computing and Distributed Systems (CLOUDS) Laboratory, School of Computing and Information Systems, The University of Melbourne, Parkville, Australia - 3010. emails: sukhpal.gill@unimelb.edu.au, rbuyya@unimelb.edu.au

shows the consumption of energy by different components of the idle server such as processor, storage, memory, network and cooling as reported in [3] [71] [95]. Carbon footprints produced by CDCs is same as aviation industry [13]. In the current scenario, CDC service providers are finding alternative ways to reduce the carbon footprint of their infrastructure. The prominent cloud providers such as Google, Amazon, Microsoft and IBM have assured to attain zero production of carbon footprints and they are finding the new ways to make CDCs and cloud-based services eco-friendly [3]. Therefore, CDCs needs to provide cloud services with a minimum carbon footprint and minimum heat release in the form of greenhouse gas emissions [71].

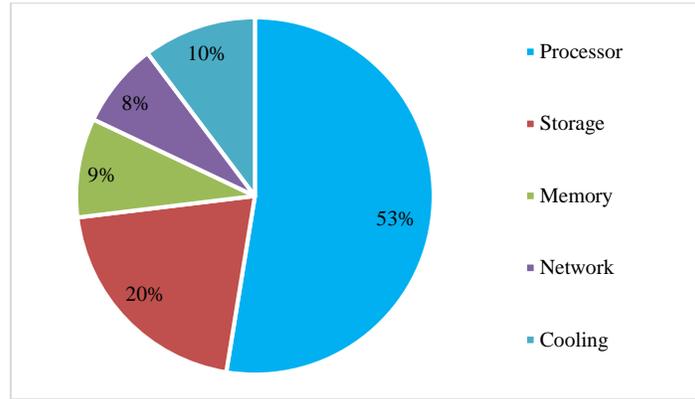

Figure 1: Energy Consumption of different Components of CDC (Data Source: [3] [71] [95])

To solve this challenge of energy-efficient cloud services, a large number of researchers proposed resource management policies, algorithms and architectures but energy-efficiency is still a challenge for future researchers. To ensure high-level of sustainability, holistic management of resources can solve new open challenges existing in the resource scheduling. There is a need of methods that harness renewable energy to decrease carbon footprints without the use of fossil fuels. Further, cooling expenses can be decreased by developing waste heat utilization and free cooling mechanisms. Location-aware ideal climatic conditions are needed for an efficient implementation of free cooling and renewable energy production techniques [91]. Moreover, waste heat recovery locations are required to be identified for an efficient implantation of waste heat recovery predictions. CDCs can be relocated based on: i) opportunities for waste heat recovery, ii) accessibility of green computing resource and iii) proximity of free cooling resources. Cloud providers such as Google, Amazon, IBM, Facebook and Microsoft are utilizing more green energy resources instead of grid electricity [31].

## 1.1 Background

The design of sustainable system is one of the greatest challenge of the 21$^{st}$ Century, the ecological transition coupled with the digital transformation. And it is scientifically difficult to decide on the benefit of a technique for improving the sustainability of a system. "Sustainability" by definition involves four areas of study, i.e. environmental, social, technological and economic spheres [7] [15] [82] as shown in Figure 2. There is a need to identify the different components of sustainable CDC to enable sustainable cloud computing economically (energy/electricity cost), socially (laws and regulations to establish cloud data centre) and technologically (data preservation, protection and retention) and environmentally (energy consumption/carbon footprints/greenhouse emissions).

This systematic review deals mostly with energy in the environmental and economic spheres and by covering issues in how to run CDCs efficiently with minimal energy (minimum cost of energy) by harnessing renewable energy powered resources through holistic management of workloads and resources. The main objectives of sustainable cloud computing are: i) to reduce energy consumption at datacenters and ii) to dispose hardware devices after their useful life. Cloud computing accelerates our economy rapidly through the use of remote servers via the Internet instead of local servers. Due to the availability of large number of datacenters, the user's data is stored, managed and processed efficiently and swiftly but it increases carbon footprints, which effects sustainability [135].

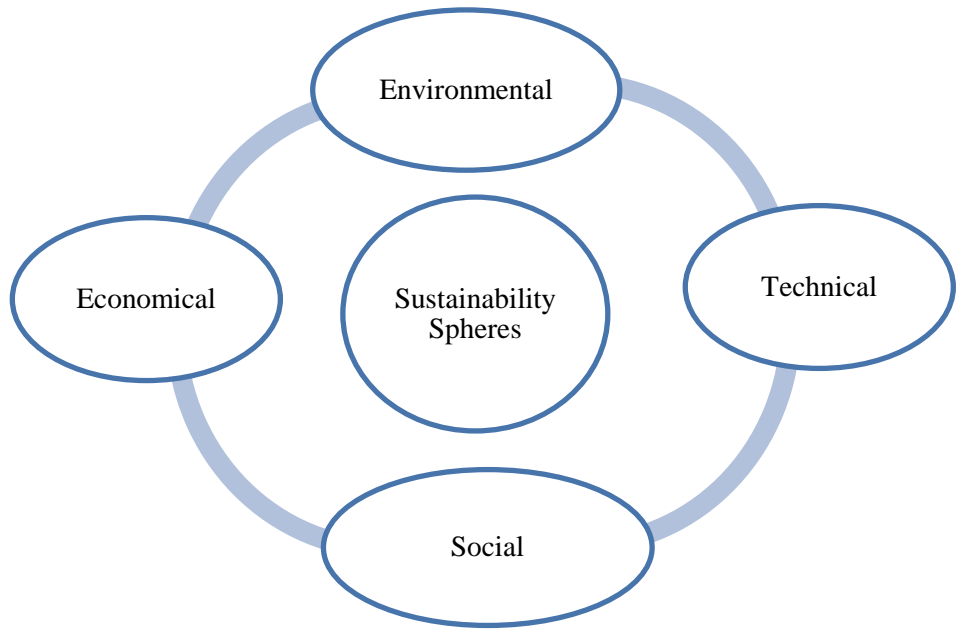

Figure 2: Types of Sustainability Spheres

Cloud computing is growing very rapidly to fulfill the demand of users. Initially, there were only few investment deals and cloud computing accounted for almost $26 million in 2005 [63] as shown in Figure 3.

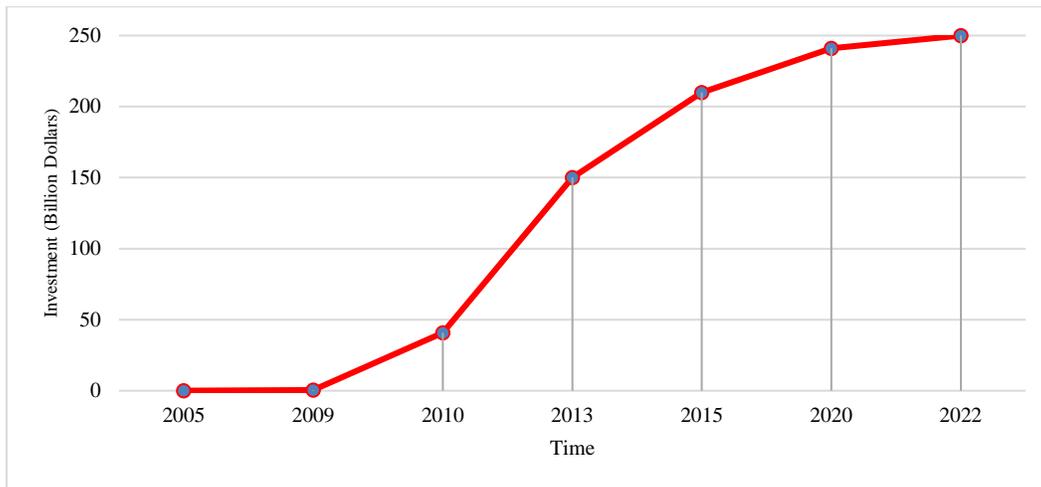

Figure 3: Investment on Cloud Computing

Further, it has reached $375 million in 2009 and attained $40.7 billion in 2010 [43]. In year 2013, it reached $150 billion and attained to $210 billion in 2015 [49]. In year 2020, cloud computing will be accounted for $241 billion and $250 billion in 2022 according to Verdantix company [136]. Cloud based solution are also used by a large number of Small and Medium Businesses corporations, mostly in North America [9] [10] [11]. Further, physical resource (bare metal as service) based services have been also replaced by virtual services, which makes the environment much more sustainable.

Cloud computing provides an efficient management of resources, which saves energy by minimizing the number of physical servers. Further, it reduces the maintenance cost and provides more flexibility and scalability for business expansion. The cloud provides the platform to conduct meetings online which also saves the time of face to face meetings and reduces in costs up to 50 percent [3]. Due to continuous use of virtualization, 31 percent of the energy consumption has been reduced [4]. The multi-tenant architecture of cloud computing makes more efficient cloud services. Energy consumption is the still main challenge of cloud computing, which is responsible for producing carbon footprints and environmental hazards. Researchers are trying to reduce carbon emissions as much as possible and by 2020, servers will avoid 85.7 million metric tons of $CO_2$ [15].

The energy consumption will be increased to 60 billion watts in 2020 [16] [17] and that will be equivalent to the energy consumption of 200K homes of Tokyo, Japan [12]. Moreover, existing energy-aware techniques mainly focus on reducing the energy consumption of the servers [3] [6] [71]. The other components (networks, storage, memory, processor and cooling systems) of CDCs are consuming huge amount of energy. Consequently, there is a need of holistic management of cloud resources to improve energy efficiency of CDCs. The identification and consideration of the important aspects of sustainable CDC (application design, sustainability metrics, capacity planning, energy management, virtualization, thermal-aware scheduling, cooling management, renewable energy and waste heat utilization) is also required to enable sustainable computing economically (energy/electricity cost) as well as environmentally (energy consumption/carbon footprints/greenhouse emissions).

## 1.2 Related Surveys and Our Work

A few works have conducted a survey on energy management techniques, but as the research has persistently grown in the field of cloud computing, there is a requirement for a systematic review to assess, update and combine the existing literature. Mastelic et al. [2] mainly focus only on energy consumption of datacenters. Shuja et al. [3] presented a survey of cloud datacenters only. Chaudhry and Ling [14] offered a survey of existing thermal-aware scheduling techniques developed for efficient management of green datacenters. Piraghaj et al. [81] reviewed energy-efficient resource management techniques at platform level. A broad review of energy-efficient datacenters is presented by Beloglazov et al. [135]. Moghaddam et al. [136] introduced a survey on energy-efficient networking solutions in cloud-based environments. Basmadjian et al. [47] reviewed energy-efficient techniques for cloud datacenters. Orgerie et al. [57] presented a survey on improving the energy-efficiency of large scale distributed systems. A broad review of energy efficiency in information systems at the hardware and at the software level is offered by Vitali and Pernici [58]. Shuja et al. [71] reviewed the techniques and architectures for designing energy-efficient datacenters and presented an overview of server cooling and storage component. Garg and Buyya [80] presented a survey on energy-efficient techniques for cloud datacenters for green cloud computing. A broad review of energy efficiency of the cooling and power supply subsystems is presented by Guitart [87]. Hameed et al. [46] reviewed the energy efficient resource allocation techniques for cloud computing systems. Shuja et al. [145] presented a review of green computing techniques among the emerging IT technologies such as Big data and IoT. Our survey is the first review paper which covers all the main characteristics (360-degree view holistically) of sustainable cloud computing. This research augments the previous surveys and presents a fresh systematic review to asses and identify the latest research issues. Table 1 shows the comparison of our survey with other survey articles based on different criteria.

**Our Focus:** Our systematic review focuses on the type and levels of consumption of energy used by cloud datacenters and identifies the reasons of large amount of energy consumption and suggests possible solutions to reduce carbon footprints and environmental problems in future. The components of CDCs such as networks, storage, memory and cooling systems are consuming huge amount of energy. To improve energy efficiency of CDC, there is a need to review energy-aware resource management techniques for management of all the resources (including servers, storage, memory, networks, and cooling systems) in a holistic manner and identifies the relationship of energy management with other related aspects of sustainable cloud computing such as application design, sustainability metrics, capacity planning, virtualization, thermal-aware scheduling, cooling management, renewable energy and waste heat utilization. The holistic management of cloud computing resources makes cloud services more energy-efficient and sustainable.

## 1.3 Our Contributions

- A comprehensive taxonomy for sustainable cloud computing is proposed.
- A broad review has been conducted to explore various existing techniques for sustainable cloud computing.
- The current research on sustainable cloud computing is organized into several categories such as application design, sustainability metrics, capacity planning, energy management, virtualization, thermal-aware scheduling, cooling management, renewable energy and waste heat utilization.

- Existing techniques have been compared and categorized based on the common characteristics and properties.
- A conceptual model for sustainable cloud computing has been proposed.
- The taxonomy and the survey results are used to find the open challenges that have not been fully explored in the research.

Table 1: Comparison of Our Survey with Other Survey Articles

| | **Criteria** | **1** | **2** | **3** | **4** | **5** | **6** | **7** | **8** | **9** | **10** | **11** | **12** | **13** | **14** | **15** |
|---|---|---|---|---|---|---|---|---|---|---|---|---|---|---|---|---|
| | Reviewed Upto | 2018 | 2014 | 2016 | 2014 | 2010 | 2013 | 2015 | 2014 | 2011 | 2012 | 2013 | 2011 | 2015 | 2012 | 2016 |
| Taxonomy | Application Design | ✓(*) | ✓ | ✓ | | ✓ | | | | | | | | | | |
| | Sustainability Metrics | ✓ | | | | | | | | | | | | | | |
| | Energy Management | ✓(*) | ✓ | ✓ | | ✓ | ✓ | | ✓ | ✓ | ✓ | ✓ | ✓ | ✓(+) | ✓ | ✓(+) |
| | Virtualization | ✓(*) | | ✓ | ✓ | ✓ | | ✓ | | | | | | | | |
| | Thermal-aware Scheduling | ✓ | | ✓ | ✓ | | | | | | | | | | | |
| | Capacity Planning | ✓ | | | | | | | | | | | | | | |
| | Cooling Management | ✓ | | | | | | ✓(+) | | | | ✓(+) | ✓(+) | ✓(+) | | |
| | Renewable Energy | ✓(*) | | | | | | ✓ | | | | | | | | |
| | Waste Heat Utilization | ✓ | ✓ | | | | | ✓ | | | | | | | | |
| Comparison | Based on Evolution and Focus of Study | ✓(*) | | | | | | | | | | | | | | |
| | Based on Taxonomy Mapping | ✓ | | | | | | | | ✓ | | | | | ✓ | |
| | Based on Objective | ✓ | | | ✓ | ✓ | ✓ | | | | | | | | ✓(+) | |
| | Based on Optimization Parameters | ✓ | | | | ✓ | ✓ | ✓ | ✓ | | | | | | ✓(+) | |
| | Based on Demerits (Open Issues) | ✓ | | | | | | | | | | | | | | |
| Future Research Directions for Every Technique# | | ✓(*) | | | | | | | | | | | | | | |
| Emerging Trends and their Impact | | ✓(*) | | | | | | | | | | | | | | |
| Proposed a Conceptual Model for Holistic Management | | ✓(*) | | | | | | | | | | | | ✓(+) | | |

1 - Our Survey (This Paper), 2- Mastelic et al. [2], 3- Piraghaj et al. [81], 4- Chaudhry and Ling [14], 5- Beloglazov et al. [135], 6- Moghaddam et al. [136], 7- Shuja et al. [3], 8-Basmadjian et al. [47], 9-Orgerie et al. [57], 10-Vitali and Pernici [58], 11-Shuja et al. [71], 12-Garg and Buyya [80], 13-Guitart [87], 14-Hameed et al. [46] and 15-Shuja et al. [145]. *Note:* (*: Means Comprehensive Discussion, +: Means Just an Overview and #: Means Summarized Open Challenges)

**1.4 Paper Structure**

The rest of the paper is organized as follows: Section 2 describes the review technique used to find and analyze the available existing research, research questions and searching criteria. Section 3 presents a proposed comprehensive taxonomy and systematic review of existing techniques for sustainable cloud computing. Further, based on the common characteristics and properties, techniques have been compared and categorized. Section 4 describes the result outcomes of this systematic review. Section 5 introduces the open challenges and future research directions along with implications of this research work in the area of sustainable cloud computing. Section 6 offers the conceptual model for sustainable cloud computing. Section 7 summarizes this research work.

**2. REVIEW METHODOLOGY**

This systematic literature review comprises of different stages, which includes formation of review framework, executing the survey, examining the review outcomes, management of review outcomes and investigation of open issues. The list of research questions is described in Table 2, which are used to plan the systematic review.

Table 2: Research Questions for different Categories

| Category | Research Questions |
|---|---|
| Application Design | 1. What is the current status of an application design?<br>2. What are the QoS requirement of different applications for sustainable cloud computing?<br>3. How to design an application architecture which can reduce coupling among different components of an application?<br>4. What is the need of green Information and Communications Technology (ICT)-based innovative applications?<br>5. What are the main challenges of an application design for recent technological developments such as IoT?<br>6. What are the different types of application design models for effective energy management?<br>7. How to decrease the execution cost and fulfill the deadline simultaneously? |
| Sustainability Metrics | 1. What are the different sustainability metrics for cloud datacenter?<br>2. What is the evolution of sustainability metrics?<br>3. How different sustainability metrics are related to each other?<br>4. How to measure the performance of CDC holistically? |
| Capacity Planning | 1. What are the conditions to change the SLA with respect to time?<br>2. What is the criteria for compensation and penalty if CDC service provider violates the SLA?<br>3. What are the different types of components for which capacity planning is required?<br>4. How to recognize and categorize the numerous workloads to design CDC successfully? |
| Energy Management | 1. How to reduce energy consumption and its impact on environment?<br>2. What is the difference between static and dynamic energy management techniques?<br>3. How to develop an energy-aware resource management technique that proficiently schedules the provisioned resources without SLA violation?<br>4. What are the configurable components for energy management?<br>5. What is the trade-off between energy consumption and execution time?<br>6. How much energy is consumed by various components (cooling, network, memory, storage and processor) of the idle server? |
| Virtualization | 1. What is the trade-off between time and energy cost for Virtual Machine (VM) migration?<br>2. What are different issues with VM migration in geographically distributed CDCs?<br>3. What type of technology is available for VM migration?<br>4. What is the optimization criteria for VM migration?<br>5. What are the different types of mechanisms for VM elasticity?<br>6. What is the difference between resource-aware and performance-aware in VM load balancing techniques?<br>7. What is the co-location criteria for VM consolidation?<br>8. What are the different types of recovery techniques for VM fault tolerance?<br>9. What are the different types of VM scheduling mechanisms? |
| Thermal-Aware Scheduling | 1. What is the different architectures for thermal-aware scheduling?<br>2. What are the different heat modeling techniques?<br>3. What different types of thermometers are using to measure temperature?<br>4. What are the different monitoring and awareness techniques?<br>5. What simulation tools are used to generate thermal-gadgets? |
| Cooling Management | 1. How to reduce the cooling cost without performance degradation?<br>2. What are the different types of cooling techniques?<br>3. What are the different temperature range for different mediums (water and air) in cooling plant?<br>4. What are the different types of heat rejection systems for cooling management? |
| Renewable Energy | 1. What are the main sources of renewable energy?<br>2. What are the different objectives of renewable energy-aware techniques?<br>3. What is the different storage devices to store renewable energy? |
| Waste Heat Utilization | 1. What are the different techniques for utilization of waste heat?<br>2. What is the different types heat transfer methods?<br>3. What are the different types of cooling methods for waste heat utilization? |

## 2.1. Sources of Information

We followed CRD guidelines [45] to perform electronic database search and manual search using different search strings as mentioned in Table 3, which retrieved 470 research articles. The following electronic databases have been used for searching:

- ACM Digital Library (<www.acm.org/dl>)
- HPC (<www.hpcsage.com>)
- Wiley Interscience (<www.Interscience.wiley.com>)
- Taylor & Francis Online (<www.tandfonline.com >)
- IEEE eXplore (<www.ieeexplore.ieee.org>)
- Google Scholar (<www.scholar.google.co.in>)
- ScienceDirect (<www.sciencedirect.com>)
- Springer (<www.springerlink.com>)

The review technique used in this systematic review is described in Figure 4.

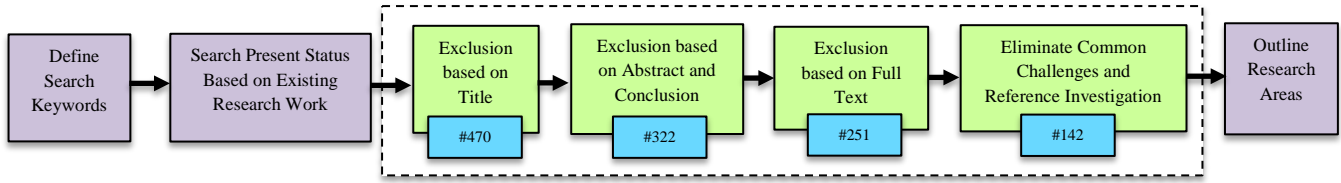

Figure 4: Review Technique used in this Systematic Review

### 2.2. Search Criteria

The keyword "sustainable cloud computing" is involved in the abstract of every research article in every search. Table 3 describes the different search strings used to conduct this detailed survey.

Table 3: Search String

| Sr. No. | Keywords | Synonyms | Dates | Content Type |
|---|---|---|---|---|
| 1 | Sustainability | Sustainable Cloud Computing, Sustainable Computing | 2010-2018 | Journal, Conference, Symposium, Workshop, Book Chapter, PhD Thesis Magazine, White paper and Transactions |
| 2 | Applications for Sustainable Cloud Computing | QoS, Models, Workloads, Architecture | | |
| 3 | Energy-aware Sustainable CDC | Static, Dynamic, Resource Management, Consolidation, Power, Configurable Components, Carbon Emission, DVFS, C-States | | |
| 4 | Capacity Planning for Sustainable CDC | IT Device, Cooling, Power Infrastructure, Autoscaling | | |
| 5 | Virtualization for Sustainable Computing | VM Migration, VM Technology, VM Elasticity, VM Load Balancing, VM Consolidation, VM Fault Tolerance, VM Scheduling, Checkpoint, Recovery, Failure Masking, Container Management | | |
| 8 | Metrics for Sustainable Cloud Computing | Sustainability Metrics, CDC Metrics, Sustainability Parameters | | |
| 9 | Thermal-aware Sustainable Cloud | Architecture, Heat Modeling, Thermometer, Scheduling, Monitoring and Awareness, Simulator | | |
| 10 | Cooling in Sustainable Cloud Computing | Cooling Plants, Temperature, Heat Rejection System, Location, Mechanical Equipment, Type of Cooling | | |
| 11 | Renewable energy and Sustainable Cloud Computing | Workload Scheduling, Energy Sources, Location of Sources, Storage Devices | | |
| 12 | Waste Heat Utilization for Sustainable Cloud Computing | Heat Model, Heat Transfer Method, Cooling Technique, Location-aware | | |
| 13 | Holistic Management for Cloud | Energy-efficiency, Cloud Resources, Sustainability, Sustainable Cloud Computing | | |

Our paper contains both qualitative and quantitative research papers from year 2010 to 2018. Literature reported that the basic research started in the area of sustainable cloud computing in 2008. This systematic review contains research work from magazines, white papers (technical reports as well as industry research work), workshops, symposiums, conferences and journals as described in Section 4 (see Table 26). To cross check the e-search, an individual search has been applied on some journals of Science Direct, Taylor & Francis, IOS Press, Wiley, ACM,

IEEE and Springer. Figure 4 shows the exclusion criteria used at different stages of this systematic review. Initially, 470 research papers were selected based on their titles, which were reduced to 322 based on their abstracts and conclusion and 251 research papers were retrieved based on their full text. Further, research papers with common challenges (based on exclusion and inclusion criterion) have been eliminated and the references of 251 research papers have been investigated thoroughly to identify a final set of 142 research papers.

**2.3. Quality Assessment**

To find suitable articles for this systematic review, the criterion of inclusion and exclusion is used to implement the quality assessment on the outstanding research papers. We have investigated several different conferences and journals related to sustainable cloud computing. Further, we used CRD guidelines given by [45] to explore the internal and external validation of results to find high-quality sustainable cloud computing research papers.

**2.4 Data Extraction**

Table 4 describes the data extraction guidelines, which were followed to include 142 research articles in this systematic review.

Table 4. Data Items Extracted from All Papers

| Data item | Description |
|---|---|
| Bibliographic data | Author, year, title, source of research paper. |
| Type of article | Conference, workshop, symposium, journal. |
| Study context | What are the research focus and its aim? |
| Study Plan | Classification of sustainable cloud computing techniques, evolution, taxonomy, comparison based on taxonomy. |
| What is the Sustainable Cloud Computing? | It explicitly refers to the sustainable cloud computing and their categories |
| How was comparison carried out? | Compare various traits such as objective, metrics, optimization parameter etc. |
| Data Collection | How the data of sustainable cloud computing was collected? |
| Data analysis | How to analyze data and extracted research challenges? |
| Simulation tool | It refers to tool used for validation. |
| Research challenges | Open challenges in the area of sustainable cloud computing. |

Further, research questions have been designed for different categories of sustainable cloud computing. When systematic review commenced, we faced a number of difficulties such as extraction of suitable data. To find out the in-depth knowledge of research work (142 papers), we have contacted various authors. The data extraction procedure used in this systematic review is described below:

- After in-depth review, first author extracted data from 142 research papers.
- Second author cross checked the review results using random samples.
- The compromised meeting was called to resolve the conflict during cross checking.

**2.5 Acronyms**

A glossary of important acronyms used in this systematic review can be found in Table 5. *Note:* The abbreviations for different techniques are mentioned in their corresponding category.

**3. SUSTAINABLE CLOUD COMPUTING: A TAXONOMY**

The ever-increasing demand of cloud computing services that are deployed across multiple cloud datacenters harness significant amount of power, resulting in not only high operational cost, but also high carbon emissions. In sustainable cloud computing, the CDCs are powered by renewable energy resources by replacing the conventional fossil fuel based grid electricity or brown energy to effectively reduce the carbon emissions [2]. Employing energy efficiency mechanisms also makes cloud computing sustainable by reducing carbon footprints to a great extent [144]. Waste heat utilization from heat dissipated through servers and employing mechanisms for free cooling of the servers make the CDCs sustainable [3]. Thus, sustainable cloud computing covers the following elements in making the datacenter sustainable [4]: i) using renewable energy instead of grid energy generated from fossil fuels ii) utilizing the waste heat generated from heat dissipating servers iii) using free cooling mechanisms and iv) using energy efficient mechanisms [145]. All of these factors contribute in reducing the carbon footprints, the operational cost, and the energy consumption to make CDCs more sustainable [146].

Table 5: List of Important Acronyms

| Acronym | Definition |
|---|---|
| PDU | Power Distribution Unit |
| QoS | Quality of Service |
| SLA | Service Level Agreement |
| CDC | Cloud Datacenters |
| IoT | Internet of Things |
| TWh | Terawatt Hours |
| CUE | Carbon Usage Efficiency |
| PUE | Power Usage Efficiency |
| ICT | Information and Communications Technology |
| DVFS | Dynamic Voltage and Frequency Scaling |
| HPC | High-Performance Computing |
| HTC | High-Throughput Computing |
| DVS | Dynamic Voltage Scaling () |
| DFS | Dynamic Frequency Scaling |
| RC | Resistor–Capacitor |
| CFD | Computational Fluid Dynamics |
| VM | Virtual Machine |
| OSs | Operating Systems |
| LAN | Local Area Network |
| WAN | Wide Area Network |
| CPU | Central Processing Units |
| IT | Information Technology |
| HFC | Hydrogen Fuel Cells |
| CoP | Coefficient of Performance |
| SaaS | Software as a Service |
| PaaS | Platform as a Service |
| IaaS | Infrastructure as a Service |
| PMU | Power Management Unit |
| DRAM | Dynamic Random-Access Memory |
| ATS | Automatic Transfer Switch |

Figure 5 describes various elements that impacts or supports sustainable cloud computing (360 Degree View), which have been categorized into nine categories: application design, sustainability metrics, capacity planning, energy management, virtualization, thermal-aware scheduling, cooling management, renewable energy and waste heat utilization based on the existing literature. Table 6 describes mapping of aspects of sustainable CDC to types of sustainability spheres based on Figure 3 and Figure 5.

### 3.1 Application Design

In sustainable cloud computing, the design of an application plays a vital role and the efficient structure of an application can improve energy efficiency of CDCs. Resource manager and scheduler follows different approach for application modelling [25] [26]. For example, scheduling algorithm for Map Reduce model follows different approach compared to other models like workflow, web application, streaming application, graph processing etc. To make the infrastructure sustainable and environment eco-friendly, there is a need of green ICT-based innovative applications [142]. Effective design of cloud applications contains APIs or services. Applications (e.g. web applications) designed following three-tiered architectures contain user interfaces, application processing and database [27] [42]. The functionality of each tier should be independent to run at different providers to improve its performance, simplicity and reliability. The components of applications should have minimum dependency i.e. loosely coupled. Applications can be ported from one server to another server without affecting its execution [43] [44]. At software level, cloud user can utilize application in a flexible manner, which are running on cloud datacenters.

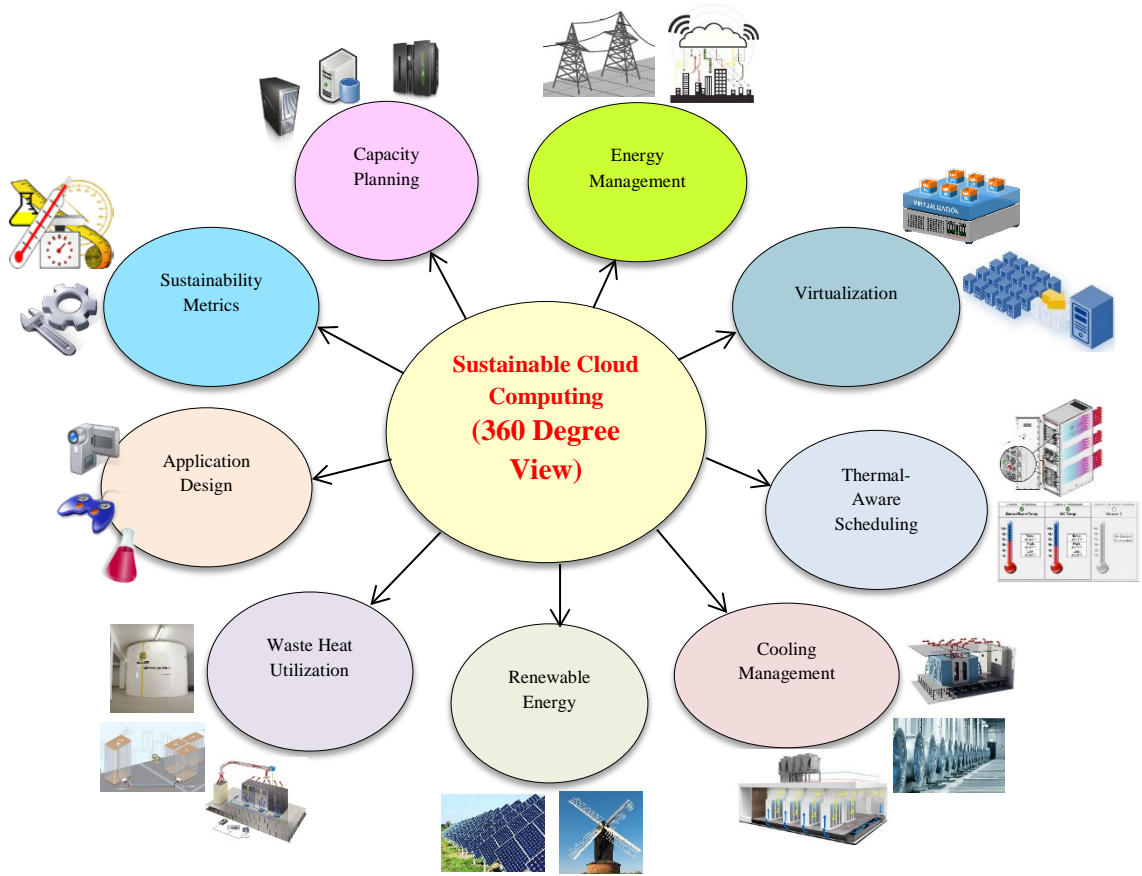

Figure 5: Taxonomy of Sustainable Cloud Computing (360-Degree View)

Recent technological developments such as IoT and software defined clouds based applications are creating new research areas for sustainable cloud computing [1]. The emerging IoT based applications such as smart cities, healthcare services etc. are increasing, which needs appropriate application design model for fast data processing that improves the performance of computing systems [3]. However, these applications are facing large delay and response time because computing systems need to transfer data to the cloud and then cloud to an application, which affect its sustainability of cloud computing [2].

Table 6: Mapping of Aspects of Sustainable CDC to Types of Sustainability Spheres

| Aspects of Sustainable CDC | Types of Sustainability Spheres | | | |
|---|---|---|---|---|
| | Economic | Environmental | Social | Technical |
| Application Design | ✓ | | ✓ | ✓ |
| Sustainability Metrics | | | | ✓ |
| Capacity Planning | ✓ | ✓ | | ✓ |
| Energy Management | ✓ | | ✓ | |
| Virtualization | ✓ | | ✓ | |
| Thermal-Aware Scheduling | | ✓ | | |
| Cooling Management | ✓ | ✓ | | |
| Renewable Energy | | ✓ | | |
| Waste Heat Utilization | ✓ | ✓ | ✓ | |

Due to a large amount of data processing at the cloud, computing system does not process at the required speed which leads to communication failures. Moreover, data security is also a high priority requirement of sustainable computing to protect a critical information from attackers in case of e-commerce applications. There is a need of re-evaluation of existing application models of cloud computing to address research issues such as energy efficiency, sustainability, privacy, security and reliability.

### 3.1.1 Related Studies

Bifulco et al. [20] discussed the concept of ICT-based smart sustainable cities to decrease energy consumption while providing services to households. In this research work, Bag of Tasks (BoT) application model driven Component Based Lifecycle (CBL) is proposed to design an application, which estimates the energy consumption for household activities. Further, it is recommended that a smart sustainable city can be designed by optimizing the power using Dynamic Voltage and Frequency Scaling (DVFS) technique, which can save power while devices are idle. Moreover, natural resources can be managed easily by locating datacenters at proper places, so that it has minimum impact on environment. Cappiello et al. [26] proposed a Green Computing Based Model (GCBM) using BoT application model, which distribute the user data into interrelated tasks and improves the utilization of resources and reduces energy consumption and $CO_2$ emissions during the deployment of applications. Park et al. [42] developed Cloud Based Clustering Simulator (CBCS) for desktop resource virtualization to choose cluster for efficient and sustainable computing. It uses BoT application model to design an application for selection of energy-efficient cluster. But utilization of resources is done only based on network infrastructure and time without considering storage, memory and CPU.

Bossche et al. [25] designed a Data Mining Based Architecture (DMBA) based application, which uses Map-Reduce model to extract the useful information from unstructured data by eliminating useless data. Further, it helps to form sustainable clusters by reducing the execution time of processing. Fu et al. [78] designed and implemented the Agricultural Information Service (AIS) application to manage the data regarding fresh products in knowledge base and enable communication among different components using Map-Reduce model. Further, Hadoop cloud computing platform is used to analyze the application of agriculture for economic cooperation and energy efficiency. Bradley et al. [24] proposed an IoT based Application Design (IoT-AD) using graph processing model to reduce the cost of data management. Further, a machine learning method is used to generate sustainable value, which improves environmental sustainability and energy efficiency. NoviFlow [35] designed a Green-Software Defined Network (G-SDN) based application using graph processing model to provide sustainable solutions, which further reduces the carbon emissions for making network infrastructure efficient. Green-SDN based network reduces energy consumption of CDCs, which improves sustainability. NoviFlow [35] reported that existing models are mainly focused on cost and quality of service to deliver sustainable services.

Pesch et al. [38] proposed a task-based Thermal-Aware Scheduling (TAS) architecture, which schedule resources to execute the user applications by optimizing energy utilization to improve sustainability of CDCs. TAS saves the energy consumption up to 40%. Juarez et al. [138] proposed a Dynamic Energy-Aware Scheduling (DEAS) technique for execution of task-based applications parallelly to estimate the energy consumption. To provide energy-efficient and sustainable cloud service, DEAS technique generates appropriate energy consumption profile automatically. Further, a trade-off between performance and energy savings is presented. Gill et al. [56] proposed an IoT based Agriculture Service (Agri-Info) model to manage agriculture data in an efficient manner, which is coming from different preconfigured devices. Moreover, Agri-Info uses task-based application model to design an application and simulated cloud environment is used to validate the Agri-Info model in terms of energy-efficiency and other QoS parameters. Dabbagh et al. [137] designed Energy Efficient Technique (EET) based application using thread model to reduce monthly expenses of datacenters by delaying the non-urgent workloads, which also decreases the execution of workload. Further, efficient peak control policy has been designed to predict the demands of datacenters such as storage, power, memory etc. for coming requests and real traces of a Google datacenter is used to validate EET. Garg et al. [54] proposed Environment Conscious Application Scheduling (ECAS) framework to design a thread-based application model for execution of High Performance Computing (HPC) applications on cloud resources. Further, performance parameters such as cost, carbon emission rate and CPU power efficiency have been optimized during execution of applications.

Gmach et al. [44] proposed a stream processing model based Power Profiling Technique (PPT) for datacenters to improve the energy utilization and its effect on environment and both demand and supply of power should be co-managed to reduce consumption of water and greenhouse gas emissions. Based on the availability of power, user data is processed without compromising with the quality of service and energy efficiency, which provides sustainability. Park and Cho [79] proposed a Mobile Building Information Modeling (BIM) Platform (MBIMP) based tracking system using stream processing based application model, which processes the user data using different streams to improve energy efficiency. Based on virtual BIM view and communication, MBIMP system tracks the required real-time information and improves the coordination among different components during the data extraction. Charr et al. [29] proposed an Online Frequency Selecting (OFS) algorithm for heterogenous environment, which uses DVFS for message passing iterative applications to reduce energy consumption. Figure 6 shows the evolution of application design techniques along with their Focus of Study (FoS) across the various years.

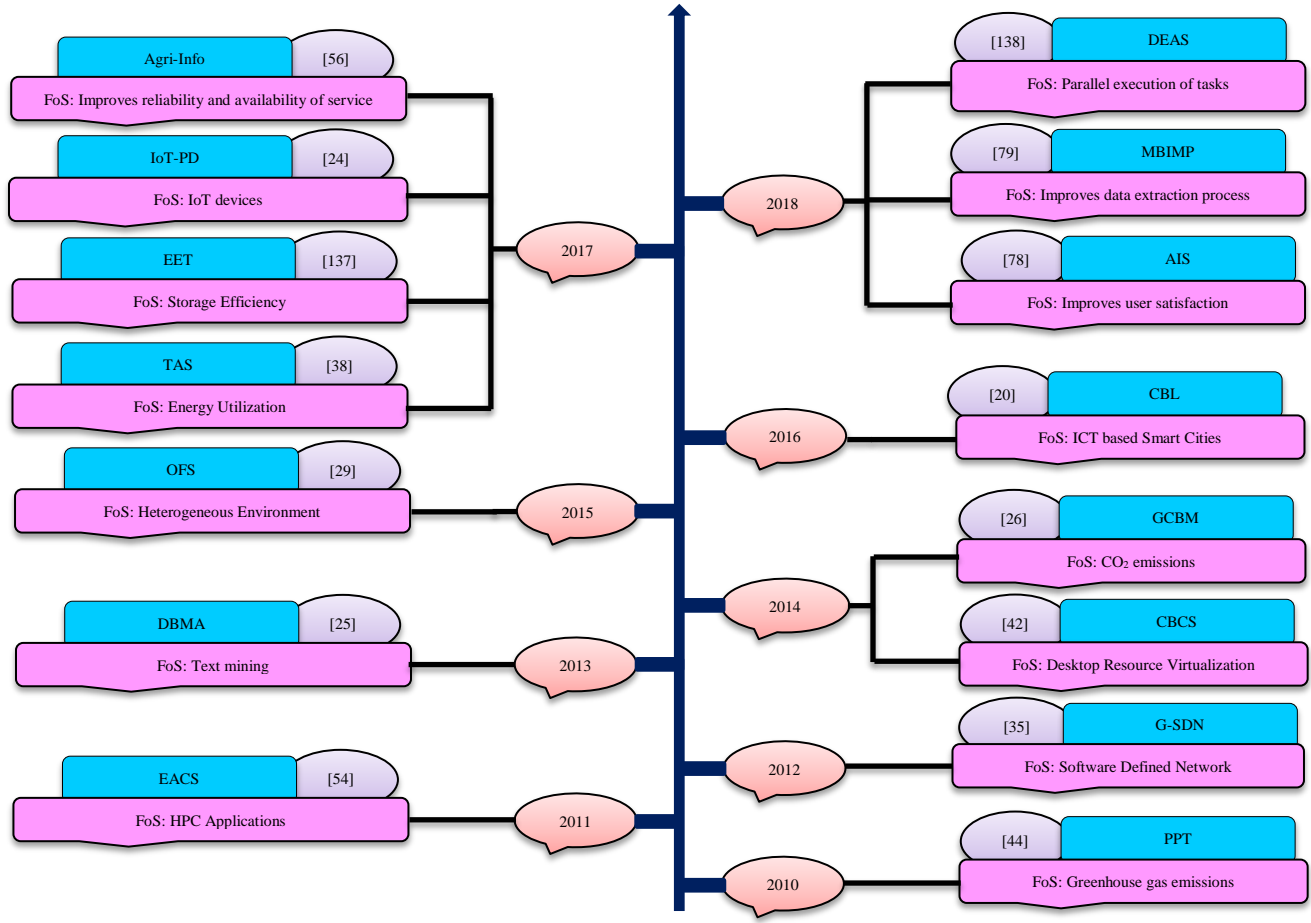

Figure 6: Evolution of Application Design Techniques

A summary of these related techniques (application design) and their comparison based on objective, optimization parameter and metric along with open research challenges is given in Table 7.

**3.1.2 Application Design based Taxonomy**

Different types of application are running in cloud environments such as computation intensive or data-intensive. To improve the performance of cloud computing systems, it is mandatory to execute applications parallelly. Based on the requirements of the cloud user, QoS parameters for every application are identified and provisions the resources for execution. The components of the application design taxonomy are: i) QoS parameter, ii) application model, iii) workload type and iv) type of architecture as shown in Figure 7. Each of these taxonomy elements are discussed below along with relevant examples. The comparison of existing techniques based on our application design taxonomy is given in Table 8.

*3.1.2.1 Quality of Service (QoS):* Different application have their different QoS requirements. There are five main types of QoS parameters for sustainable computing as identified from literature [20] [33] [39] [58] [59], such as execution cost, time, energy consumption, security and throughput. *Execution cost* is total money that can be spent in one hour to execute the application successfully. *Execution time* is the amount of time required to execute application successfully.

Table 7: Comparison of Existing Techniques (Application Design) and Open Research Challenges

| Technique | Organization | Objective | Metric | Optimization Parameter | Citations | Open Research Challenges |
|---|---|---|---|---|---|---|
| OFS [29] | University of Franche-Comté, France | Reduce energy consumption | Energy | Energy consumption | 7 | 1. Does not include power consumed by memory and network resources 2. Secure communication is required |
| DEAS [138] | Workflows and Distributed Computing Group, Spain | Improve energy-efficiency | Energy-efficiency | Energy consumption | 10 | Trade-off between energy saving and reliability is an open research area |
| MBIMP [79] | Georgia Institute of Technology, USA | Improve real-time information tracking | Correctness | Data accuracy | 1 | High response time and low workload utilization level |
| AIS [78] | Huazhong Agricultural University, China | Improve throughput | User satisfaction | Memory Usage | - | Security and privacy of cloud service is an open issue |
| Agri-Info [56] | The University of Melbourne, Australia | Improve throughput | Throughput, Response time, latency and cost | Execution time, cost, latency and security | 6 | Need autonomic management of cloud resources. |
| CBL [20] | University of Naples, Italy | Reduce energy consumption | Security and Energy | Energy consumption | 38 | Power is required to be managed automatically |
| IoT-AD [24] | University of Kentucky, USA | Reduce cost | Cost | Execution cost | 3 | Secure communication is required |
| DMBA [25] | University of Antwerp, Belgium | Reduce execution time | Response Time | Execution time | 128 | Longer Response time and no QoS guarantee |
| GCBM [26] | Polytechnic University of Milan, Italy | Improve throughput | Energy and Cost | Energy Utilization | 9 | Need to improve reliability of applications |
| OFS [29] | Huazhong University of Science and Technology, China | Reduce energy consumption | Cost | Execution cost | 18 | Communication complexity |
| G-SDN [35] | NoviFlow Inc, Canada | Reduce execution time | Energy | Energy utilization | - | No cost optimization involved |
| TAS [38] | Cork Institute of Technology, Ireland | | Correctness | Data Accuracy | 1 | Not feasible for non-time-sensitive systems |
| CBCS [42] | Seoul National University of Science and Technology, Korea | Improve throughput | Availability | Availability | 12 | Difficult to obtain optimal scheduling decisions with dynamic workloads |
| PPT [44] | HP Lab, USA | Reduce energy consumption | Energy | Energy consumption | 43 | Suitable for homogeneous CDCs only |
| EET [137] | Oregon State University, USA | Reduce cost | | | 1 | Under-utilization of resources |
| ECAS [54] | The University of Melbourne, Australia | Improve CPU power efficiency | Application transaction rate | Response time | 314 | Lacks applicability in a virtualized cloud environment |

*Energy* is the amount of electricity expended by resource to complete the execution of an application. *Security* is an ability of the computing system to protect the system from malicious attacks. *Throughput* is the ratio of total number of tasks of an application to the amount of time required to execute the tasks. Other QoS requirements of cloud service can be reliability, availability, scalability and latency [4].

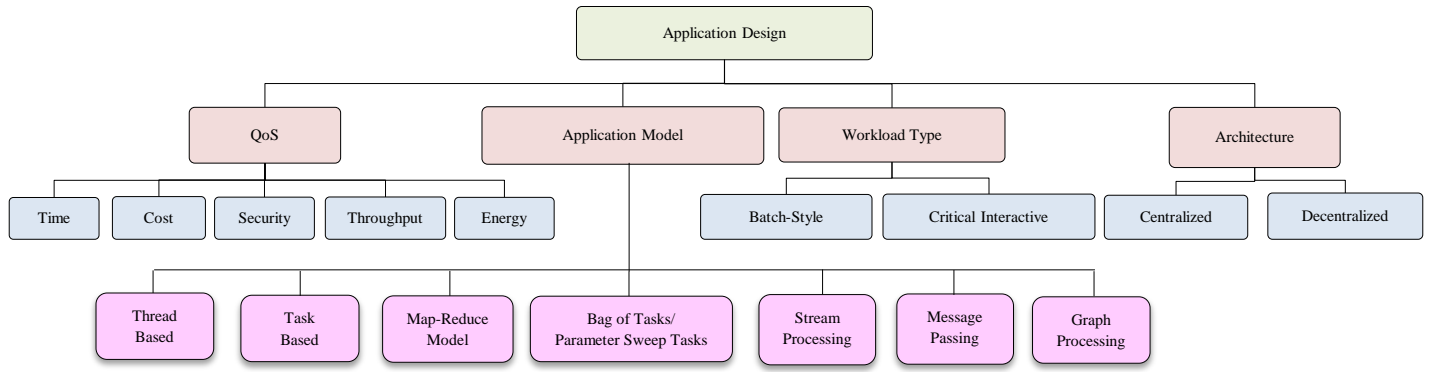

Figure 7: Taxonomy Based on Application Design

***3.1.2.2 Application Models:*** The complexity of applications is increasing day by day and the cloud platform can be used to handle user applications. The different types of application models are developed for wide range of domains to satisfy the different types of customers for sustainable computing [25] [42-44] [125]. There are seven types of *application models* as identified from literature [47] [48]: 1) thread-based, 2) task-based, 3) Map-Reduce model, 4) bag of tasks or parameter sweep tasks, 5) stream processing, 6) message passing and 7) graph processing. In *thread-based* model, one process is divided into multiple threads, which executes concurrently and sharing resources such as memory, network, processor to complete execution. In *task-based* model, a large task is divided into small tasks and execute them parallelly on different non-sharable cloud resources. *Map-reduce tasks* split the input data-set into independent chunks and parallel way of execution, which is used to execute the mapped tasks. Further, the outputs of the maps are sorted and used as an input to the reduce tasks. *Bag-of-tasks* or *parameter sweep tasks* refers to the jobs that are parallel among which there are no dependencies and are identical in their nature and differ only by the specific parameters used to execute them, for example: video coding and encoding. *Stream processing* is a processing of small-sized data (in Kilobytes) generated continuously by thousands of data sources (geospatial services, social networks, mobile or web applications, online gaming and video streaming), which typically send data records simultaneously and an example of stream processing model can be video processing application. *Message Passing* interface provides a communication functionality between a set of processes, which are mapped to nodes or servers in a language-independent way and it encouraged development of portable and scalable large-scale parallel applications. *Graph processing* involves the process of analyzing, storing and processing graphs to produce effective outputs.

***3.1.2.3 Workload Types:*** For workload management in sustainable cloud computing, there are mainly two types of IT workloads that are considered for sustainable computing: batch-style and critical interactive [32] [33]. *Batch style* workloads are those workloads which are submitted to a job queue, and it will be executed when resources become available. Multiple batch jobs are often submitted without any deadline constraint together and are executed with maximum resource utilization. The workloads that need immediate response but its execution should be completed before their deadline are called *critical interactive* workloads.

***3.1.2.4 Architecture:*** The architecture is an important component of sustainable cloud computing and there are basically two types of architectures: centralized and decentralized [46] [55] [58] [137]. In *centralized* architectures, there is a central controller, which manages all the tasks that are required to be executed, and further it executes the task using scheduled resources. The central controller is responsible for the execution of all tasks. In *decentralized* architectures, resources are allocated independently to execute the tasks without any mutual coordination. Every resource is responsible for their own task execution.

The performance of QoS parameters of different cloud applications is measured using different metrics as discussed in *Section 3.2*.

Table 8: Comparison of Existing Techniques based on Taxonomy of Application Design

| Technique | Author | Application Model | QoS Parameter | Workload Type | Architecture |
|---|---|---|---|---|---|
| OFS | Charr et al. [29] | Message Passing Interface | Energy | Critical Interactive | Centralized |
| DEAS | Juarez et al. [138] | Task-based | Energy | Batch Style | Centralized |
| MBIMP | Park and Cho [79] | Stream Processing | Cost | Batch Style | Centralized |
| AIS | Fu et al. [78] | Map-Reduce | Time | Batch Style | Decentralized |
| Agri-Info | Gill et al. [56] | Task-based | Execution time, cost and security | Critical Interactive | Decentralized |
| IoT-AD | Bradley et al. [24] | Graph Processing | Cost | Critical Interactive | Centralized |
| EET | Dabbagh et al. [137] | Thread based | Cost | Batch Style | Decentralized |
| TAS | Pesch et al. [38] | Task-based | Energy | Critical Interactive | Centralized |
| CBL | Bifulco et al. [20] | Bag-of-Tasks or Parameter Sweep Tasks | Security | Batch Style | Centralized |
| SIA | Xia et al. [29] | Message Passing | Energy | Critical Interactive | Decentralized |
| GCBM | Cappiello et al. [26] | Bag-of-Tasks or Parameter Sweep Tasks | Time | Critical Interactive | Centralized |
| CBCS | Park et al. [42] | Bag-of-Tasks or Parameter Sweep Tasks | Throughput | Batch Style | Decentralized |
| DMBA | Bossche et al. [25] | Map-Reduce Tasks | Time | Batch Style | Centralized |
| G-SDN | NoviFlow [35] | Graph Processing | Energy | Batch Style | Decentralized |
| ECAS | Garg et al. [54] | Thread-based | Cost and energy | Batch Style | Decentralized |
| PPT | Gmach et al. [44] | Stream Processing | Energy | Critical Interactive | Centralized |

## 3.2 Sustainability Metrics

As use of cloud infrastructure is growing exponentially, it is important to monitor and measure the performance of cloud data center regularly. We have identified different types of metrics from the literature [14] [22] [23] [28] [30] [32] [34] [36] [51] [52] [60] [67] [83] [85] [86] [125] and presents a taxonomy of metrics for different categories for sustainable cloud computing based on the core operations of CDC. Figure 8 shows the taxonomy of metrics for application design, capacity planning, energy management, virtualization, thermal-aware scheduling, cooling management, renewable energy and waste heat utilization.

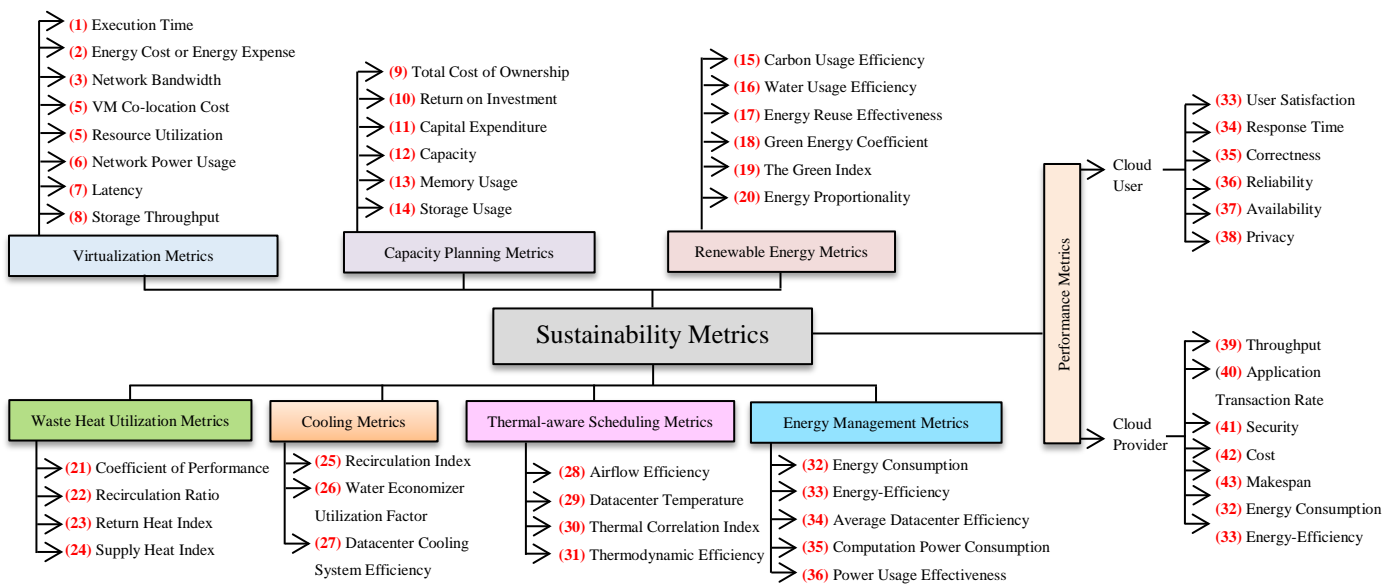

Figure 8: Taxonomy of Metrics for Sustainable Cloud Computing

Table 9 shows the year-wise use of sustainability metrics in different categories of sustainable cloud computing to measure the performance of numerous infrastructure components of CDC.

Table 9: Year-wise use of Sustainability Metrics in different categories

| Year | Sustainability Metrics |
|---|---|
| 2010 | (32) |
| 2011 | (40) |
| 2012 | (11) (20) (32) (36) |
| 2013 | (3) (10) (15) (22) (30) (32) (34) (42) |
| 2014 | (5) (9) (18) (24) (29) (35) (41) |
| 2015 | (2) (14) (16) (21) (27) (28) (33) (39) |
| 2016 | (1) (4) (13) (17) (22) (26) (31) (32) (34) (43) |
| 2017 | (6) (8) (12) (15) (19) (23) (27) (30) (36) (38) |
| 2018 | (7) (11) (18) (21) (25) (26) (27) (28) (32) (35) (36) (37) |

Table 10 presents the brief definition of sustainability metrics as described in Figure 7. The detailed description of metrics for sustainable cloud computing can be found in [36].

Table 10: Sustainability Metrics and their Definition

| Metric | Definition |
|---|---|
| Execution Time | Execution time is the amount of time required to execute application successfully. |
| Energy Cost | It is combination of monetary and non-monetary costs related to generation, transmission and usage of energy. |
| Network Bandwidth | It is the number of bits transferred/received in one second. |
| VM Co-Location Cost | It is the total cost of VM migration from one cloud datacenter to another. |
| Resource Utilization | It is a ratio of execution time of a workload executed by a particular resource to total uptime of that resource. |
| Network Power Usage | It is the amount of electricity usage on the network. |
| Latency | It is the delay before the transfer of user request for processing. |
| Storage Throughput | It is the amount of time taken for storage system to execute required operation per second. |
| Total Cost of Ownership | It is the addition of direct cost (purchase cost of CDC) and indirect cost (operational cost of CDC) |
| Return on Investment | It is the ratio of net profit to the cost of investment for CDC. |
| Capital Expenditure | It is the amount of money, which is used to obtain, upgrade, and maintain physical components related to CDC. |
| Capacity | It is the capability of an application to execute user tasks using available resources such as power infrastructure, IT devices etc. |
| Memory Usage | It is the total usage of main memory (RAM) to execute user tasks. |
| Storage Usage | It is the total usage of secondary memory (hard disk) to execute user tasks. |
| Carbon Usage Efficiency | It is a ratio of total CO2 emissions produced by total CDC energy to energy of IT equipment. |
| Water Usage Efficiency | It is a ratio of annual water usage to energy of IT equipment. |
| Energy Reuse Effectiveness | It is the ratio of energy (reused) consumed by Cooling, Lighting and IT devices to the total energy consumed by IT devices. |
| Green Energy Coefficient | It is the ratio of green energy consumed by CDC to total energy consumption of CDC. |
| The Green Index | It is used to measure the economic growth of company with the environmental consequences of that growth. |
| Energy Proportionality | It is the relationship between power consumed and resource utilization. |
| Coefficient of Performance | It is calculated by dividing quantity of removed heat to total work done for removal of heat. |
| Recirculation Ratio | It is the amount of waste-water that flows through the advanced pretreatment component divided by the amount of waste-water that is sent to the final treatment and dispersal component. |
| Return Heat Index | It is the measure of net level of recirculation air in data center. |
| Supply Heat Index | It is the measure of suppling heat from outside for recirculation. |
| Recirculation Index | It is used to measure the amount of water saved while circulating the hot water to the water heating system. |
| Water Economizer Utilization Factor | It is the percentage hours in a year that the water side economizer system is utilized to provide the required cooling to the CDC. |
| Datacenter Cooling System Efficiency | It is the amount of cooling capacity per unit of energy it consumes to maintain the working of CDC. |
| Airflow Efficiency | It is the amount of airflow a ceiling fan can produce per minute. |
| Datacenter Temperature | It is the operating temperature of CDC. |
| Thermal Correlation Index | It is used to measure the relation between the heat flux and the thermodynamic driving force for the flow of heat. |
| Thermodynamic Efficiency | It is the amount of heat utilized by heat utilization system based on the amount of heat received. |
| Energy Consumption | It is the amount of electricity expended by resource to complete the execution of an application. |
| Energy-efficiency | It is a ratio of number of workloads executed to total energy consumed by CDC to execute those workloads. |
| Average Datacenter Efficiency | It is the ratio of IT equipment power by total facility power. |
| Computation Power Consumption | It is the amount of energy consumed during the execution of compute-intensive tasks. |
| Power Usage Effectiveness | It is the ratio of the energy consumed by ICT devices to the energy consumed by all the devices including ICT and cooling devices. |
| User Satisfaction | It is used to measure how cloud services of cloud provider fulfills user QoS requirements. |
| Response Time | It is the length of time taken for a system to react to a user request |
| Correctness | It is the degree to which the cloud service will be provided accurately to the cloud customers. |
| Reliability | It is the capability of an application to sustain and produce correct results in case of occurrence of faults such as network, hardware or software related faults. |
| Availability | It is the amount of time (hours) a specific application will be available for use per day. |
| Privacy | It is a parameter through which user and provider can store their information privately using authorization and authentication. |
| Throughput | Throughput is the ratio of total number of tasks of an application to the amount of time required to execute the tasks. |
| Application Transaction Rate | It is a ratio of number of applications coming for processing per unit time. |
| Security | Security is an ability of the computing system to protect the system from malicious attacks. |
| Cost | It is total money that can be spent in one hour to execute the application successfully. |
| Makespan | It is the time difference between the start and finish of a sequence of tasks of an application. |

Effective capacity planning in the cloud era demands some resource flexibility due to changing application requirements and hosting infrastructure, which is discussed in *Section 3.3*.

### 3.3 Capacity Planning

Cloud service providers must involve an effective and organized capacity planning to enable sustainable computing. The capacity planning can be done for power infrastructure, IT devices and cooling. The capacity of cloud datacenter can be planned effectively by considering the devices of end users, for example, encoding techniques for video on-demand application [109]. SLA should be there for important parameters such as backup and recovery, storage and availability to improve user satisfaction, which attracts more customers in future. There is a need to consider important utilization parameters per application to maximize the utilization of resources through virtualization by finding the applications, which can be merged. Merging of applications improves resource utilization and reduces capacity cost, which makes cloud infrastructure more sustainable. For efficient capacity planning, cloud workloads should be analyzed before execution to finish its execution for deadline-oriented workloads [11]. To manage power infrastructure effectively, VM migration should be provided for migration of workloads or machines to successfully complete the execution of workloads with minimum usage of resources, which improves the energy efficiency of cloud datacenters. An effective capacity planning can truly enable the sustainable cloud environment.

#### 3.3.1 Related Studies

Ghosh et al. [109] proposed a Stochastic Model based Capacity Planning (SMCP) technique for virtual infrastructure to serve the user requests and execute their workloads within specified deadline and budget. To improve the capacity of CDC, an optical network is deployed for enabling sustainability in CDCs to save energy consumption. Kouki et al. [110] designed a SLA-aware Technique (SLAT) for improving capacity planning for cloud based applications and their QoS requirements. Further, the trade-off between user satisfaction and profit has been identified using a queueing network to plan the configuration of the required cloud datacenter autonomously. Jaing et al. [111] designed Cloud Analytics for Immediate Provisioning (CAIP) of virtual machines and planning capacity for cloud datacenters. Predication error is calculated based on asymmetric and heterogeneous metrics and this error value is used for effective capacity planning, which improves the quality of cloud services and reduces the emissions of carbon dioxide. This technique uses IBM data traces and is effective in reducing time of VM provisioning and reducing overhead to improve energy efficiency. Sousa et al. [112] aimed to develop an approach for Capacity Planning for Cloud Infrastructure (CPCI) by considering two important parameters such as cost and dependability. A stochastic model generator is used to identify the dependability of servers while developing the cost-effective capacity plan. Further, Moodle [3] hosted on a Eucalyptus-based environment to test CPCI, which shows that CPCI is cost and energy efficient capacity planning technique.

Kong et al. [113] proposed a methodology for Selection of Optimal Energy Source (SOES) for effective capacity planning to design an energy-efficient green datacenter. Further, this research focuses on: i) criteria to choose energy source, ii) plan the capacity for cloud infrastructures with minimum energy cost and carbon emissions and iii) find and optimize the requirements of datacenters to improve service availability. Further, SOES is effective in reducing lifetime total cost, which includes operational costs as well as capital costs. Carvalho et al. [114] proposed a Capacity Planning Framework (CPF) for cloud market using different price schema, such as on-demand, reservation and spot, to serve user requests. CPF helps to identify the sustainable price schema required to execute current workloads. It has been concluded that spot price schema [139] is cost effective without degrading the quality of cloud service. Menasce et al. [115] performed experiments to test the capacity planning model [113] using three different cloud providers such as Microsoft Azure, Google's App Engine and Amazon EC2. Further, SLA is defined based on four QoS parameters [140] such as throughput, response time, availability and cost. Dorsch et al. [116] investigated the correlation between environmental sustainability and economic efficiency to enable exchange of extra capacity between different cloud infrastructures to execute cloud workloads. This method can improve resource utilization and reduces cost of energy and response time. Triantafyllidis et al. [149] developed an Integrated Optimization Platform (IOP) to evaluate the effectiveness of capacity planning for infrastructure and resources. The cost-optimal solutions are identified using a mixed-integer linear programming in IOP based on the optimization of objective functions. The performance of proposed platform is evaluated with respect to energy, system cost, network topology and emission flow. Figure 9 shows the evolution of capacity planning techniques along with their Focus of Study (FoS) across the various years.

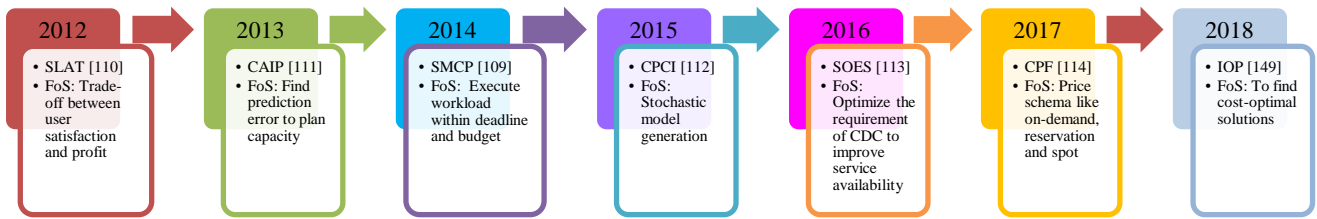

Figure 9: Evolution of Capacity Planning Techniques

A summary of these related techniques (capacity planning) and their comparison based on objective, optimization parameter and metric along with open research challenges is given in Table 11.

**3.3.2 Capacity Planning based Taxonomy**

Capacity planning is done based on: i) component, ii) IT workload, iii) model, iv) autoscaling and v) utility function as shown in Figure 10. Each of these taxonomy elements are discussed below along with relevant examples. The comparison of existing techniques based on our capacity planning taxonomy is given in Table 12.

***3.3.2.1 Component:*** Capacity planning is required for every *component* of a cloud datacenter such as IT devices, cooling and power infrastructures [109] [110]. *IT devices* are an important component, which are required to execute the operations of CDCs. Due to consumption of huge amount of energy, an efficient planning of *cooling* is required to maintain the temperature of cloud datacenter. The planning of *power infrastructure* is the most important element of a CDC to run it every time i.e. *24 X 7*.

Table 11: Comparison of Existing Techniques (Capacity Planning) and Open Research Challenges

| Technique | Organization | Objective | Metric | Optimization Parameter | Citations | Open Research Challenges |
|---|---|---|---|---|---|---|
| SMCP [109] | Duke University, USA | Improve energy efficiency | Capital Expenditure | Energy consumption | 35 | Combining of applications improves resource utilization and reduces capacity cost. Cloud workloads should be analyzed before execution to finish its execution on time because some workloads are deadline-oriented. |
| SLAT [110] | INRIA, France | Reduce SLA violation | Total Cost of Ownership | Energy cost | 20 | VM migration should be provided for migration of workloads or machines to successfully complete the execution of workloads with minimum usage of resources which improves the energy efficiency of cloud datacenters. |
| CAIP [111] | Florida International University, Florida | Reduce $CO_2$ emissions | Return on Investment | Data Center Cost | 51 | There is a need of an effective data management policy to store data effectively at lower cost and data can accessed easily for modification, deletion etc. |
| CPCI [112] | Federal Rural University, Brazil | Identify the dependability of server | Capacity | Cost per dependability | 19 | Configuration of cloud should be examined for future execution of applications with or without migration of data. |
| SOES [113] | McGill University, Canada | Design energy-efficient green CDC | Storage Usage | Storage cost | 10 | There is a requirement of strong plan in case of disaster management, so that data can be recovered successfully at later stage with minimum cost. |
| CPF [114] | Federal University of Campina Grande, Brazil | Improve quality of service | Memory Usage | Memory Cost | 1 | User requirements should be considered clearly while preparing its technical design to achieve maximum user satisfaction. |
| IOP [149] | Imperial College London, UK | Improve resilience to deal with climate change and poverty | Return on Investment and Capital Expenditure | Energy, system cost, network topology and emission flow | 2 | IOP can be extended to support uncertainty in the modelling formulation for other parameters such as magnitude of demand in resources or capital and operational expenditure. |

***3.3.2.2 IT Workload:*** There are mainly two types of *IT workloads,* which are considered for capacity planning: batch-style and critical interactive as described in *Section 3.1.2.3*.

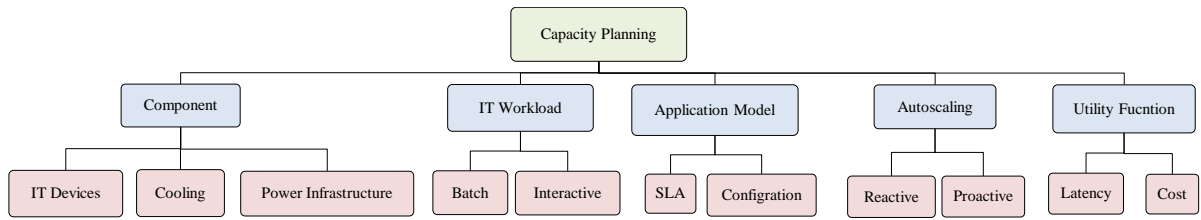

Figure 10: Taxonomy Based on Capacity Planning

***3.3.2.3 Application Models:*** There are two types of design *models* for effective capacity planning: SLA based and Configuration based [111] [114]. In the *SLA*-based model, capacity of CDCs is planned based on the QoS requirements of the workloads without SLA violations but *configuration*-based model focuses on the configuration of CDC such as processor, memory, network devices, cooling and storage, which are required to execute the workloads effectively.

***3.3.2.4 Autoscaling:*** The capacity of a CDC is also planned for *autoscaling*, which may be proactive or reactive [109] [113]. *Reactive* autoscaling works based on feedback methods and manages the requirements based on their current state to maintain its performance. *Proactive* autoscaling manages the capacity requirements based on the prediction and assessment of performance in terms of QoS values. Based on previous data, predictions have been identified and plan their required action to optimize CDC performance.

***3.3.2.5 Utility functions:*** Latency and cost based *utility functions* are defined to measure the aspects of capacity planning [113] [114]. *Cost* is defined as the amount of money that can be spent to design a CDC with required configuration. *Latency* is the amount of execution delay with a particular configuration of CDC.

Table 12: Comparison of Existing Techniques Based on Taxonomy of Capacity Planning

| Technique | Author | Component | IT Workload | Application Model | Autoscaling | Utility Function |
|---|---|---|---|---|---|---|
| IOP | Triantafyllidis et al. [149] | Power Infrastructure and cooling | Batch style | Configuration | Proactive | Cost (Energy) |
| CPF | Carvalho et al. [114] | Power Infrastructure | Interactive critical | SLA | Proactive | Cost |
| SOES | Kong et al. [113] | IT device | Interactive critical | SLA | Proactive | Latency |
| CPCI | Sousa et al. [112] | Power Infrastructure | Batch style | Configuration | Proactive | Latency |
| SMCP | Ghosh et al. [109] | IT device | Batch style | SLA | Reactive | Latency |
| CAIP | Jaing et al. [111] | Power Infrastructure | Batch style | Configuration | Reactive | Cost |
| SLAT | Kouki et al. [110] | cooling | Interactive critical | SLA | Proactive | Cost |

To manage power infrastructure for capacity planning, energy management of resources is required for execution of workloads, which improves the energy efficiency of cloud datacenters.

### 3.4 Energy Management

Energy management in sustainable computing is an important issue for cloud service providers. Ficco et al. [5] reported that more than 2.4% of electricity is consumed by CDCs with a large economic impact of $30 billion globally. The energy requirement to manage the CDCs is also rising in proportion to the operational cost. IBM spends 45% of total expenses on electricity bills of CDCs and the consumption of electricity will be increased to 101.5 Billion Kilo Watt Hour (kWh) by 2022 [10]. Sustainable cloud services are attracting more cloud customers and making it more profitable [62] [60]. Improving energy utilization, which reduces electricity bills and operational costs to enable sustainable cloud computing. The important requirements of sustainable cloud datacenters are optimal software system design, optimized air ventilation and installing temperature monitoring tools for adequate resource utilization, which improves energy efficiency [67] [74]. There are mainly three levels where energy consumption can be optimized such as software level (efficient use of registers, buffers etc.), hardware level (transistors, voltage supply, logical gates and clock frequency) and intermediate level (energy-aware resource provisioning techniques) [80].

### 3.4.1 Related Studies

Domdouzis [15] studied the existing technologies of green computing and its effects on sustainable cloud computing. Further, QoS-based energy-aware scheduling techniques have been discussed and identified the impact of cloud infrastructure on environment. Moreover, technologies related to sustainable cloud computing such as web services, capacity planning and application design have been discussed. Authors [5] stated that sustainable cloud computing optimizes the performance of different applications such as education, healthcare, agriculture etc. in terms of power consumption. Abbasi [16] studied existing literature of sustainable cloud computing and stated that sustainability and energy efficiency of CDCs is ensured to reduce the impact on environment. To solve this problem, an architecture has been proposed for management of workloads through multi-layer server which consists of upper and lower level. The lower level focuses on the management of energy aspects such as energy consumption, cooling power efficiency and server consolidation, while the upper level focuses on carbon footprints as well as utility cost [32]. The proposed architecture reduced carbon footprints as well as energy costs [5]. This research work only considers homogeneous datacenters without considering the important QoS parameters such as time, throughput, security etc.

Accenture [17] suggested the ecological advantages of moving to the cloud in terms of performance and efficiency of service, which can improve reliability and sustainability of services. Further, this research identified important factors to reduce energy consumption as well as carbon emissions such as i) decrease power consumption by efficient cooling management (datacenter management), ii) improving utilization of servers by avoiding underutilization and overutilization of resources (server utilization), iii) executing a large number of requests using shared infrastructures (multi-tenancy) and iv) efficient matching of workload and resource for execution (dynamic provisioning). Moreover, business community can be benefited through cloud computing due to large number of users of cloud services including mobile applications, online gaming, social media and email. So, there is a need to make cloud services more energy-efficient and sustainable, which can fulfil user demands timely without affecting the environment [6]. Further, this study concluded that the amount of carbon footprint is largely dependent on deployment size. Microsoft deployed 60 percent large-sized deployments, 30 percent medium-sized deployments and 10 percent small-sized deployments [7].

Gill et al. [60] developed a Particle Swarm Optimization (PSO) based resource scheduling algorithm (BULLET) for effective management of energy consumption. Initially, user workloads have been classified based on QoS requirements for provisioning of resources. Further, this PSO-based scheduling algorithm is used for scheduling provisioned resources and workloads are executed successfully, using Dynamic Voltage Scaling (DVS). The proposed algorithm is effective in improving energy utilization, time and cost of resources in datacenters and enables sustainable computing. Brown et al. [61] proposed a Method for the Assessment and Analysis (MAA) of sustainability for energy-efficient cloud services. MAA has been tested using the case study of different multi-story buildings in Sweden. Further, Swedish environmental rating tool has been utilized to calculate the life-cycle cost. It has been concluded that 25% more cost is required to develop energy-efficient and sustainable cloud environment, which further improves the indoor environmental quality. Hsu et al. [62] proposed Energy Efficient and Sustainable Model (EESM) for management of transport system, which: i) provides an effective management of data (video and audio) using available storage, ii) improves the energy utilization and reliability of services and iii) identifies effective route and status based on the analysis of traffic conditions such as traffic lights, traffic flow etc. EESM helps to reduce energy consumption and fuel consumption, which enables sustainable eco-driving. Uddin et al. [64] investigated that large amount of energy that is required to run CDCs in an efficient manner and identified the issues like backup and recovery, low carbon emissions and huge energy consumption, which effects global warming. Further, a framework is proposed to decrease carbon emissions by dividing datacenters into different resource pools for efficient management of energy consumption and green metric i.e. PUE [68], which is used to measure the effectiveness of a datacenter. Singh et al. [108] analyzed the QoS-aware autonomic computing systems and identified that energy is one of the important parameters, which requires optimization to enable sustainable cloud computing.

Giacobbe et al. [65] reviewed the cost-effective energy-aware resource management technique in cloud federation and investigated that cost saving is an important factor for sustainable computing. Most of the energy-saving techniques are based on cloud federation, and utilization of resources in datacenters is also affected by large energy consumption. It has been reported that the dynamic migration of computational resources reduce energy consumption during the interaction of IoT devices [73]. Kramers et al. [66] proposed the genetic algorithm based Energy-Aware Resource Allocation (EARA) technique for allocation of virtual machines on heterogenous CDCs. Further, a new metric named POWERMARK has been designed to find the energy efficiency of CDCs and reduce VM co-location cost and bandwidth cost to fulfil cloud user requirements. Gill et al. [67] proposed a QoS-aware scalable resource management policy (CHOPPER) for effective scheduling of resources by considering the self-optimization for improving energy utilization, self-protection against cyber-attacks, self-configuration of energy-efficient cloud resources and self-healing by managing unexpected failures. Further, energy consumption of resources has been reduced using an autonomic system to execute workloads in sustainable datacenters. Chen et al. [69] investigated the trade-off between throughput and energy in cloud-based radio access networks. In this network, renewable energy is used to provide the power to multiple base stations to exchange information between senders and receivers. It has been concluded that energy consumption can be reduced by selecting optimal paths to transfer data in the network. Further, data is managed efficiently at sender as well as receiver side using cloud repository, which improves the throughput and sustainability of datacenters [70]. Singh et al. [83] developed an energy-aware resource scheduling technique (EARTH) through resource consolidation, which considers fuzzy logic to take scheduling decisions for the execution of cloud workloads. Dynamic Frequency Scaling (DFS) based proposed technique [77] is effective in improving energy efficiency and resource utilization due to the timely decisions of resource scheduler.

Dandres et al. [74] proposed an Energy based Resource Management (ERM) approach to reduce greenhouse emissions of datacenters and it has been identified that server load migrations affect the energy consumption of datacenters. Xu et al. [75] identified that DVFS manages the datacenter effectively but these techniques failed in case of overloaded data. Authors used Brownout to find out the overloads effectively and manages the load by deactivating the free resources to save energy, which improves sustainability of CDCs. Wang et al. [76] recognized that data coming towards a cloud in a short time are difficult to manage and this issue is solved by incorporating multiple mobile sinks for data gathering. Further, a time adaptive schedule policy is used to reduce latency triggered by arbitrary allocation of tasks. The optimization of latency and the effective gathering of data reduce energy consumption, which creates sustainable cloud. Garg et al. [80] proposed the Green Cloud Framework (GCF) to minimize carbon footprints and to measure the impact of carbon emissions to the environment. In this research work, it is mentioned that there is a need of holistic management of energy to make CDCs more sustainable by minimizing their overall power consumption. Mardani et al. [82] investigated multi-objective based decision-making techniques for sustainable and renewable energy to identify the parameters (social, political, environmental, technical and economical), which are affecting environment. Further, it has been suggested that energy should be managed efficiently at different levels such as social, political, environmental, technical and economical to make clouds more sustainable in the future. Singh et al. [125] proposed a SLA-aware resource allocation mechanism (STAR) for efficient management of resources by considering scalable components such as processor, storage and memory. Further, this technique uses DVFS for power management of cloud resources in sustainable CDCs. Experimental results prove that STAR is effective in decreasing energy consumption of resources, which can effectively provide the sustainable cloud environment, without violation of SLA. Battistelli et al. [41] proposed a Cloud based Energy-aware Service Automation (CESA) technique for sustainable cloud datacenter to reduce energy consumption during the execution of cloud resources. Authors used MAPE-k loop to offer automation of cloud service and using an efficient communication path to transfer data from source to destination with the maximum value of energy-efficiency. The experimental results show that CESA technique is effective in providing the required cloud service with maximum value of energy-efficiency. Figure 11 shows the evolution of energy management techniques along with their Focus of Study (FoS) across the various years.

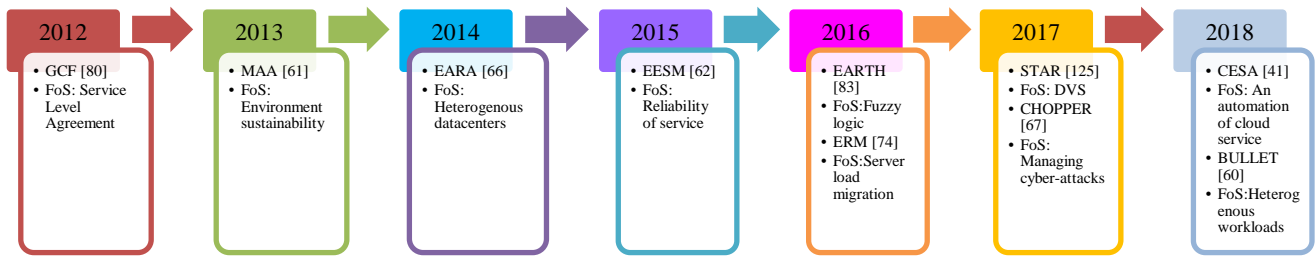

Figure 11: Evolution of Energy Management Techniques

A summary of these related techniques (Energy Management) and their comparison based on objective, optimization parameter and metric along with open research challenges is given in Table 13.

Table 13: Comparison of Existing Techniques (Energy Management) and Open Research Challenges

| Technique | Organization | Objective | Optimization Parameter | Metric | Citations | Open Research Challenges |
|---|---|---|---|---|---|---|
| CESA [41] | RWTH Aachen University, Germany | Improve energy efficiency | Energy consumption | Energy Consumption | 1 | Trade-off between energy consumption and reliability is an open research challenge. |
| BULLET [60] | Thapar University, India | Reduce energy consumption | Execution time | Energy Consumption | 3 | Resource utilization of CDCs is affected during workload execution |
| MAA [61] | Royal Institute of Technology, Sweden | Improve energy utilization | Energy cost | Energy Consumption | 53 | Switching of resources between high scaling and low scaling modes increases response time |
| EESM [62] | Yuan Ze University, Taiwan | Improve energy efficiency | Energy consumption | Energy-efficiency | 34 | Cannot handle the dynamic nature of tasks and SLA violation can be reduced. |
| EARA [66] | Royal Institute of Technology, Sweden | Improve energy efficiency | VM co-location cost and bandwidth | Computation Power Consumption | 16 | Putting servers in sleeping mode or turning on/off servers affects the reliability of the storage component. |
| CHOPPER [67] | Thapar University, India | Improve energy utilization | Execution time and cost | Energy-efficiency | 6 | Resources are reserved in advance, but resource requirement is less than resources available, which increases cost |
| EARTH [83] | Thapar University, India | Improve resource utilization | Energy | Energy Consumption | 25 | The large amount of clock speed is wasted while waiting for the data because of speed gap between processor and main memory |
| ERM [74] | University of Montreal, Canada | Reduce SLA violation | Energy consumption | Average Datacenter Efficiency | 1 | Energy consumption can be saved by reducing the processor frequency through the manipulation of supply voltage |
| GCF [80] | The University of Melbourne, Australia | Reduce carbon footprints | Power cost | Power Usage Efficiency | 133 | A large number of workloads are waiting for execution due to unavailability of sufficient amount of resources |
| STAR [125] | Thapar University, India | Reduce greenhouse emissions | Execution cost | Power Usage Efficiency | 17 | Switching of resources between high scaling and low scaling modes increases service delay |

### 3.4.2 Energy Management based Taxonomy

Energy management has two important components (static and dynamic) as shown in Figure 12. Each of these taxonomy elements are discussed below along with relevant examples. The comparison of existing techniques based on our energy management taxonomy is given in Table 14.

***3.4.2.1 Static Energy Management:*** Static energy management is more engineering oriented approach, in which circuitry systems are considering more by offline energy management [60]. During design time, the whole optimization happens at system level and it deals with factorization, path balancing, transistor sizing, instruction sets, redesigning of architectures, circuit manipulation and processing centres [67].

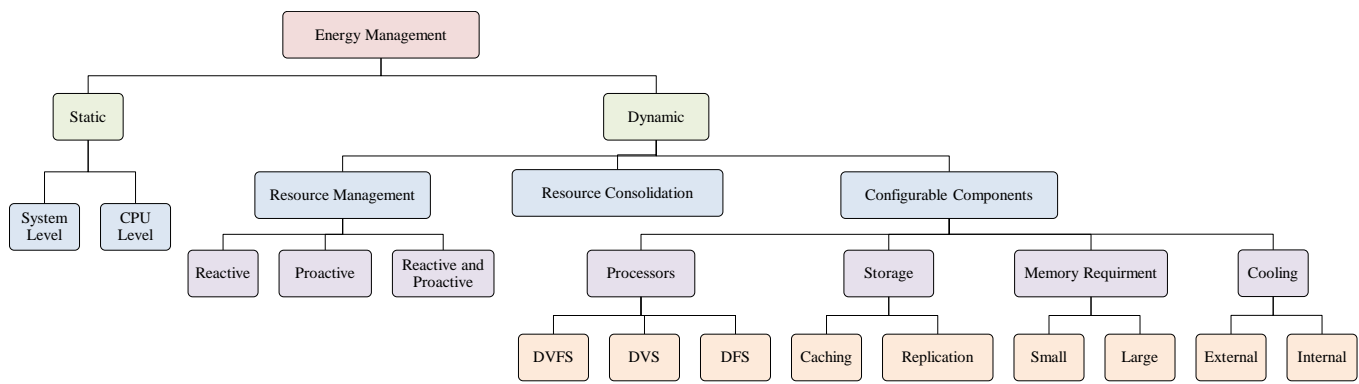

Figure 12: Taxonomy Based on Energy Management

The low power usage components are used in this management to consume energy consumption as minimum as possible. Static energy management performs at two levels: System level and CPU level. Existing studies [125] [83] found that *CPU* offers big scope of optimization of energy consumption because computing components of the CPU consumes 35-40% of energy [3] [71]. The optimization of energy at *CPU level* can be performed at instruction set level or register level. Researchers designed different number of instruction set architectures to improve resource utilization like reduced bit-width architecture at instruction set level. On the other hand, the activities of register transfer level are optimizing to decrease energy consumption. Energy cost and $CO_2$ emission for static and dynamic energy management techniques [122] [144] [145] is shown in Figure 13.

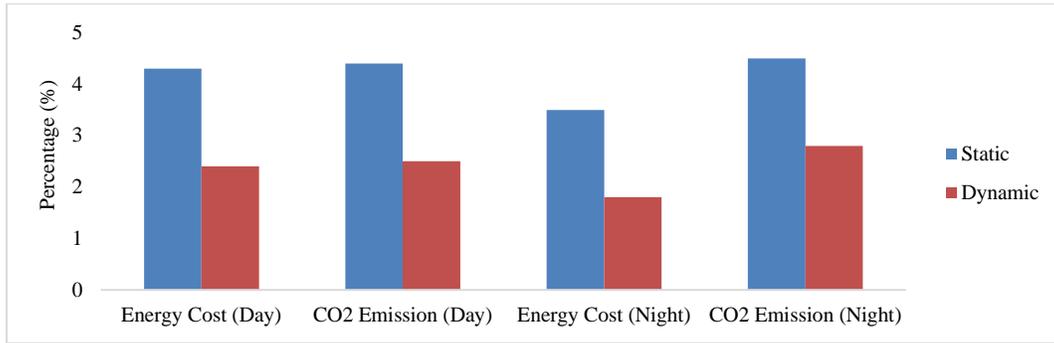

Figure 13: Energy Cost and $CO_2$ Emission for Static and Dynamic Energy Management Techniques (Data Source: [122] [144] [145])

It is clearly shown that static energy management techniques consume more energy as well as produce more CO2 emissions as compared to dynamic energy management techniques. Other components of *system* along with CPU are software systems, network facility and memory components that consume large amounts of energy [143]. Researchers proposed different management techniques to optimize the power consumption of these components based on the setup techniques used to design a system. At design time, it is very difficult to select the right components to design a cloud system with maximum synchronization among the components [61] [62] [66]. Other important challenges during system design can be: i) type of application and software, ii) selection of operating system and iii) placement of severs to reduce delay. Gordan and Fast Array of Wimpy Nodes (FAWN) [3] architecture has been designed to improve the performance of cloud systems by balancing the input-output activities and computation processes by coupling datacenter powering systems and local flash storage with low power CPUs. Energy consumption can be reduced by proper distribution of resources geographically and the selection of suitable network topologies and components with maximum compatibility.

In order to save energy when the CPU is idle, the CPU can be transferred into a low-power mode, which is called C-states or C-modes [61] [62]. The main motive of C-states is to cut the clock signal and power from idle components inside the Central Processing Unit (CPU). The more units stop (by cutting the clock), the voltage is needed or even CPU becomes completely shut down, the more energy can save, but more time is required for the CPU to "wake up" and again be 100% operational [66]. There are main seven cases: i) C0 mode (Operating State) - CPU is fully turned on and it consumes the maximum amount of energy, ii) C1 mode (Halt) - CPU main internal clocks are stopped via software and it consumes less amount of energy as compared to C0 mode but needs little time to make CPU 100% operational, iii) C2 mode (Stop Grant and Clock) - CPU main internal and external clocks are stopped via hardware and it consumes less amount of energy as compared to C1 mode but needs more time to make CPU 100% operational as compared to C1 mode, iv) C3 mode (Sleep) - Stops all internal and external clocks of the CPU, v) C4 mode (Deeper Sleep) - Stops all internal clocks of the CPU and Decreases voltage of CPU, vi) C5 mode (Enhanced Deeper Sleep) - Decreases voltage of the CPU even more and put off the memory cache and vii) C6 mode (Deep Power Down) - Decreases the internal voltage of the CPU to any value, including 0V and this state consumes the minimum amount of energy (depends on the voltage value), but this state needs the maximum time to make CPU 100% operational.

*3.4.2.2 Dynamic Energy Management:* Software based policies are used by dynamic energy management to improve energy utilization. There is a different dynamic power range for every component. During low activity modes, CPU consumes 30% of the peak value of its energy consumption and it can be scaled up and down upto 70% [80]. Dynamic range of energy consumption for disk drives is 50%, for memory is 25% and for network devices such as routers and switches is 15% [83]. To improve energy utilization, the number of components can be scaled up or down based on the range of dynamic power. Dynamic energy management is divided into three categories based on the reduction of the dynamic power range: i) configurable components, ii) resource consolidation, and iii) resource management.

*Configurable components* are those components like *CPU,* which supports low activity modes at component level. Dynamic energy management can be used to control CPU. CPU is the main source of energy consumption. So, existing research work mainly focused on optimization of energy consumption in CPU or processor and memory. There is a relationship between power supply, voltage and operational frequency [66] [62]: ($Power_{Dynamic} = Utilization_{CPU} \times$ Frequency $\times Voltage^2$ ). Based on the different values of voltage and operational frequency, CPU can run in different activity modes or C-modes in advance processor architectures. As supply voltage increases, the energy consumption increases quadratically in CMOS (Complementary Metal Oxide Semiconductor) circuits [3]. The values of linear relations can be exploited by changing operation frequency (DFS), voltage (DVS) or both simultaneously (DVFS) [60].

DVFS is an energy optimization power management technique, which is basically the adjustment of frequency settings of the computing devices in order to optimize the resource allotment for tasks, and if resources are not required then DVFS minimizes the CPU frequency thus by supplying lower voltage to CPU, which maximize the power savings [60]. DVFS technique is used for virtual machines hosted by physical machines along with the algorithm or scheduling mechanism to reduce the energy. DVFS system and workload planning can be joined in two different manners: (1) workload scheduling, and (2) slack reclamation. In the schedule generation, DVFS-empowered processors are used to (re)schedule tasks of the graph in a worldwide cost function including both makespan (execution time) and energy saving to meet both time and energy limitations in the meantime [3]. In slack reclamation, which fills in as post-preparing method on the yield of planning calculations, DVFS mechanism is utilized to limit the power utilization of undertakings in a timetable created by a different scheduler [25]. The current strategies in light of DVFS strategy, have two noteworthy inadequacies: (1) a large portion of them centre around schedule age and do not adopt enough the *slack reclamation* strategies into record to spare extra energy, and (2) the current slack reclamation techniques utilize just a single frequency for every task among all distinct arrangement of processor's frequencies. Utilizing one frequency generally brings about revealed slack time where processor and different devices just waste energy [66]. DVFS based energy management techniques reduce energy consumption, but response time and service delay are increased due to the switching of resources between high scaling and low scaling modes.

There are number of methods proposed to control energy consumption by scaling down the high voltage supply, but the best way is to exploit the *stall time*. The large amount of clock speed is wasted while waiting for the data because of the speed gap between processor and main memory. Energy may be saved by reducing the processor frequency through manipulation of supply voltage. For different devices, semiconductor chip vendors optimizing energy consumption use different frequency scaling policies. Eight different kinds of operational frequencies are available in Intel's Woodcrest Xeon Processor [3]. Two CPU throttling technologies developed by AMD are PowerNow and CoolnQuiet [3] [125]. Another, SpeedStep CPU throttling technology is developed by Intel to control energy consumption [62]. The *cooling* can be *internal* (fans) or *external* (as discussed in Section 3.7) for a cloud datacenter.

The management of *storage* devices like disk drives is handled by scalable storage systems to reduce energy consumption because disk drives consume significant amounts of energy. The storage of data can be managed using either *replication* or *caching*. Mechanical operations of storage components consume one third of the total electricity provided to the CDCs, and disks also consume one tenth during standby mode. The need of storage components is increasing by 60% annually [66] [67], so this is a serious research issue to control energy consumption. Disk drive utilizes only 25% of their storage space and it remains underutilized in large CDCs [3] [71]. The power usage can be minimized by reducing underutilization by switching-off the unnecessary disks. Large number of mechanisms are proposed to improve the energy efficiency of disk drives [2]. In large scale CDCs, *memory* component may be considered to decrease power usage, but it is a least addressed component by researchers. Memory consumes 23% of energy consumption to run specific workload [83] [125]. The dynamic range for memories is 50% as discussed above, so there is a chance to improve energy consumption in this component [61] [62]. DVFS is also applicable to memory components by reducing frequency and voltage. Storage arrays are the most important components of DRAMs in which power consumption can be reduced. It is challenging to develop energy-aware memory components in cloud computing to reduce power consumption without degradation of performance. Also, it is difficult to manufacture energy-efficient memory devices with lesser cost. Existing memory management infrastructures can minimize energy consumption up to 70% [6].

*Resource consolidation* is a technique for effective utilization of resources (processor, memory or network devices) to minimize the number of resources and the location of servers, which a cloud company requires to serve user requests [71]. Resource scheduler allocates resources to execute workloads dynamically to avoid over-utilization and under-utilization of resources.

*Resource management* is an important challenge because: i) heterogenous resources, ii) priced differently, iii) applications with varying requirements (compute, data, network, memory) and iv) user QoS requirements. An effective resource management includes resource allocation, resource scheduling and resource monitoring to effective utilization of resources [32]. There are many questions that are required to be answered [32] [71]:

a) How to allocate the resources in an energy-efficient manner for the execution of workloads?
b) When to migrate workloads from one machine to another to save energy consumption?
c) Which devices need to be switched off to save energy consumption without degradation of performance?

Based on existing research, the number of existing techniques above discussed issues to improve energy utilization. These techniques are classified into following categories: a) Proactive, b) Reactive and c) Proactive and Reactive.

*Proactive* management manages the resources based on the future prediction of the performance of the system instead of its current state. The resources are selected based on the previous executions of the system in terms of reliability, energy consumption, throughput etc. The predictions are required to be identified based on previous data, and plan their appropriate action to optimize energy consumption during resource execution. *Reactive* management works based on feedback methods and manages the resources based on their current state to optimize energy. There is a need of continuous monitoring of resource allocation to find whether the energy is consumed less than its threshold value or not (threshold value can be based on energy as well as resource utilization). If power usage is higher than threshold value then corrective action will be taken to optimize the energy consumption. The accuracy of the monitoring module improves the productivity of reactive management. In case of underutilization of resources, energy consumption can be reduced through VM consolidation or migration as discussed in *Section 3.5*. Increase in energy consumption also requires effective cooling management because temperature is increasing due to large amounts of heat. *Reactive and proactive* management manages the resources with minimum value of power usage and maximum value of resource utilization to handle every situation by i) monitoring the resource execution continuously and ii) performing the actions based on predicted failures. In real time, it is challenging to accurately forecast the behavior of a system in proactive management. In reactive management, there is a larger overhead which causes unnecessary delay as well as energy inefficiency [60].

Table 14: Comparison of Existing Techniques Based on Taxonomy of Energy Management

| Technique | Author | Resource Management | Processor | Storage | Memory Requirement | Resource Consolidation | Cooling |
|---|---|---|---|---|---|---|---|
| CESA | Battistelli et al. [41] | Proactive | DVFS | Caching | Large | Yes | Internal |
| BULLET | Gill et al. [60] | Proactive | DVS | Caching | Large | Yes | Internal |
| CHOPPER | Gill et al. [67] | Proactive and Reactive | DVFS | Caching | Large | No | Internal |
| STAR | Singh et al. [125] | Reactive | DVS | Caching | Large | Yes | Internal |
| EARTH | Singh et al. [83] | Reactive | DFS | Replication | Large | Yes | Internal |
| ERM | Dandres et al. [74] | Proactive | DFS | Replication | Small | No | Internal |
| EESM | Hsu et al. [62] | Reactive | DFS | Replication | Small | Yes | Internal |
| EARA | Kramers et al. [66] | Reactive | DVS | Caching | Large | No | Internal and External |
| MAA | Brown et al. [61] | Proactive | DVS | Replication | Small | No | Internal |
| GCF | Garg et al. [80] | Reactive | DVFS | Caching | Small | No | Internal and External |

A virtualization technology reduces the number of physical machines or resources and executes the workloads using virtual resources, which leads to a reduction in energy consumption.

## 3.5 Virtualization

Virtualization technology is an important part of sustainable cloud datacenters to support energy-efficient VM migration, VM elasticity, VM load balancing, VM consolidation, VM fault tolerance and VM scheduling [88]. Operational costs can be reduced by using VM scheduling to manage cloud resources using efficient dynamic provisioning of resources [102]. During the execution of workloads, VM load balancing is required to balance the load effectively due to decentralized CDCs and renewable energy resources. Due to the lack of on-site renewable energy, VM techniques migrate the workloads to the other machines distributed geographically. VM technologies also offer migration of workloads from renewable energy based CDCs to the CDCs utilizing the waste heat at another site [105]. To balance the workload demand and renewable energy, VM based workload migration and VM consolidation techniques provide virtual resources using few physical servers. VM fault tolerance creates and maintains the identical secondary VM for the replacement of Primary VM in failover situation without affecting the availability of cloud service. VM elasticity maintains the performance of the computing system by providing the dynamic adaptation of computing resources or capacity to fulfill the changing requirements of workloads. Waste heat utilization and renewable energy resources alternatives are harnessed by VM migration techniques to enable sustainable cloud computing [104]. It is a great challenge for VM migration techniques to improve energy savings and network delays while migrating workloads between resources distributed geographically.

### 3.5.1 Related Studies

VM migration is the process of migrating VMs (where a VM is the software implementation of the computer that runs a number of applications and operating system) to another physical server without interrupting the running application operation. A virtualization takes place when the server is underutilized or over-utilized or in case of temperature overpass, which improves energy efficiency of CDC. Also, the virtualization reduces the carbon emission by moving the workload to location having renewable energy supply. Wang et al. [99] proposed Green-aware VM migration (GVM) policy for the efficient utilization of energy coming from grid sources and optimizing the power consumption of cooling and IT devices. Further, a statistical searching approach is used to find out the destination of migration and post-copy technique is used in GVM. Wang et al. [88] proposed an Extended version of GVM (E-GVM) by replacing grid energy with renewable energy to make datacenters sustainable, which further reduces power cost and carbon emissions. Further, VM consolidation-based resource allocation is improved using IT-enabled virtualization [100], which measures the variation of energy requirements during pre-copy and post-copy of data from one server to another. Bolla et al. [103] analyzed the migration time of live migration of a number of VMs between physical machines. The migration time is amount of data copied from one physical machine to another. Further, it also estimates the interference effects during live migration. Kernel-based Virtual Machine (KVM) is used to test the proposed technique to prove its effectiveness to calculate the migration time more accurately. Khosravi et al. [106] developed a technique to improve the usage of renewable energy for Online Virtual Machine Migration (OVMM), from one physical server to another, to enable sustainable cloud computing. In this research work, an optimal offline algorithm and an online algorithm are proposed for VM migration [141]. The offline algorithm is effectively working when future information of renewable energy is known a priori and the online algorithm is used when future information is unavailable.

Ranjbari and Torkestani [159] proposed a Learning Automata-based VM Consolidation (LAVMC) algorithm to reduce energy consumption and SLA violation rate. Further, LAVMC algorithm decreases the number of migrations, and shuts down idle servers to decrease the energy consumption of the CDC. Similarly, Ashraf and Porres [160] proposed an Ant Colony Optimization (ACO) based VM Consolidation (ACOVMC) technique to reduce the number of VM migrations. Dabbagh et al. [101] proposed the VM Predication and Migration (VMPM) mechanism for overcommitted clouds to reduce the usage of physical machines through predication of virtual machines and run the datacenter sustainably, thus improving the energy efficiency of CDCs. Further, load balancing mechanism distributes the load effectively on physical machines. This technique uses Google workload traces and it is effective in reducing overload, energy consumption and improving resource utilization. Dastagiraiah et al. [162] proposed a VMware Off-Loading (VMOL) based dynamic load balancing technique to distribute the load effectively on virtual nodes to minimize energy consumption of resources. Giacobbe et al. [102] developed a VM based Resource Allocation (VMRA) technique (VM scheduling), which schedules virtual resources efficiently to decrease carbon footprints. In this technique, the best green destination is identified to migrate a VM from one server to another with minimum emissions of carbon dioxide. Rybina et al. [105] proposed a VM-based Assessment Technique (VMAT) for the investigation of network functions during VM scheduling, which need to be optimized for efficient consumption of energy and reduction of carbon footprints. In this research work, virtual Evolved Packet Core [83] is used to estimate the processing latency of VM migration with minimum emissions of carbon footprints. A large amount of energy consumption reduces energy efficiency, which affects the environment and increases power cost [104]. Further, there is a need to define the sustainability level of environmental impact to improve sustainability of CDCs.

Zhao et al. [37] proposed a Performance-guaranteed and Power-aware VM Placement (PPVMP) technique to investigate the relationship between CPU utilization and power consumption to develop fault tolerance mechanism using forward error recovery technique. Further, ACO-based power-aware technique is proposed for VM placement with minimum power consumption. Chinnathambi et al. [157] proposed a Scheduling and Checkpointing Optimization (SCO) algorithm for efficient management of VMs for mission critical applications. Further, SCO algorithm reduces energy consumption and crash failure errors during execution of applications. Sotiriadis et al. [158] proposed a Self-managed VM Scheduling (SVMS) technique for virtual resource management using Bin-packing algorithm [74], which increases infrastructure capacity to enable VM elasticity and maximizes VMs real CPU utilization during workload execution. Beechu et al. [161] proposed an Energy-efficient VM Fault Tolerance (EVMFT) technique, which focuses on core mapping based on the application core graph and reduces failure using restart recovery mechanism.

Mishra et al. [40] proposed a Task-based VM-Placement (TVMP) algorithm to perform mapping of tasks to VMs and VMs to physical machines for optimization of an energy consumption. VM elasticity based TVMP algorithm performs searching of optimal VM and physical machine to fulfil the resource requirement for execution of user tasks. The proposed algorithm is implemented using CloudSim toolkit [61] and it reduces the task rejection rate and makespan. Al-Dhuraibi et al. [163] proposed a Docker Containers based VM Elasticity (DCVME) technique, which scales up and down both CPU and memory assigned to each container according to the application workload autonomously and experimental results show that DCVME technique performs better than Kubernetes elasticity mechanism [164] in terms of energy efficiency and resource utilization. Metzger et al. [107] suggested that the use of virtualization technology in educational universities can reduce energy consumption as well as carbon footprints. For an efficient resource utilization of CDC, the hypervisor multiplexes VMware based Operating Systems (OS) to the primary hardware resources (storage, memory and processor). Nevertheless, co-hosted VM interference reduces the performance of an application if virtual machines are isolated poorly. Due to an increase in requirement of datacenters, energy consumption is also increasing, which effects the sustainability of CDCs. To solve this problem, virtualization technology can be used and transfer the data from one server to other using VM migration and consolidation. Figure 14 shows the evolution of virtualization technology along with their Focus of Study (FoS) across the various years.

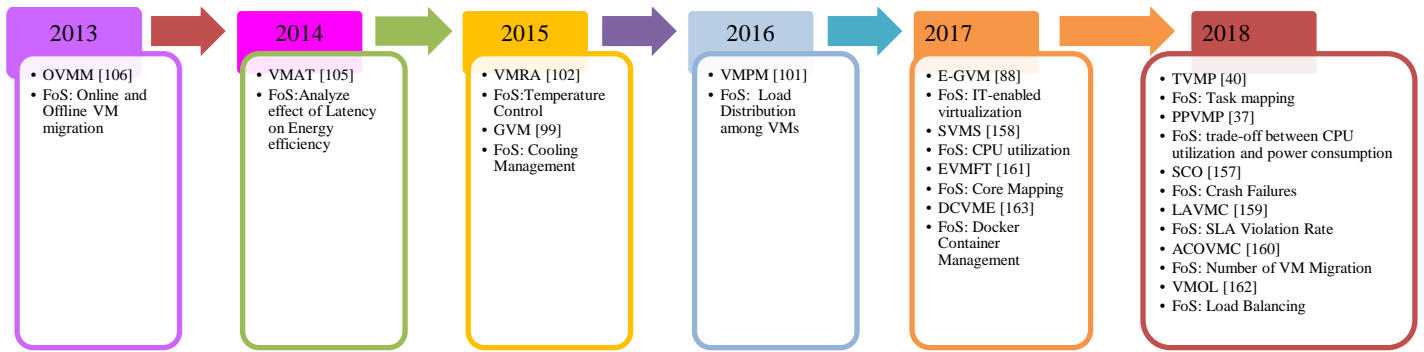

Figure 14: Evolution of Virtualization Technology

A summary of these related techniques (virtualization technology) and their comparison based on objective, optimization parameter and metric along with open research challenges is given in Table 15.

Table 15: Comparison of Existing Techniques (Virtualization Technology) and Open Research Challenges

| Technique | Organization | Objective | Optimization Parameter | Metric | Citations | Open Research Challenges |
|---|---|---|---|---|---|---|
| TVMP [40] | NIT, Rourkela, India | Optimal searching of VM | Task rejection rate and makespan | Execution Time | 1 | During the execution of workloads, VM fault tolerance is required. |
| LAVMC [159] | Islamic Azad University, Iran | Reduce energy consumption | SLA violation rate | Energy Cost | 2 | Dynamic energy-aware technique can be adopted to reduce energy consumption in case of idle server. |
| ACOVMC [160] | Abo Akademi University, Finland | Reduce no. of VM migrations | VM Migration Cost | VM Co-Location Cost | 6 | VM fault tolerance is required to deal with different types of faults. |
| VMOL [162] | K L University, India | Improve energy utilization | Power consumption | Energy cost | 2 | VM load balancing mechanism can be improved using nature or bio-inspired optimization algorithms. |
| SCO [157] | Coimbatore Institute of Technology, India | Reduce number of failures | Number of VM migration | Network Bandwidth | - | Replication technique can improve the VM fault tolerance mechanism. |
| SVMS [158] | University of Toronto, Canada | Increase infrastructure capacity | Resource utilization | Resource Utilization and Energy Cost | 1 | VM consolidation mechanism is required to reduce number of VM migrations. |
| EVMFT [161] | NIT, Goa, India | Improve fault tolerance | Number of failures per VM migration | Reliability | - | VM load balancing mechanism can be improved using resource prediction technique. |
| DCVME [163] | University of Lille, Spain | Improve resource utilization and energy efficiency | CPU utilization and Energy consumption | Resource Utilization and Energy Cost | - | Container startup can be accelerated by optimizing the storage driver. |
| PPVMP [37] | Xidian University, China | Reduce power consumption | CPU utilization | Resource Utilization and Energy Cost | 1 | VM scheduling mechanism is required for effective management of virtual resources. |
| GVM [99] | Tsinghua University, China | Improve energy utilization | Power consumption | Energy cost | 7 | During the execution of workloads, VM load balancing mechanism is required to balance the load effectively due to decentralized CDCs and renewable energy resources. |
| E-GVM [88] | Qinghai University, China | Reduce carbon emissions | Power cost | Network Power Usage | 1 | To balance the workload demand and renewable energy, VM based workload migration and consolidation techniques provide virtual resources using few physical servers. |
| VMPM [101] | Oregon State University, USA | Analyze variation in energy consumption | Resource utilization | Resource Utilization and Energy Cost | 13 | It is the great challenge for VM elasticity techniques to control the cost in terms of energy consumption and network delay while migrating workloads between distributed resources geographically. |
| VMRA [102] | University of Messina, Italy | Reduce carbon emissions | Energy consumption | VM Co-Location Cost | 15 | Increasing the size of VM creating another challenge for VM consolidation which consumes more energy and resultant in service delay. |
| VMAT [105] | Technical University, Dresden | Improve energy-efficiency | Latency | Network Bandwidth and Latency | 20 | WAN based VM migration requires storage migration which can be an overhead for cost-effective migration. |
| OVMM [106] | The University of Melbourne, Australia | Improve renewable energy utilization | Throughput | Storage Throughput | 4 | Mainly focuses on minimizing the carbon footprint and ignores performance metrics. |

### 3.5.2 Virtualization based Taxonomy

Based on literature, virtualization consists of the following components: i) VM migration, ii) VM elasticity, iii) VM load balancing, iv) VM consolidation, v) VM fault tolerance and vi) VM scheduling as shown in Figure 15. Each of these taxonomy components are discussed below along with their sub-components and relevant examples.

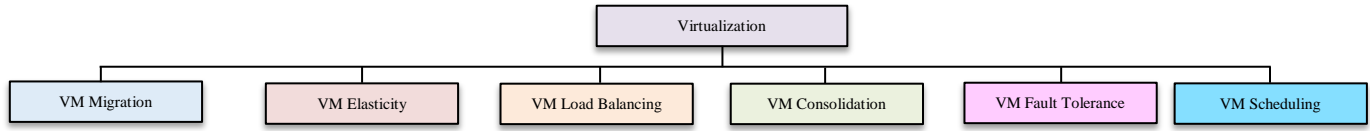

Figure 15: Taxonomy Based on Virtualization

The comparison of existing techniques based on our virtualization taxonomy is given in Table 16 (VM Migration, VM Elasticity and VM Load Balancing) and Table 17 (VM Consolidation, VM Fault Tolerance and VM Scheduling).

#### 3.5.2.1 VM Migration based Taxonomy

VM migration is a process of relocation of a running VM from one physical machine to another without affecting the execution of user application. Based on literature [101] [102] [106] [164], VM migration consists of the following components: i) technique, ii) VM technology, iii) optimization criteria, iv) network technology and v) storage migration as shown in Figure 16.

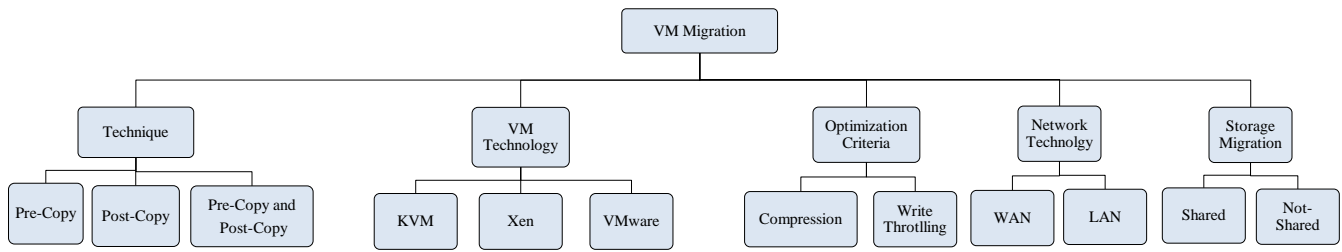

Figure 16: Taxonomy Based on VM Migration

*3.5.2.1.1 Technique:* VMs can be migrated from one place to another for better utilization of resources and it reduces the under-utilization and over-utilization of resources [99]. Three types of techniques have been proposed for VM migration: 1) Pre-copy, 2) Post-copy, 3) and Pre-copy and Post-copy. There are two different phases of *pre-copy* technique: a) warm-up and b) stop and copy. In *warm-up* phase, the hypervisor copies the state from source server to destination server, which contains the information about the memory state and the CPU state. *Stop and copy* phase copies the pending files (if any file is modified during the warm-up phase) from source to destination servers and starts the execution at destination server [88]. In *post-copy*, it stops the VM at the source server, transfers all the details like CPU state and memory state to the destination server and starts execution. Some VM migration mechanisms use both *pre-copy and post-copy* together to transfer states from one server to another.

*3.5.2.1.2 VM technology:* There are three different types of technology that are available in literature for VM migration: KVM, Xen and VMware. *KVM* is a Kernel-based VM, which permits many Operating Systems (OSs) to share single resource or hardware. *Xen* is working based on microkernel design to share the same resources to run multiple OSs. *VMware* can be used for application consolidation to provide services through virtualization [101].

*3.5.2.1.3 Optimization criteria:* It is identified that optimization criteria for virtualization technology can be compressed or write throttling. ESXi is an independent hypervisor, which offers memory *compression* cache to increase the performance of VMs and it further increases the capacity of the CDC [105]. *Write throttling* is used to perform write and incoming copy operations, which limit the transfer of data [106].

*3.5.2.1.4 Network technology:* There are two different type of network technologies are used for VM migration: i) WAN and LAN. *Wide Area Network (WAN)* is used to migrate VM geographically using wireless connection, while *Local Area Network (LAN)* is used to migrate VM from one server to another within limited area.

*3.5.2.1.5 Storage migration:* In this technique, storage from one running server to another can be migrated without affecting the workload execution of VMs. Storage migration can also be used to upgrade storage resources or transfer service [101] [32]. The distributed files systems can be used to provide shared storage space.

In virtualized cloud environments, VM migration is the mostly used mechanism to achieve energy efficiency at runtime [88]. The value of the migration time rapidly increases as the value of available network bandwidth reduces and also size of VM; but migration time can be decreased by consolidation of VMs, which improves resource utilization and effective utilization of resources can reduce energy consumption [101]. Furthermore, the energy consumption depends on the size of the VM, which is migrated. Remarkably, for the similar set of VMs, different orders of migrations lead to different migration-time based on VM size [102] [106]. So, there is a requirement to investigate the trade-off between energy cost and migration time of VMs.

The followings are the important issues with VM migration in geographically distributed datacenters [88] [99] [105] [106]:

- It is challenging to transfer VMs on shared bandwidth while maintaining the SLA of an application because memory size of VM is dynamic, which varies from 1 GB to 50+ GBs.
- WAN based VM migration has many issues such as higher packet drop ratio, larger communication distances, heterogeneous network architecture design, unpredictable network behavior, greater latencies and limited network bandwidth, which can increase the possibility of SLA violation.
- VM migration complexity increases with the increase of WAN-based storage migration because live storage migration needs asynchronous and synchronous communication modes to transmit storage blocks from one CDC to another.
- Moreover, existing VM migration techniques are not able to distribute load dynamically in a coordination manner among various CDCs.
- Due to longer communication distances, secure VM migration on heterogeneous CDCs is a challenging task. Consequently, hijackers acquire hardware states, application sensitive data, currently hosted applications and OS kernel states for malicious activities.

Container as a Service (CaaS) such as Kubernetes, Docker etc. uses resource isolation features of Linux kernel such as Control groups (Cgroups) and kernel namespaces to allow independent containers to run within a single Linux instance to avoid substantial overhead of starting virtual machines on hypervisors [156]. The shifting to container-based deployments can reduce overheads related to deployment of containers. Further, CaaS can increase realization by supporting real-time workloads [40]. Container-based deployments outperforms VMs transition based deployment due to following reasons [37] [99]: a) containers start up very quickly and their launching time is less than a second and b) containers have tiny memory footprint and consume a very small amount of resources. Moreover, containers permit the host to support more instances simultaneously as compared to VM based cloud testbed.

### 3.5.2.2 VM Elasticity based Taxonomy

VM elasticity enables the automatic provisioning and de-provisioning of computing resources to fulfill the changing demand of workloads at runtime. Based on literature [40] [37] [163] [164], VM elasticity consists of the following components: i) scope ii) policy, iii) objective and iv) mechanism as shown in Figure 17.

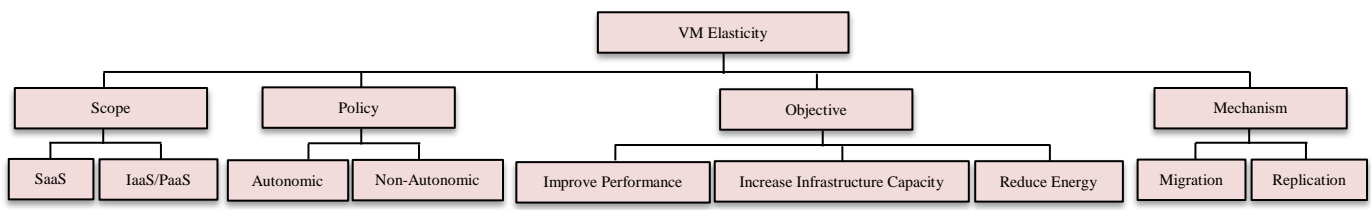

Figure 17: Taxonomy Based on VM Elasticity

*3.5.2.2.1 Scope:* It defines the location, where the elasticity actions are managed, which can be application (SaaS) or platform level (PaaS) and infrastructure level (IaaS) [62]. At *IaaS level,* the elasticity controller monitors the application execution and perform different decisions based on resource (hardware) scalability. At *SaaS or PaaS level*, the elasticity controller is implanted in application or within the execution platform, which performs the dynamic scalability of cloud resources.

*3.5.2.2.2 Policy:* There are two types of policies for the execution of elasticity actions: autonomic and non-autonomic [66]. In *autonomic* policy, cloud system or application controls the elasticity actions and performs action based on the SLA constraints. In *manual* policy, user monitors the virtual environment and performs the elasticity actions accordingly.

*3.5.2.2.3 Objective:* VM elasticity techniques have three main objectives: 1) improve performance, 2) increase infrastructure capacity and 3) reduce energy [67]. The main objective of VM elasticity techniques to improve *performance* such as optimal searching of VM, reduce task rejection rate and makespan. The second objective is to *reduce energy consumption* of CDC during execution of workloads. The third objective is to *improve the infrastructure capacity* by adding different resources at runtime to execute workloads within their specified budget and deadline.

*3.5.2.2.4 Mechanism:* There are two different mechanisms for VM elasticity as identified form literature [83]: migration and replication. To *migrate* the VM from one physical machine to another for effective utilization of application load using deconsolidation and consolidation of resources. *Replication* refers to elimination and removal of instances (application modules, containers, VMs) from virtual environment.

### 3.5.2.3 VM Load Balancing based Taxonomy

VM load balancing refers to the optimization of utilization of VMs to reduce resource wastage due to underloading and overloading of resources. It helps to achieve QoS and maximize resource utilization to improve performance of cloud service. Based on literature [88] [101] [162], VM load balancing consists of the following components: i) resource-aware and ii) performance-aware as shown in Figure 18.

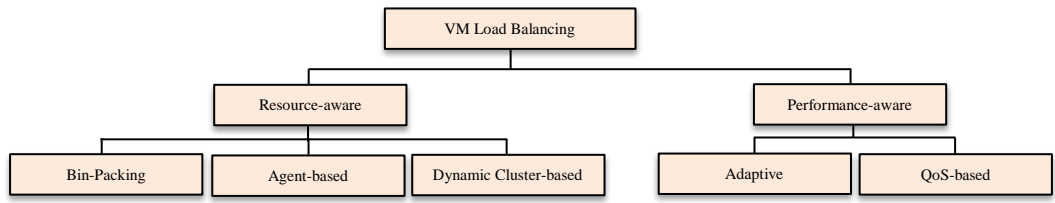

Figure 18: Taxonomy Based on VM Load Balancing

*3.5.2.3.1 Resource-aware:* CDC require different type of resources (memory, processor, cooling, storage, networking etc.) to execute user workloads [74]. *Resource-aware* load balancing algorithm execute workloads and it also monitors and analyses the different performance parameters related to resources such as energy consumption, degree of resource capacity imbalance and resource utilization to perform load balancing. There are three different types of resource-aware load balancing algorithms: bin-packing, agent-based and dynamic cluster-based. In *bin-packing*, different bins are used to pack objects of different capacities and it uses minimum number of bins to provide the same capacity in a balanced way. In *agent-based*, a software agent is used to monitor the performance of different components such as network devices, storage device and processor and balances the load effectively. In *dynamic cluster-based*, resources are categorized automatically based on requirement and availability of resources. Further, categorized resources are allocated for execution of workloads with maximum resource utilization and minimum energy consumption.

*3.5.2.3.2 Performance-aware:* In *performance-aware* load balancing algorithms, different performance parameters are analyzed to make decisions for effective load balancing of VMs [80]. There are two different types of performance-aware load balancing algorithms: adaptive and QoS-based. In *adaptive*, performance is maintained using dynamic computing environment for execution of workloads with changing behavior. In *QoS-based*, resources are provisioned and scheduled for workload execution by fulfilling the QoS requirements of applications such as energy efficiency, makespan, execution cost and response time.

### 3.5.2.4 VM Consolidation based Taxonomy

VM consolidation refers to the effective usage of VMs to improve resource utilization and reduce energy consumption [49]. Based on literature [99] [102] [159] [160], VM consolidation consists of the following components: i) resource assignment policy, ii) architecture, iii) co-location criteria and iv) migration triggering point as shown in Figure 19.

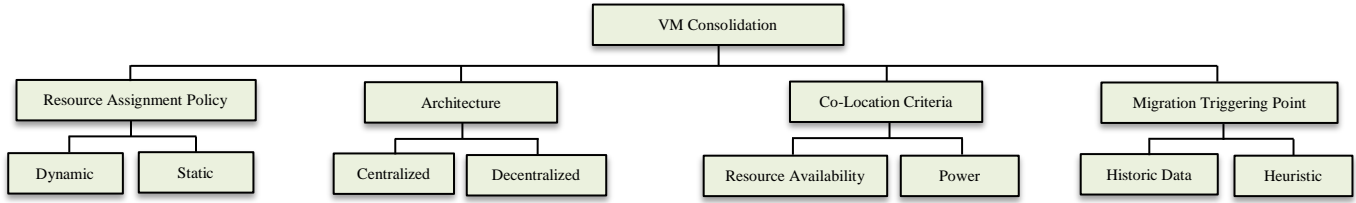

Figure 19: Taxonomy Based on VM Consolidation

*3.5.2.4.1 Resource assignment policy:* It defines the mechanism to select resources for virtual machines within a CDC [125] and resource assignment policy can be static or dynamic. In *dynamic* approach, VMs are reconfigured using dynamic attributes proactively based on the demand of workloads. In *static* approach, maximum resources are pre-assigned to VM for workload execution.

*3.5.2.4.2 Architecture:* There are two different types of architectures, which are used in VM consolidation techniques: centralized and decentralized as described in *Section 3.1.2.4.* There is no risk of a single failure point in *decentralized* architectures, while *centralized* architectures are prone to single failure point.

*3.5.2.4.3 Co-location criteria:* There are two main types of co-location criteria in VM consolidation techniques, which is considered based on: resource availability and power [71]. VMs can be co-located from one CDC to another: i) if there is less number of resources are *available* in current CDC or ii) if there is unavailability of adequate *power* to run CDC.

*3.5.2.4.4 Migration triggering point:* VMs can be migrated from one CDC to another for consolidation and target CDC is identified using two different approaches [67]: historic data and heuristic based. In *historic data* based approach, VM can be migrated to the most efficient CDC based on their historic data of previous performance. In *heuristic* based approach, the most efficient CDC can be identified based on their performance parameters such as resource utilization, energy consumption and response time.

### 3.5.2.5 VM Fault Tolerance based Taxonomy

VM Fault Tolerance supports primary VM by maintaining the identical secondary VM to provide continuous availability of cloud service in case of VM failure. Based on literature [105] [106] [157] [161], VM fault tolerance consists of the following components: i) redundancy, ii) failure semantics, iii) recovery and iv) failure masking as shown in Figure 20.

*3.5.2.5.1 Redundancy:* In case of resource failure, redundancy provides redundant components to maintain the performance of computing system, which can be software or hardware [125]. For *hardware* components, physical redundancy technique adds redundant hardware components to tolerate failures, which support computing system to continue its service in an efficient manner. For *software* components, two different types of processes are created: active (primary) and passive (backup). The backup process is identical to the primary process and backup process will be active during the failure of primary process to maintain the performance of the system.

*3.5.2.5.2 Failure semantics:* It refers to the selection of failure tolerance method based on the two types of failure modes [83]: arbitrary errors and crash failure errors. An *arbitrary error* occurs when communication service to lose or delay messages or message may be corrupted. A *crash failure error* occurs when system suddenly stops processing of instructions. To deal with both type of failures, computing system needs duplicate processor.

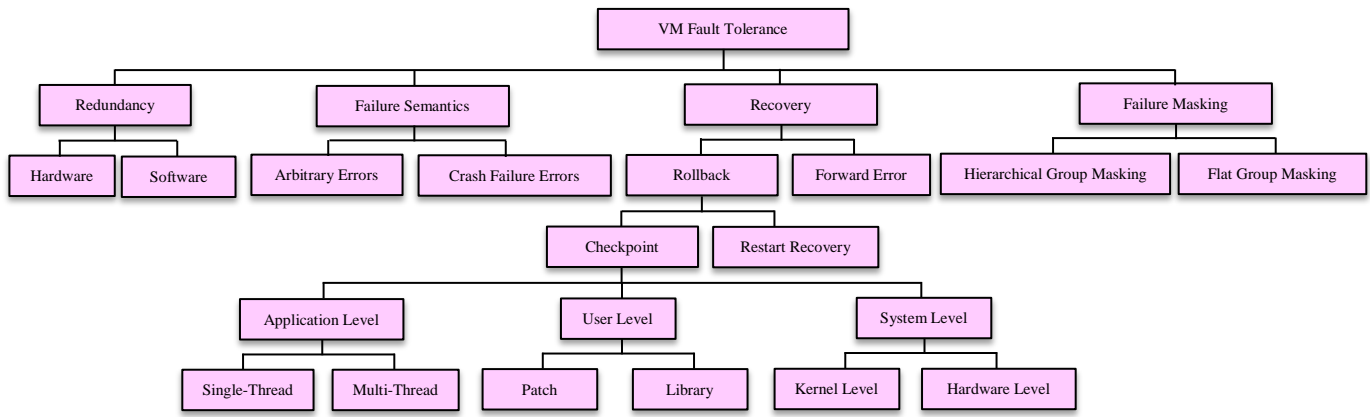

Figure 20: Taxonomy Based on VM Fault Tolerance

*3.5.2.5.3 Recovery:* This mechanism replaces the erroneous state with stable state using different recovery mechanisms [122], which are Forward Error Recovery (FER) and backward (or rollback). *FER mechanism* tries to correct the errors to move the system into new correct state and this mechanism is effective when there is a need of continue service. *Backward Error Recovery (BER)* or Rollback-recovery is widely used fault tolerance mechanism, which consists of two different methods: checkpoint and restart recovery. *Restart recovery* mechanism performs the process of rebooting to recover or restore the system to correct state. To incorporate fault tolerance into system, a snapshot of the application's state is saved, so that system can reboot from that point in case of system crash, this process is called checkpointing. *Checkpoint* can be performed at three different levels: application, user and system. In *application-level*, a checkpointing code is inserted automatically into the application code if failure is occurred and checkpointing code can be write using single-thread or multi-thread programming. In *user level*, an application program is linked to the *library* and Condo [20] and Esky [66] are library implementations. Further, user can use *patch* to perform user level checkpointing. In *system* level, the process of checkpoint can be performed at OS kernel level and hardware level. The digital hardware is used in *hardware level checkpointing* to modify a group of commodity hardware. *OS kernel* level checkpointing installs the available package for a particular OS.

*3.5.2.5.4 Failure masking:* Failure masking technique ensures the availability of cloud service during failure of nodes without the user observing any interruption [144] [145]. There are two types of masking techniques: flat and hierarchical group masking. In *flat group masking*, individual workers are appearing as a single worker and hidden from the clients and a new worker will be selected using voting process [14] in case of failure. In *hierarchical group masking*, a central coordinator controls the activities of different workers and coordinator selects the new worker in case of failure.

### 3.5.2.6 VM Scheduling based Taxonomy

VM scheduling algorithm schedules the virtual resources (local or remote) effectively for workload execution. Based on literature [60] [61] [40] [157] [158], VM scheduling consists of the following components: i) application type, ii) operating environment and iii) objective function as shown in Figure 21.

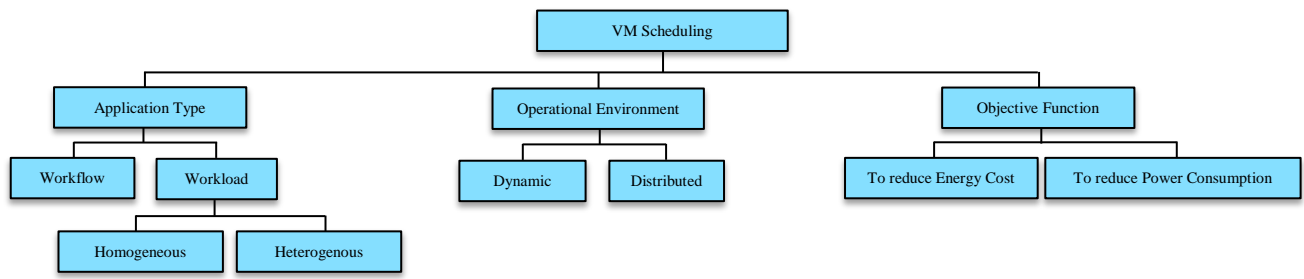

Figure 21: Taxonomy Based on VM Scheduling

*3.5.2.6.1 Application type:* Cloud application consist of different tasks, which needs computing resources for their execution [95] and it is of two types: workload and workflow. A *workload* is the execution of set of instances to achieve desired output and it can be either *homogeneous* (same QoS requirements) or *heterogenous* (different QoS requirements). *Workflow* is a combination of interrelated tasks, which distributes on different resources to achieve a single objective.

*3.5.2.6.2 Operational environment:* There are two types of operational environment: dynamic and distributed or it can be both [143]. In *dynamic* environment, VMs are scheduled for workload execution to reduce resource wastage and energy consumption. In *distributed* environment, optimized VMs are scheduled from different CDC, which are distributed geographically to improve resource utilization for workload execution.

*3.5.2.6.3 Objective function:* Literature reported that there are two types of objective functions for VM scheduling: 1) to reduce energy cost and 2) to reduce power consumption. The *energy cost* is a combination of monetary and non-monetary costs associated with energy usage for scheduling of VMs [3]. The *power consumption* is the amount of electricity expended by resource to complete the execution of an application [62].

Table 16: Comparison of Existing Techniques Based on Taxonomy of VM Migration, VM Elasticity and VM Load Balancing (**NA:** Means Not Applicable)

| Technique | Author | VM Migration | | | | | VM Elasticity | | | | VM Load Balancing | |
|---|---|---|---|---|---|---|---|---|---|---|---|---|
| | | Technique | VM Technology | Optimization Criteria | Network Topology | Storage Migration | Scope | Policy | Objective | Mechanism | Performance-aware | Resource-aware |
| TVMP | Mishra et al. [40] | NA | VMware | NA | NA | NA | IaaS/PaaS | Non-autonomic | Improve performance | Migration | QoS-based | Bin-Packing |
| LAVMC | Ranjbari and Torkestani [159] | NA | VMware | NA | NA | NA | NA | NA | NA | NA | QoS-based | NA |
| ACOVMC | Ashraf and Porres [160] | Pre-copy | VMware | Write Throttling | LAN | Shared Storage | NA | NA | NA | NA | Adaptive | NA |
| VMOL | Dastagiraiah et al. [162] | Post-copy and Pre-copy | Xen | Write Throttling | LAN | Shared Storage | NA | NA | NA | NA | Adaptive | Dynamic Cluster-based |
| SCO | Chinnathambi et al. [157] | NA | KVM | NA | NA | NA | SaaS | Autonomic | Reduce energy | Migration/Replication | NA | NA |
| SVMS | Sotiriadis et al. [158] | NA | VMware | NA | NA | NA | SaaS | Non-autonomic | Increase Infrastructure Capacity | Migration | QoS-based | Bin-Packing |
| EVMFT | Beechu et al. [161] | NA | KVM | NA | NA | NA | NA | NA | NA | NA | NA | Agent-based |
| DCVME | Al-Dhuraibi et al. [163] | NA | KVM | NA | NA | NA | IaaS/PaaS | Non-autonomic | Reduce energy | Migration/Replication | QoS-based | Dynamic Cluster-based |
| PPVMP | Zhao et al. [37] | Post-copy and Pre-copy | VMware | Compression | WAN | Shared Storage | SaaS | Autonomic | Reduce energy | Migration/Replication | Adaptive | Dynamic Cluster-based |
| E-GVM | Wang et al. [88] | Post-copy and Pre-copy | Xen | Write Throttling | LAN | Shared Storage | NA | NA | NA | NA | NA | NA |
| VMPM | Dabbagh et al. [101] | Post-copy | VMware | Write Throttling | WAN | Shared Storage | IaaS/PaaS | Autonomic | Improve performance | Replication | Adaptive | Dynamic Cluster-based |
| GVM | Wang et al. [99] | Post-copy | KVM | Compression | LAN | Not-Shared Storage | NA | NA | NA | NA | QoS-based | Agent-based |
| VMRA | Giacobbe et al. [102] | NA | NA | NA | NA | NA | IaaS/PaaS | Autonomic | Improve performance | Migration | Adaptive/QoS-based | Dynamic Cluster-based |
| VMAT | Rybina et al. [105] | Pre-copy | KVM | Compression | WAN | Shared Storage | IaaS/PaaS | Autonomic | Reduce energy | Replication | Adaptive | Dynamic Cluster-based |
| OVMM | Khosravi et al. [106] | Post-copy and Pre-copy | KVM | Compression | LAN | Not-Shared Storage | NA | NA | NA | NA | QoS-based | Agent-based |

For effective management of virtualized cloud datacenters, thermal-aware scheduling is required to execute workloads on energy-efficient computing resources, which further reduce the heat recirculation and therefore the load on the cooling systems.

**3.6 Thermal-aware Scheduling**

Cloud datacenters consist of chassis and racks to place the servers to process the IT workloads. To maintain the temperature of datacenters, cooling mechanisms are used to reduce heat [86]. So, there is a need of effective management of temperature to run the cloud datacenter efficiently. Servers produce heat during execution of IT workload, so cooling-management is required to keep the stable temperature of room [92]. The processor is an important component of a server that consumes the most electricity. Sometimes the heat generation of processors is higher than the threshold because servers are organized in a compact manner [93]. Both cooling and computing mechanisms are consuming a huge amount of electricity. It would be better to reduce the energy consumption instead of improvement of cooling mechanism [90]. To solve the heating problem of CDCs, thermal-aware scheduling is designed to minimize cooling setpoint temperature, hotspots and thermal gradient. Thermal-aware scheduling is better than heat modelling [142]. Thermal-aware scheduling based on heat modelling performs computational scheduling of workload. Thermal-aware monitoring and profiling module monitors and assess the distribution of heat in CDCs while profiling maintains the details of computational workload, microprocessors and heat emission of servers. With the use of renewable energy, the load of cooling can be decreased to enable sustainable CDCs.

Table 17: Comparison of Existing Techniques Based on Taxonomy of VM Consolidation, VM Fault Tolerance and VM Scheduling (**NA:** Means Not Applicable)

| Technique | VM Consolidation | | | | VM Fault Tolerance | | | | VM Scheduling | | |
|---|---|---|---|---|---|---|---|---|---|---|---|
| | Resource Assignment Policy | Architecture | Co-Location Criteria | Migration Triggering Point | Redundancy | Failure Semantics | Failure Masking | Recovery | Application Type | Operational Environment | Objective Function |
| TVMP [40] | Static | Centralized | Resource Availability | Historic Data | NA | NA | NA | NA | NA | NA | NA |
| LAVMC [159] | Static | Centralized | Resource Availability | Heuristic | Software | Arbitrary Errors | Hierarchical Group Masking | - Rollback<br>-- Checkpoint<br>---User Level<br>----Patch | Workload (Heterogenous) | Dynamic | To reduce Power Consumption |
| ACOVMC [160] | Autonomic | Centralized | Resource Availability | Heuristic | NA | NA | NA | NA | NA | NA | NA |
| VMOL [162] | NA | NA | NA | NA | NA | NA | NA | NA | Workload (Homogenous) | Distributed | To reduce Energy Cost |
| SCO [157] | NA | NA | NA | NA | Hardware | Crash Failure Errors | Flat Group Masking | - Rollback<br>-- Restart Recovery | Workload (Homogenous) | Distributed | To reduce Energy Cost |
| SVMS [158] | NA | NA | NA | NA | NA | NA | NA | NA | Workflow | Dynamic and Dynamic | To reduce Power Consumption |
| EVMFT [161] | NA | NA | NA | NA | Hardware | Crash Failure Errors | Flat Group Masking | - Rollback<br>-- Checkpoint<br>---System Level<br>----Hardware Level | Workload (Homogenous) | Distributed | To reduce Energy Cost |
| DCVME [163] | NA | NA | NA | NA | Hardware | Crash Failure Errors | Flat Group Masking | - Rollback<br>-- Checkpoint<br>---User Level<br>----Library | NA | NA | NA |
| PPVMP [37] | Autonomic | Centralized | Power | Heuristic | Software | Arbitrary Errors | Hierarchical Group Masking | Forward Error Recovery | NA | NA | NA |
| E-GVM [88] | Static | Decentralized | Power | Historic Data | NA | NA | NA | NA | Workload (Heterogenous) | Dynamic | To reduce Energy Cost |
| VMPM [101] | NA | NA | NA | NA | Hardware | Crash Failure Errors | Flat Group Masking | - Rollback<br>-- Checkpoint<br>---Application Level<br>----Single-Thread | Workload (Heterogenous) | Dynamic | To reduce Power Consumption |
| GVM [99] | NA | NA | NA | NA | NA | NA | NA | NA | Workload (Heterogenous) | Distributed | To reduce Power Consumption |
| VMRA [102] | NA | NA | NA | NA | Hardware | Crash Failure Errors | Flat Group Masking | - Rollback<br>-- Checkpoint<br>---Application Level<br>----Multi-Thread | Workflow | Dynamic | To reduce Power Consumption |
| VMAT [105] | Autonomic | Decentralized | Power | Historic Data | NA | NA | NA | NA | Workload (Homogenous) | Dynamic | To reduce Energy Cost |
| OVMM [106] | Static | Centralized | Resource Availability | Heuristic | Software | Arbitrary Errors | Hierarchical Group Masking | - Rollback<br>-- Checkpoint<br>---System Level<br>----Kernel Level | Workload (Homogenous) | Distributed | To reduce Power Consumption |

### 3.6.1 Related Studies

Chaudhry and Ling [14] explored the existing thermal-aware scheduling techniques developed for efficient management of green datacenters. Authors stated that the microprocessor is the most important part of the server, which consumes large amounts of electricity and produces heat continuously. Further, efficient cooling management is required to avoid overheating of servers. Moreover, thermal monitoring and profiling approaches provide different techniques for measuring both lower level (microprocessor) and upper level (datacenter) temperatures to reduce generation of heat [142]. Authors did not identify the effect of thermal-aware scheduling on sustainability and energy efficiency of CDCs. Oxley et al. [84] analyzed that a sufficient amount of cooling is required to maintain the temperature for smooth working of sustainable datacenters. Further, cloud workloads should be executed before deadline using available electricity but sharing of resources creates problems for the execution of different tasks on different cores of available resources [95]. Co-location, power and thermal-aware resource allocation techniques have been proposed to execute a number of tasks. Experimental results proved that this approach is effective in executing tasks before a deadline and under a temperature constraint.

Cupertino et al. [85] proposed the Energy-Efficient Workload Management (EEWM) approach for thermal-aware CDCs. Further, workloads and applications are managed effectively by creating their profile. Moreover, power-aware resource scheduling mechanism has been proposed to execute the workloads, while optimizing energy consumption. Sun et al. [86] extended EEWM by adding a Heat Distribution Matrix (EEWM-HDM) to control the energy effect of servers to other servers in a CDC. Further, fuzzy-based priority policy is used to make trade-off between power and cooling and execute user workloads in a sustainable cloud environment. Guitart et al. [87] discussed a thermal-aware resource management framework to minimize carbon footprints of CDCs. They identified that power usage in CDC is increasing with the increase in number of IT devices. Guo et al. [89] proposed the Thermal Storage based Power and Network (TSPN) aware workload management framework which provides integration with green energy and reduces bandwidth costs between CDC and cloud users. Further, stochastic cost reduction procedure uses the Lyapunov optimization [14] to create trade-off between workload delay and energy cost. Fu et al. [92] proposed a Temperature-Aware Resource Management (TARM) mechanism by extending TSPN to reduce energy consumption in CDCs. Further, TARM focused on the reliability of cloud services using a soft server temperature constraint. Shamalizadeh et al. [90] proposed the Thermal-Aware Workload Distribution (TAWD) model to reduce heat recirculation in sustainable datacenters. Firstly, the cooling and computing power requirement is assessed to control the servers in the cloud for the execution of user workloads without violation of SLAs.

Han et al. [91] proposed the DVFS-based Thermal-Aware Energy-Efficient (TAEE) resource scheduling policy to execute workloads while focusing on the reduction of AC and computation energy usage. TAEE assumes a linear relationship between CPU frequency and computation energy usage to find out the thermal correlation among the servers of CDCs. TAEE improves the energy-efficiency of sustainable CDCs. Dou et al. [93] proposed Carbon-Aware Resource Management (CARM) approach, which focuses on cost reduction of electricity to run sustainable CDCs. Further, time-varying system states have been analyzed to identify the trade-off between workload delay and electricity usage to run CDC. Singh et al. [94] extended EARTH [83] and proposed a thermal-aware autonomic resource (SOCCER) management policy for the execution of heterogeneous cloud workloads and improves energy efficiency. Arroba et al. [97] proposed a DVFS-based Dynamic Consolidation for Allocation of Resources (DCAR) to execute cloud workloads and a thermal-aware VM allocation technique [96] that is designed to control the temperature of a cloud datacenter during the execution of cloud resources. Chien et al. [98] proposed Thermal-Aware Scheduling of Resources (TASR) for multicore architectures, which divides applications into small threads using the concept of dynamic programming and execute those threads using different cores of the processor. Further, the temperature of the datacenter is controlled using both proactive and reactive scheduling procedures to analyze the variation of temperature with increasing number of threads. Oxley et al. [53] proposed a Rate-based Thermal-aware Resource Management (RTRM) technique for heterogenous cloud datacenters to execute user different workloads within their respective deadlines. Further, RTRM technique satisfies the power and temperature constraints while a linear regression technique based co-location interference model co-locating the tasks with minimum amount of energy consumption. Experimental results show that proposed technique performs effectively in terms of temperature and power as compared to greedy and genetic algorithm. Damme et al. [50] proposed an Optimized Thermal-aware Job Scheduling (OTJS) technique to analyze the cloud datacenters and reduce consumption of energy. OTJS technique schedules the jobs effectively on cloud resources, so that it maintains the temperature of the system below its threshold value, which reduces the required amount of cooling. The performance of OTJS technique is tested under varying workload conditions and experimental results show that proposed technique reduces temperature of cloud datacenter. Figure 22 shows the evolution of thermal-aware scheduling techniques along with their Focus of Study (FoS) across the various years.

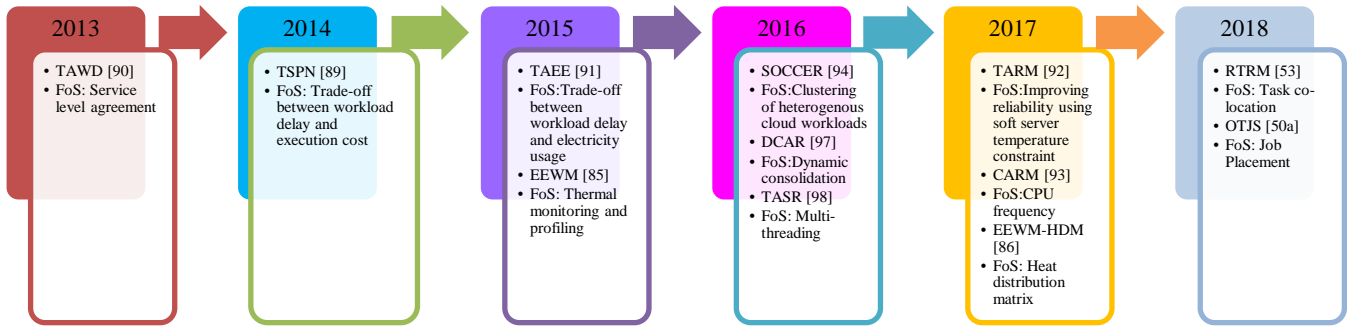

Figure 22: Evolution of Thermal-aware Scheduling Techniques

A summary of these related techniques (thermal-aware scheduling) and their comparison based on objective, optimization parameter and metric along with open research challenges is given in Table 18.

Table 18: Comparison of Existing Techniques (Thermal-aware Scheduling) and Open Research Challenges

| Technique | Organization | Objective | Optimization Parameter | Metric | Citations | Open Research Challenges |
|---|---|---|---|---|---|---|
| RTRM [53] | Colorado State University, USA | Execute workloads within their deadline | Temperature and Power | Datacenter Temperature and Energy efficiency | 5 | The relationship between varying optimal temperature distributions and computational capacity can be developed to reduce the response time of jobs. |
| OTJS [50] | University of Groningen, Netherlands | Reduce cooling cost and energy consumption | Temperature | Datacenter Temperature | 1 | Server consolidation can reduce number of active racks, which can further reduce consumption of energy. |
| EEWM [85] | University of Toulouse, France | Improve energy efficiency | Energy consumption | Energy efficiency and Datacenter Temperature | 24 | If scheduling is performed based on deferent thermal aspects like inlet temperature and heat contribution then admission control mechanism at processor level and server level contradict each other. |
| EEWM-HDM [86] | University of Toulouse, France | Improve cooling | Power | Energy efficiency and Thermodynamic efficiency | 21 | Datacenter temperature is required to be optimized. |
| TSPN [89] | University of Florida, USA | Reduce cost | Bandwidth | Network Bandwidth and Thermal Correlation Index | 61 | Large amount of heat concentration and dissipation affects the performance of CDC. |
| TARM [92] | Inner Mongolia University of Technology, China | Reduce energy consumption | Temperature | Datacenter Temperature and Thermal Correlation Index | 1 | A large variation of temperatures also increases the complexity of scheduling and monitoring. |
| TAWD [90] | University of Aveiro, Portugal | Improve heat recirculation | Energy | Airflow efficiency and Energy efficiency | 12 | Dynamic thermal profiles are required for continuous updation of temperature. |
| TAEE [91] | Oakland University, USA | Improve energy efficiency | Computation Power consumption | Energy efficiency | 4 | TAEE technique focused on reducing datacenter temperature, but reduction in temperature may not improve airflow efficiency. |
| CARM [93] | Xi'an Jiaotong University, China | Reduce electricity cost | Energy cost | Energy efficiency and Thermal Correlation Index | 3 | There is a need of recirculation coefficient matrix to identify the heat circulation values for every node. |

| SOCCER [94] | Thapar University, India | Improve energy efficiency | Energy consumption | Energy efficiency | 8 | It requires the coordination between processor level and server level workload schedulers. |
|---|---|---|---|---|---|---|
| DCAR [97] | University of Madrid, Spain | Reduce energy consumption | Execution time | Energy efficiency and Thermal Correlation Index | 2 | The energy consumption of CDCs can be minimized by activating those servers which are adjacent to each other in rack or chassis but power density increases which creates heat concentration. |
| TASR [98] | National Chung Cheng University, Taiwan | Analyze variation of temperature | Temperature | Datacenter Temperature | 1 | Reducing cost makes hardware reliability is an open challenge. |

**3.6.2 Thermal-aware Scheduling based Taxonomy**

The components of thermal-aware scheduling are: i) architecture, ii) heat modelling, iii) thermometer, iv) scheduling, v) monitoring and awareness and vi) simulator as shown in Figure 23. Each of these taxonomy elements are discussed below along with relevant examples. The comparison of existing techniques based on our thermal-aware scheduling taxonomy is given in Table 19.

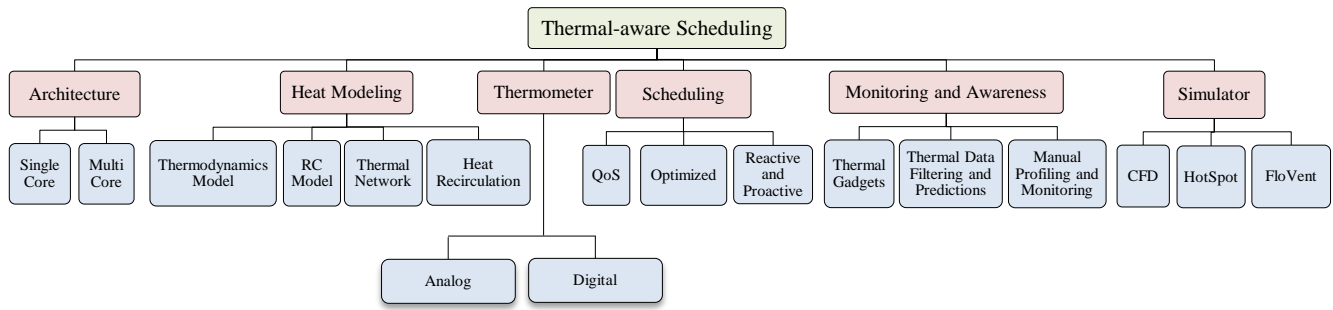

Figure 23: Taxonomy Based on Thermal-aware Scheduling

*3.6.2.1 Architecture:* Thermal-aware scheduling techniques have been designed based on different architectures: i) single core and ii) multicore [86] [92] [93]. Thermal-aware scheduling techniques execute workloads based on their priorities at different speed of processor for *single-core* architecture and for execution of high priority workload, current workload can be preempted. Generally, high priority workloads are running at high speed and the temperature of processor can be reached to its threshold value. To optimize the temperature of the processor, low priority workloads are running at lower speed to cool down the processor. To improve the execution of thermal-aware scheduling, *multi-core* processor is using in which a task is divided into number of threads and independent threads are running on different cores based on their priorities. Multi-core processors are designed with thermal-aware aspects such as intelligent fan control, clock gating and frequency scaling and these aspects are working in coordination to control the temperature within its operating limits. If one core is getting hot then a thread can be transfer to another cooler core to maintain the temperature.

*3.6.2.2 Heat-modelling:* It is an effective mechanism in thermal-aware scheduling to develop a relationship between eventual heat dissipation and energy consumed by computing devices. The scope of heat models is defined based on evaluation of environmental variables like temperature, air pressure and power. The selection of the heat model also affects the energy efficiency. The types of heat models used in existing literature are: i) thermodynamics model, ii) RC model, iii) thermal network and iv) heat recirculation.

*Thermodynamics model* is used to explore the heat exchange mechanisms in CDCs. The value of heat is quantified using law of energy conservation [89] [90]. Thermodynamic process produces the details about heat emissions, energy consumption and pass cold air to remove heat from the datacenter and researchers are still working on this process for further optimization. *RC model* is basically a Resistor–Capacitor (RC) circuit to make a relationship between electrical phenomena of RC circuit and heat transfer. Temperature difference between two surfaces and energy consumption is used to find out the value of *R* and *C* for conductance and convention. The value of RC is not changed after manufacturing of processor package. RC model is used to find out the value of various thermal parameters. *Thermal network* is based on both RC model and thermodynamics model. In thermal network, every node of CDC belongs to one of the network, which can be IT network or cooling network. A server executes workload by consuming energy and producing heat and server is part of both cooling and IT network. Thermal network is efficient for heat modelling of heterogenous equipment of a datacenter. *Heat recirculation* deals with mixing of hot air (coming from server outlets) and cold air (coming from the cooling manager). Temperature of cold air is changing with time after entering into CDCs. To maintain the temperature of CDC, it is a great challenge to provide the uniform cold air temperature every time. The resource utilization of servers which participates in heat recirculation will be reduced and performance of cloud datacenters is also affected in terms of QoS.

**3.6.2.3 Thermometer:** This is a device, which is used to measure the temperature of cloud datacenter. There are two types of thermometers have been identified from literature [14] [86] [92]: i) digital and ii) analog. The *digital* or infrared thermometer is an electronic device, which uses digital sensor to provide a digital display. Most digital thermometers are resistive thermal devices, which uses a function of electrical resistance to measure the temperature variations. The *analog* thermometer contains alcohol, which falls or rises as it contracts or expands with temperature variations and temperature value is displaying in degrees Celsius or Fahrenheit, which is marked on glass capillary tube.

**3.6.2.4 Scheduling:** The energy consumed by CDCs has been utilized for execution of workloads, but it is dissipated as heat. Lower energy is used to remove heat while workloads are scheduling using thermal-aware aspects. Thermal profiles of thermal-aware schedulers are used to find out the resource with minimum dissipation of heat in cloud datacenters. The aim of thermal-aware scheduling is to reduce dissipation of heat from active servers and minimize the active servers by turning off idle servers. Three types of thermal-aware scheduling are used in the existing literature [85] [86] [89] [92] are: i) QoS, ii) optimized, iii) reactive and proactive. *QoS based thermal-aware scheduling* schedules the energy-efficient resources to improve the performance of the cloud datacenter. The scheduler controls the temperature and reduces the load of overcooling using dynamic thermal management techniques. Further, a challenge of maintaining the SLA based on these QoS parameters is introduced and it requires the trade-off between cost saving and compensation or penalty in case of SLA violations. *Optimized thermal-aware scheduling* schedule workloads using the concept of an autonomic computing. These techniques are basically a combination of heat-recirculation and thermal-aware techniques. The main aim of server based scheduling techniques is to reduce the peak inlet temperature which is increased by heat-recirculation. Heat-recirculation can be minimized by placing lesser workloads on servers that are nearer to the floor. Processor based scheduling techniques execute the workloads by sustaining the steady core temperature called throttling. Earlier, workloads are executed using zig-zag schemes till a temperature threshold is achieved. *Reactive* management works based on feedback methods and manages the temperature based on their current state to maintain its temperature. There is a need of continuous monitoring of thermal-aware scheduling to find whether the temperature is lower than its threshold value or not. If temperature is higher than a threshold value then corrective actions will be taken to make it stable. The *proactive* approach manages the resources based on the prediction and assessment of temperature and thermal profiling. Based on previous data, predictions have been identified and plan their required action to reduce temperature during scheduling.

**3.6.2.5 Monitoring and Awareness:** Thermal monitoring and awareness is used to perform thermal-aware scheduling decisions. The thermal profile is created based on resultant heat dissipation and power consumption for thermal-awareness, which is used to rank the servers for future scheduling decisions. There are three different methods of thermal monitoring and awareness as identified form literature [14] [32] [85] [86]: i) manual profiling and monitoring, ii) thermal gadgets and iii) thermal data filtering and predictions. In *manual profiling and monitoring*, heat generation and recirculation, and power consumption of individual servers is noted manually to create thermal profile. If there is no real data available then simulation tools can be used for manual profiling. Some thermal-aware scheduling techniques [89] [92] [93] [97] estimates the thermal index to evaluate the efficiency of different cloud datacenters and perform their ranking. The *thermal gadgets* such as thermal cameras and sensors are used to generate the accurate and timely thermal information automatically. The multiple sensors can be used per unit area and both onboard and externals thermal sensors can be used to collect thermal information. In *thermal data filtering and predictions*, rise in temperature and resulting heat can be predicted for proactive thermal-aware scheduling, which helps to make effective decisions to minimize thermal gradient and peak outlet temperature. The advance prediction of temperature and heat can help to maintain the QoS during workload execution.

*3.6.2.6 Simulator:* The results of thermal simulators can be used to create thermal profiles and there are three different simulators identified from literature [14] [86] [92] [14] [32] [85] [86]: i) CFD, ii) HotSpot and iii) FloVent. *Computational Fluid Dynamics (CFD)* simulator is used to analyze and optimize airflow and heat transfer for cloud datacenter to create the thermal profile, which further helps to create thermal map. *HotSpot* is a temperature modeling tool [14], which uses thermal resistances to design the architecture of cloud datacenter based on power density and hence cooling costs, which are rising exponentially. *FloVent* simulator [14] is used to predict contamination distribution, heat transfer and 3D airflow for different types of cloud datacenters, which mainly focuses on air conditioning and ventilating systems.

Table 19: Comparison of Existing Techniques Based on Taxonomy of Thermal-aware Scheduling

| Technique | Author | Architecture | Heat Model | Scheduling | Thermometer | Monitoring and Awareness | Simulator |
|---|---|---|---|---|---|---|---|
| RTRM | Oxley et al. [53] | Multi core | Thermodynamics model | Proactive | Digital | Thermal Data Filtering and Predictions | FloVent |
| OTJS | Damme et al. [50] | Multi core | Heat recirculation | Optimized | Digital | Thermal Gadgets | CFD |
| TARM | Fu et al. [92] | Single core | Heat recirculation | QoS | Digital | Thermal Gadgets | CFD |
| CARM | Dou et al. [93] | Multi core | Thermodynamics model | Reactive and Proactive | Analog | Thermal Data Filtering and Predictions | HotSpot |
| EEWM-HDM | Sun et al. [86] | Multi core | RC model | Reactive and Proactive | Digital | Manual Profiling and Monitoring | FloVent |
| SOCCER | Singh et al. [94] | Multi core | Heat recirculation | Optimized | Analog | Thermal Gadgets | HotSpot |
| DCAR | Arroba et al. [97] | Multi core | Thermodynamics model | Proactive | Digital | Thermal Data Filtering and Predictions | CFD |
| TASR | Chien et al. [98] | Multi core | Thermal network | QoS | Analog | Thermal Gadgets | FloVent |
| TAEE | Han et al. [91] | Single core | Heat recirculation | Reactive and Proactive | Digital | Manual Profiling and Monitoring | HotSpot |
| EEWM | Cupertino et al. [85] | Single core | Thermodynamics model | QoS | Digital | Manual Profiling and Monitoring | HotSpot |
| TSPN | Guo et al. [89] | Single core | Thermal network | Optimized | Analog | Thermal Gadgets | FloVent |
| TAWD | Shamalizadeh et al. [90] | Multi core | Thermal network | QoS | Digital | Manual Profiling and Monitoring | CFD |

There is a need of effective cooling mechanisms to maintain the temperature of cloud datacenters to enable sustainable cloud computing.

**3.7 Cooling Management**

The increasing demand for computation, networking and storage expands the complexity, size and energy density of cloud datacenters exponentially, which consumes large amount of energy and produces a huge amount of heat [14]. To make cloud datacenters more energy-efficient and sustainable, we need an effective cooling management system, which can maintain the temperature of cloud datacenters [21] The heat dissipation is a critical factor to be considered for cooling management of CDC, which affects the reliability and availability of the cloud service. In cloud datacenter, high heat density causes high temperature, which needs to be controlled for smooth functioning of CDC [86]. An effective cooling management can attain a complete environmental control including pollution concentration, humidity and air temperature [92]. So, it is mandatory to discuss the existing and emerging technologies for data center cooling systems to find out the effective solution to maintain CDC's working in a safe and reliable manner.

**3.7.1 Related Studies**

Ndukaife and Nnanna [150] proposed a Split Air-Conditioning System (SACS) based cooling management technique, which investigates the characteristics of water consumption to test the performance. Further, authors vary the thickness of cooling pad to investigate its effect on the coefficient of performance. SACS technique is efficient in maintaining the adequate quantity of humidity with same amount of water consumption while cooling the cloud datacenter. SACS technique helps to reduce the energy consumption using process of cooling. Wu et al [151] proposed Weather-Aware Geo-Scheduling (WAGS) based cooling management technique, which reduces the energy consumption for cooling while distributing the load of end users among different datacenters. A workload distribution model is designed, which primarily focuses on the SLA constraints during workload execution. Moreover, trace-driven experiments have been conducted on real clouds to test the performance of WAGS technique and experimental results show that the proposed technique performs effectively in terms of energy consumption and latency. Sahana et al. [152] proposed a Server Utilization-based Smart Temperature Monitoring (SUSTM) technique, which maintains the cooling of CDC. In this technique, the concept of Mean Utilization Factor is used to find and control the amount of cool air to maintain the operating temperature in and around the servers within a CDC.

Matsuoka et al. [153] proposed a Natural Convection based Liquid Immersion (NCLI) cooling technology for saving energy consumption and space. CFD simulator [86] is used to test the performance of NCLI technique and the experimental results show that the proposed technique is effective in improving the cooling-efficiency, which further improves the value of PUE. Liu et al. [154] proposed a Cloud-Assisted Smart Temperature Control (CASTC) system, which uses IoT sensing technology to enable green data center air conditioning for cooling. CASTC system has two subcomponents: i) cloud management platform (which offers support to application layer and manage data effectively) and ii) datacenter air conditioning system (which includes ventilation and temperature control, air conditioning and environment monitoring). CASTC system effectively reduces the energy consumption of CDC without affecting the cooling management. Manousakis et al. [155] proposed an Under-provisioning Datacenter Cooling (UDC) technique to reduce the cost of cooling by under-provisioning the infrastructure of cloud datacenter. The authors developed a trade-off between cooling and QoS and they suggested that the processing capacity can be reduced in case of relaxed deadline by selecting the available cheapest provisioning. The experimental results show that UDC technique is capable to reduce cooling cost. Figure 24 shows the evolution of cooling management techniques along with their Focus of Study (FoS) across the various years.

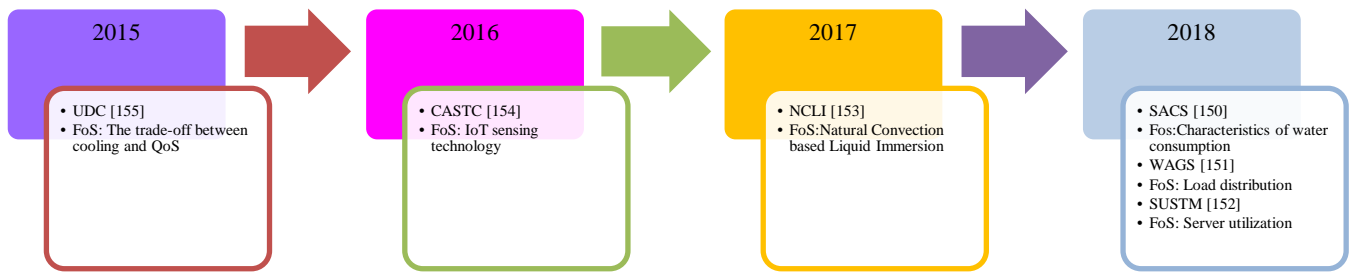

Figure 24: Evolution of Cooling Management Techniques

A summary of these related techniques (cooling management) and their comparison based on objective, optimization parameter and metric along with open research challenges is given in Table 20.

**3.7.2 Cooling Management based Taxonomy**

Based on existing literature [150-155], cooling management consists of the following components: i) cooling management techniques and ii) cooling plant as shown in Figure 25. Each of these taxonomy elements are discussed below along with relevant examples. The comparison of existing techniques based on our cooling management taxonomy is given in Table 21.

***3.7.2.1 Cooling Plant:*** The cooling plant is a system, which provides cooling to space where CDC is placed and it consists of the following components: i) medium, ii) mechanical equipment, iii) heat rejection system, iv) location, v) type and vi) temperature. The cooling system uses two different types of mediums to produce cooling: 1) water and 2) air. *Water-based* cooling system uses water pumping mechanism to generate cooling, while *air-based* cooling system uses air compressor mechanism to produce cooling. A *mechanical equipment* is used to maintain the humidity, air distribution and temperature in the CDC. There are two different types of mechanical equipment are using in cooling system: 1) Computer Room Air Conditioning (CRAC) and 2) chiller. The *Heat Rejection System (HRS)* performs the process of heat removal using two types of different ways: 1) dry cooler and 2) cooling tower. There are different types of temperature range is established for different locations in cooling systems [14] [86] [92] [150] [155] and location can be 1) chiller, 2) rack and 3) Computer Room Air Handler (CRAH). There are two types of temperature classification for three different locations with different range of temperature: 1) supply temperature and 2) return temperature. The different *types* of cooling plants are: 1) Direct Expansion (DE) air cooled systems, 2) DE glycol cooled systems and 3) chilled water systems [152] [154]. DE air cooled system contains CRAC and an air-cooled condenser as a HRS. In DE glycol cooled systems, a glycol mixture is used as heat transfer fluid from the CRAC to dry cooler. In chilled water system, a chiller provides cold water to the CRAH.

Table 20: Comparison of Existing Techniques (Cooling Management) and Open Research Challenges

| Technique | Organization | Objective | Optimization Parameter | Metric | Citations | Open Research Challenges |
|---|---|---|---|---|---|---|
| SACS [150] | Purdue University, USA | Reduce energy consumption | Water consumption | Water Economizer Utilization Factor | 1 | The trade-off between water consumption and energy consumption can be developed. |
| WAGS [151] | Shanghai Jiao Tong University, China | Reduce cooling cost | Energy consumption and latency | Datacenter Cooling System Efficiency and Latency | 1 | The nature or bio-inspired load distribution policy can be adopted for further reduction of SLA violation rate. |
| SUSTM [152] | JIS College of Engineering, India | Maintain operating temperature | Temperature | Recirculation Index | 1 | The cost of cooling can be reduced by adopting under-provisioning of infrastructure, which can further reduce the temperature of CDC. |
| NCLI [153] | Osaka University, Japan | Improve cooling-efficiency | Power Utilization Efficiency | Datacenter Cooling System Efficiency | 1 | The trade-off between PUE and energy consumption can be developed. |
| CASTC [154] | Beijing Jiaotong University, China | Reduce energy consumption | Energy efficiency | Water Economizer Utilization Factor | 26 | Thermal-aware scheduling can improve temperature control. |
| UDC [155] | Rutgers University, USA | Reduce cooling cost | Cooling cost | Datacenter Cooling System Efficiency | 10 | Under-provisioning of resources can violate SLA, which affects the QoS. |

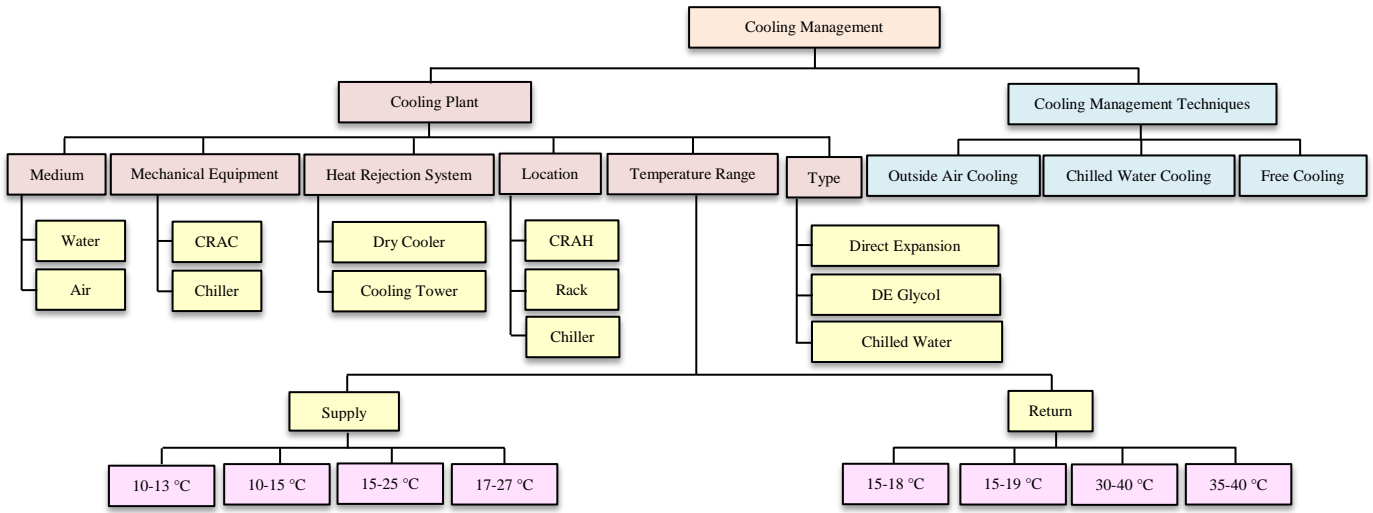

Figure 25: Taxonomy Based on Cooling Management

***3.7.2.2 Cooling Management Techniques:*** Literature [14] [86] [92] [150] [155] identified three different types of cooling management techniques: i) outside air cooling, ii) chilled water cooling and iii) free cooling. In *outside air cooling*, the cooler is used to bring the fresh air from outside and, cooled and pushed it through the CRAC, which is better than air recirculation mechanism. In *chilled water cooling* system, electricity is used to freeze water at night and circulate this water throughout the CRAC unit during the day. In *free cooling*, air is passed into chamber, which performs cooling through water evaporation [14].

Table 21: Comparison of Existing Techniques Based on Taxonomy of Cooling Management

| Technique | Cooling Plant | | | | | | | Cooling Management Techniques |
|---|---|---|---|---|---|---|---|---|
| | **Medium** | **Mechanical Equipment** | **Heat Rejection System** | **Location** | **Temperature Range (°C)** | | **Type** | |
| | | | | | Supply | Return | | |
| SACS [150] | Air | CRAC | Dry cooler | CRAH | 15-25 | 30-40 | DE | Outside air cooling |
| WAGS [151] | Water | Chiller | Cooling tower | Chiller | 10-13 | 15-18 | Chilled water | Chilled water cooling |
| SUSTM [152] | Water | Chiller | Cooling tower | CRAH | 10-13 | 15-19 | Chilled water | Chilled water cooling |
| NCLI [153] | Air | CRAC | Dry cooler | CRAH | 15-25 | 30-40 | DE | free cooling |
| CASTC [154] | Air | CRAC | Dry cooler | Rack | 17-27 | 35-40 | DE Glycol | Outside air cooling |
| UDC [155] | Water | Chiller | Cooling tower | Chiller | 10-13 | 15-18 | Chilled water | Chilled water cooling |

There is a need to maximize the utilization of renewable energy for cooling, which further reduce the carbon footprints and environmental problems.

## 3.8 Renewable Energy

Sustainable computing needs the energy-efficient workload execution by utilizing renewable energy resources to reduce carbon emissions [117]. Fossil fuels such as oil, gas and coal generate brown energy and it produces carbon-dioxide emissions in large extent but green energy resources such as sun, wind and water generate energy with nearly zero carbon-dioxide emissions [121]. One type of green energy is hydroelectricity, which is produced using hydraulic power. Wind and solar energy can be purchased from off-site companies or can be generated using on-site equipment [118]. In the next decade, the cost/watt will be reduced by half for renewable energy due to following reasons: i) government organizations provide monetary incentives for the incorporation of resources of renewable energy, ii) the storage capacity of rechargeable batteries will be increased and iii) advancement in technology to improve capacity of materials like photovoltaic arrays [124]. Workload migration and energy-aware load balancing techniques addressed the issue of unpredictability in the supply of renewable energy. To make the 100% availability of cloud services, it is recommended to adopt hybrid designs of energy generation, which use energy from renewable resources and grid resources [117]. Mostly, sites of commercial CDCs are located away from abundant renewable energy resources. Consequently, portable CDCs are placed nearer to the renewable energy sources to make it cost effective.

### 3.8.1 Related Studies

Running CDCs using renewable sources of energy like wind, solar, water etc. instead of energy generated from fossil fuels (grid electricity) result in lower carbon emissions and lower operational costs. Renewable energy results in almost zero carbon footprint, as compared to brown energy, which results reduction in metric tons of carbon. Grid electricity contributes to high operational costs and high energy usage. Balasooriya et al. [18] identified that business operations of different cloud providers such as Microsoft, Google, Facebook, Intel, Amazon are consuming large amounts of electricity continuously, which are increasing the environmental impact on society. To overcome this impact, green cloud computing is required, which can provide more sustainable cloud services. Renewable resources of energy (wind or solar) can be utilized to produce electricity, which reduces cost due to grid electricity [8]. To make an effective use of renewable energy, datacenters should be located nearer to the source of energy to adopt green cloud computing. Even the use of mixed energy (renewable and non-renewable) generated will also increase cost by 13%. Accenture [17] identified that datacenters produced 116 million tons carbon dioxide equivalent (mtCO$_2$e), telecom services produced 307 mtCO$_2$e and IT devices produced 407 mtCO$_2$e. Due to the intermittent and unpredictable nature of renewable power supply, some mechanisms should be adopted for making the renewable supply constant as the most workload is dynamic and time sensitive in nature. The two categories of managing renewable energy are: i) dynamic load balancing and ii) renewable energy-based workload migration.

***3.8.1.1 Dynamic load balancing technique:*** In this technique, a cloud datacenter is powered by two sources of energy: Green energy and Grid energy [118]. There is a server rack powered by renewable energy supply and server rack powered by grid electricity; switching of the workload between these two racks take place based on the energy supply. There is a dynamic load balancing between Green electricity and Grid electricity according to i) renewable supply based on weather data and ii) workload demand based on workload traces [119]. Renewable energy supply is exploited if it is available even for deadline sensitive workloads. In case of non- availability of renewable energy, workloads are served by grid supply instead of bringing the server in low power state [124]. The design implements hybrid grid-renewable supply design, where cloud datacenters are powered by either of the energies, VM migration is used to shift workload between the servers [121].

***3.8.1.2 Renewable energy-based workload migration:*** The workload is migrated between the geographically located cloud datacenter's according to the availability of renewable energy [119] [126]. This approach requires the availability of a number of cloud datacenters at different locations so the maximum amount of renewable energy is exploited according to the availability. A dynamic request routing mechanism is used to route the user request or workload towards the datacenter with abundant supply.

Raza et al. [117] proposed the Renewable Energy-Aware (REA) approach that chooses energy source (hydrogen fuel cell, lithium polymer battery and lead acid cell) to store power. This study identified that hydrogen fuel cell is effective in providing long life to stored energy. Further, a sustainability index is designed based on factors such as environment, biological, technical and economic to identify the energy source based on user requirements to make it more eco-friendly. Pierie et al. [118] proposed an Industrial Metabolism Approach (IMA) to find out green gas production pathways for effective use of renewable energy to generate power for the smooth working of CDCs in a sustainable manner. Further, a decentralized energy system is simulated using Material Flow Analysis [18] to calculate indirect material and energy requirements. Further, temporal dynamics identify energy consumption, carbon footprint and environmental impact to find the sustainable pathway. Toosi et al. [119] proposed a Renewable-Aware Technique (RAT) for sustainable datacenters to perform load balancing of web applications geographically. Based on the availability of renewable resources, load balancing of web application requests is distributed geographically and processed at different sites of resources. Real traffic of Wikipedia is used as workload to test the proposed technique and experimental results prove that RAT is effective in utilization of green resources. Petinrin et al. [120] investigated and analyzed available renewable energy-aware approaches for the sustainability of CDCs available in Malaysia and identified the available renewable energy sources. Further, it has been also concluded that the use of renewable energy sources reduces the carbon emissions and creates an eco-friendly environment.

Stein et al. [121] proposed the Multi-Criteria Model (MCM) to evaluate the feasibility of renewable energy and considered four types of energy sources like wind energy, solar energy, hydropower and biomass. Further, different aspects of sustainability are considered such as economic, social and ecological aspects for development of renewable energy. Andrae et al. [122] analyzed the use of renewable and electrical energy for communication technology like social networking websites, mobile applications, banking websites etc. This study describes that 4G devices consume more energy as compared to 3G and 2G. It is challenging to quantify the sustainability of renewable energy resources. To measure sustainability, grey rational analysis based an indicator is developed [123] by considering important aspects (economic, social and ecological) of sustainability. Liu et al. [124] proposed a Cooling and Renewable Aware (CRA) approach for a sustainable cloud infrastructure to manage cloud workloads effectively. In this research work, three important aspects of sustainable computing are considered such as IT workload, power infrastructure and cooling for the execution of user requests and IT workload that can be batch style or critically interactive. CRA manages workloads effectively and reduces power consumption using a cooling-aware technique, which leads to sustainable cloud infrastructures. Mardani et al. [126] proposed a Multi-Criteria Decision Making (MCDM) technique for sustainable and renewable energy and found that biomass and hydropower are much effective energy sources than solar and wind energy sources due to use of extra land. Xu et al. [127] examined the assessment of synchronization of renewable energy and grid energy to enable a sustainable cloud infrastructure. Three important factors such as power generation, power distribution and power transmission have been assessed along with economic operation, stability and security of power systems to generate a sustainability index. Khosravi et al. [148] proposed a Short-Term Prediction (STM) technique using Gaussian mixture model, which aids to forecast the future energy level. Experimental results show that STM technique is able to predict 15 minutes ahead with 98% accuracy.

Figure 26 shows the evolution of renewable energy techniques along with their Focus of Study (FoS) across the various years.

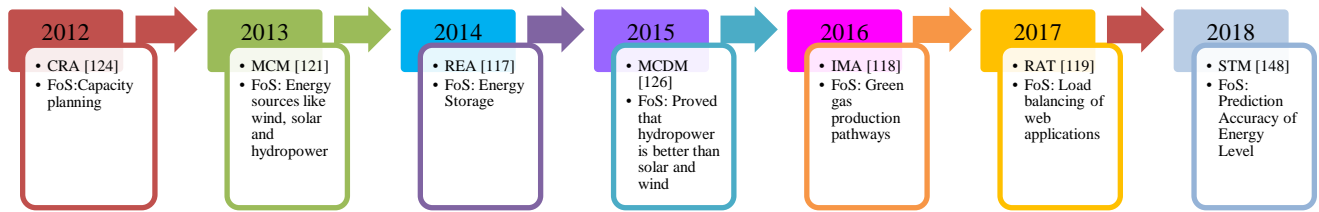

Figure 26: Evolution of Renewable Energy Techniques

A summary of these related techniques (renewable energy) and their comparison based on objective, optimization parameter and metric along with open research challenges is given in Table 22.

Table 22: Comparison of Existing Techniques (Renewable Energy) and Open Research Challenges

| Technique | Organization | Objective | Metric | Optimization Parameter | Citations | Open Research Challenges |
|---|---|---|---|---|---|---|
| REA [117] | Masdar Institute of Science and Technology, UAE | Choose effective energy source | Carbon Usage Efficiency and Water Usage Efficiency | Sustainability Index | 31 | Larger capital cost of green resources. |
| IMA [118] | Hanze Research Centre Energy, Netherlands | Reduce carbon footprints | Green Energy Coefficient | Energy consumption | 19 | There is a need of hybrid design of energy generation which uses energy from renewable resources and grid resources. |
| RAT [119] | The University of Melbourne, Australia | Improve resource utilization of green resources | Energy Proportionality Coefficient | Energy Utilization | 8 | A huge of amount of power is lost during transmission from renewable source to CDC site. |
| MCM [121] | Great Valley School of Graduate Professional Studies, USA | Evaluate the feasibility of energy sources | The Green Index | Energy Cost | 112 | CUE can be optimized by adding renewable energy resources. |
| CRA [124] | California Institute of Technology, USA | Improve cooing management | Energy Reuse Effectiveness | Power Usage | 294 | There is need to investigate capacity planning in terms of energy cost to cover other important aspects of sustainable computing. |
| MCDM [126] | Universiti Teknologi Malaysia, Malaysia | Reduce carbon footprints | Carbon Usage Efficiency and The Green Index | Energy Utilization | 53 | The issue of unpredictability in supply of renewable energy and demand of cloud datacenter is required to be addressed effectively. |
| STM [148] | The University of Melbourne, Australia | Predict energy level | Green Energy Coefficient | Energy Prediction | 1 | Cloud provider can use STM technique to perform online VM migrations |

### 3.8.2 Renewable Energy based Taxonomy

Based on existing literature, renewable energy consists of the following components: i) workload scheduling, ii) focus, iii) source of energy, iv) location-aware and v) storage devices as shown in Figure 27. Each of these taxonomy elements are discussed below along with relevant examples. The comparison of existing techniques based on our renewable energy taxonomy is given in Table 23.

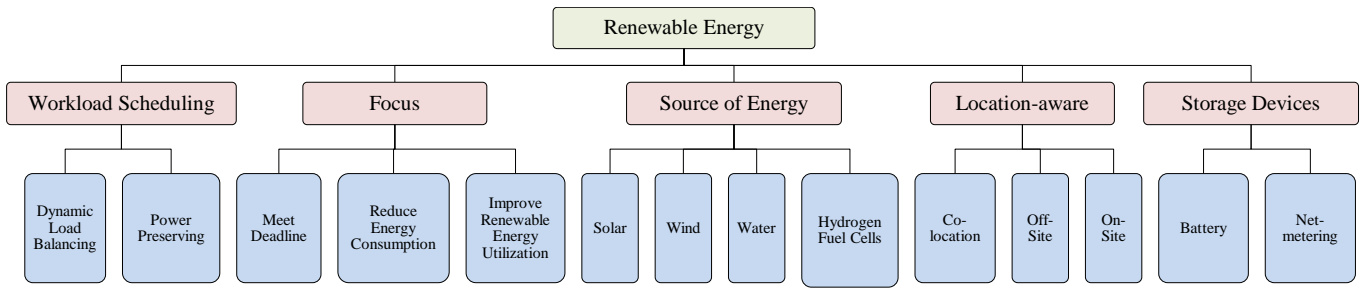

Figure 27: Taxonomy Based on Renewable Energy

***3.8.2.1 Workload scheduling:*** The scheduling of workloads in renewable energy-aware techniques has been done in two ways: i) dynamic load balancing and ii) power preserving. *Dynamic load balancing* is the most famous approach to make a balance between renewable energy and grid energy. These techniques supply renewable energy to execute workloads efficiently and predict the amount of energy that can be produced to run a CDC and the amount of energy that is needed to execute the workloads using that energy at the demand side. There is a large requirement of renewable energy for deadline oriented workloads [121] [124]. On the other side, workloads are scheduled using server *power-preserving* techniques. These techniques use power transition and voltage scaling to run the suitable workload using available renewable energy to balance demand-supply. Further, DVFS based power-preserving techniques are also designed to control the energy based on operational frequency along with voltage scaling. In these approaches [117] [118] [119], the workload (web application) requests are distributed to the specific datacenter by matching the workload demand with the available renewable energy across geo-dispersed CDCs by using a load balancing algorithm. This approach follows two level of load balancing i) at local level; redirects the request within web servers in a datacenter known a local load balancing and ii) at global level; redirects the requests among local load balancers related with a CDC known as global load balancing. Each datacenter has autoscaler in addition to local load balancer that add/removes web servers dynamically in response to the request [124]. The incoming request is distributed among a geo-dispersed CDC based on the place that has a higher availability of renewable energy, so that maximum renewable energy is utilized for making datacenter sustainable. In case of not having enough renewable energy the request is redirected to the location having cheap brown electricity. The global load balancer, uses 'weighted round robin' load balancing algorithm to redirect the requests.

***3.8.2.2 Focus:*** There are three main objectives of renewable energy-aware techniques to: i) meet deadline, ii) reduce energy consumption and iii) improve renewable energy utilization [121] [126]. SLA is an important component and workload should be executed without violation of SLA. Cloud providers are mainly focused on the *deadline* of the workload during execution. Other renewable energy-aware techniques focus on minimizing *power usage* of CDCs to execute workloads. Further, renewable energy can be utilized effectively while placing the cloud datacenter nearer to the source of *renewable energy* to save more energy and utilized to process more work.

***3.8.2.3 Source of energy:*** There are three different kinds of energy sources as identified from literature [118] [117]: i) solar, ii) wind, iii) water and iii) hydrogen fuel cells. The renewable energy can be generated using sun light or it can be generated using *wind* to run generator to produce electricity. Some techniques used the combination of both solar and wind. Other sources of renewable energy can be *water* as well as *hydrogen fuel cells* [117] [126].

***3.8.2.4 Location-aware:*** In renewable energy generation, energy can be stored using three different localities [121] [124]: i) on-site, ii) off-site and iii) co-location. In *on-site* locality, utilization of renewable energy is done at same place where energy is produced. In *off-site*, place of utilization of renewable energy is different than the place generating energy, means energy can be transported to off-shore site. On the other hand, CDCs are *co-located* from different places to sites where the chances of utilization of renewable energy are existing.

***3.8.2.5 Storage devices:*** There are two main storage devices used by renewable energy-aware techniques to store energy [118] [119]: Battery and Net-metering. Lithium ion *batteries* are using to store energy effectively. *Net-metering* is another device that can be used to store the generated energy for future.

Table 23: Comparison of Existing Techniques Based on Taxonomy of Renewable Energy

| Technique | Author | Workload Scheduling | Focus | Source of Energy | Location-aware | Storage Device |
|---|---|---|---|---|---|---|
| STM | Khosravi et al. [148] | Dynamic load balancing | Improve accuracy of prediction | Solar and wind | On-site | Batteries |
| RAT | Toosi et al. [119] | Dynamic load balancing | Reduce energy consumption | Wind | On-site | Batteries |
| IMA | Pierie et al. [118] | Dynamic load balancing | Reduce energy consumption | Solar | Off-site | Batteries |
| MCDM | Mardani et al. [126] | Dynamic load balancing | Meet deadline | Hydrogen Fuel Cells (HFC) | On-site | Net-metering |
| REA | Raza et al. [117] | Dynamic load balancing | Meet deadline | Solar and Water | Co-location | Net-metering |
| MCM | Stein et al. [121] | Power-preserving | Improve renewable energy utilization | HFC | On-site | Net-metering |
| CRA | Liu et al. [124] | Power-preserving | Improve renewable energy utilization | Solar and HFC | On-site | Net-metering |

The waste heat can be another source of renewable energy, which can be used in an efficient manner that generates electricity or can be used for heating the houses and greatly reduce electricity costs and carbon emissions.

### 3.9 Waste Heat Utilization

Reuse of waste heat is becoming a solution for fulfilling the demand of energy in energy conservation systems because fossil fuel deposits are being reduced swiftly. There is a need of cooling management to maintain the temperature of CDC in operational range due to generation of large amounts of heat during energy consumption. The cooling mechanism of CDC consumes large amount of electricity i.e. 40-50% [3] [71]. Power densities of servers are increased by using stacked and multi-core server designs, which further increases the cooling costs. The energy efficiency of CDC may be improved by reducing the energy utilized in cooling. There is a need to change the location of CDCs to reduce cooling costs and it can be done through placing the CDCs to that area which has availability of free cooling resources. Due to consumption of large amount of energy, CDCs are acting as a heat generator [130] [131]. The vapor-absorption based cooling systems of CDCs can use waste heat, then it takes away the heat while evaporating. Vapor-absorption based free cooling mechanisms can make the value of PUE ideal by neutralizing cooling expenses. Low temperature areas can use the heat generated by cloud datacenters for heating facilities. Literature reported [3] [71] [131] that there are two main solutions to control temperature of CDCs: 1) relocation of CDCs to nearby waste heat utilization recovery places, 2) vapor-absorption based cooling systems.

### 3.9.1 Related Studies

The waste heat dissipated from various electronic components (servers utilizes about 40-50% of the energy to cool down in CDCs) [132]. Nowadays a vast amount of waste heat is generated by the cloud datacenters due to rising demand of cloud based services and performance. The concept of waste heat utilization is used to capture and reuse the waste heat (to heat power plant) thus reducing the load on cooling equipment's that is used to cool the servers. It also contributes in reducing the carbon emissions and the operational costs. The waste heat recovery is better done on-site as the heat generated is not of good quality and large quantity of heat is wasted during the transfer to off-site heat recovery sites. Thus, on site Waste Heat Utilization (WHU) is more beneficial. The broad classification of waste heat utilization techniques is: i) Air Recirculation and ii) Power Plant Co-location.

***3.9.1.1 Air recirculation*:** This technique enables the reuse of waste heat in a CDC. In this technique, the servers are place with their front ends facing each others (known as cold aisles), so the back end of servers face each other (known as hot aisles). The cool air from the CRAC (Computer Room Air Conditioning) unit is supplied to the cold aisles either through raised floor design or through diffused ceiling [143]. The warm air produced in the hot aisles is captured and transferred to the CRAC unit. The chiller absorbs the heat in CRAC and outlet the cool air which is then supplied back to the cold aisles [134].

*3.9.1.2 Power plant co-location:* One of the common techniques for waste heat utilization is to co-locate the datacenter with power plant powered with fossil fuels. It helps in indirect power generation of the plants. The heat dissipated from the datacenter is used to heat the power plant and contributes in reduction of fossil fuels and carbon footprints [133]. The low-quality heat is used to preheat the boiler feed water. As a condenser counter flow heat exchanger transports the datacenter heat to the water of power plant. Further, power plant efficiency is improved and reduces the cost in the form of carbon tax and coal.

Heat can be recovered form hot streams using an energy recovery heat exchanger to utilize this heat to generate energy, which can be used to run cloud infrastructures without affecting the environment [129]. Markides [63] analyzed the existing techniques of waste heat utilization to enable sustainable computing in UK. Further, fossil fuel-based energy generators are compared with renewable energy sources and the role of power schemes, combining heat and pumped heat is discussed. Load factor (heat-to-power demand ratio) calculates the efficiency of different WHU technologies such as power schemes, combined heat and pumped heat. It has been found that pumped heat is effective way to transfer waste heat to useful work as compared to others. Chae et al. [128] designed a technique to assess the available techniques for the utilization of waste heat in an eco-industrial park to make it eco-friendly and economical. Due to continuous increase in oil price globally, it is a challenging to fulfil the demand of energy consumption of multinational companies to run CDCs. This study remarked that quantity of waste heat and energy cost can be decreased by developing more eco-industrial parks, which will be helpful to maintain sustainable environment. Karellas et al. [130] proposed a Biomass Co-location Aware (BCA) waste heat utilization technique to produce energy from waste heat. This technique is working by utilizing computational fluid dynamics at high temperature. Freeman et al. [131] proposed a Solar Organic Rankine Cycle (SORC) based indirect power generation technique for waste heat utilization. Thermal energy storage devices are used to store the generated heat and SORC has high electrical work output per unit storage volume as compared to BCA. Du et al. [132] proposed a Piezoelectric and Thermoelectric based Direct Power Generation (PTDPG) technique for waste heat utilization where a fabric device is used to store energy for future use. This technique uses a thermoelectric power generator, which is flexible and air-permeable and it can be effective in high speed wind areas. Helm et al. [133] proposed the Absorption Chiller based Solar Heating and Cooling (ACSHC) model to utilize waste heat and seasonal energy efficiency ratio. The proposed technique is effective as compared to PTDPG. Latent heat storage is used to generate energy at low temperature. Ayompe et al. [134] proposed a Water Heating System (WHS) using stand-alone wind turbine generators to produce energy. WHS reduces consumption of fuels by keeping the constant temperature of water flow of pumps. Experimental results prove that WHS performs better in energy saving and cost reduction. Oro et al. [147] proposed a Dynamic Energy model for Heat Reuse (DEHR) analysis, which uses liquid cooled CDCs for waste heat reuse. Further, a case study of indoor swimming pool is used to validate the proposed model and experimental results show that DEHR model reduces operational expenses by 18%. Figure 28 shows the evolution of waste heat utilization techniques along with their Focus of Study (FoS) across the various years.

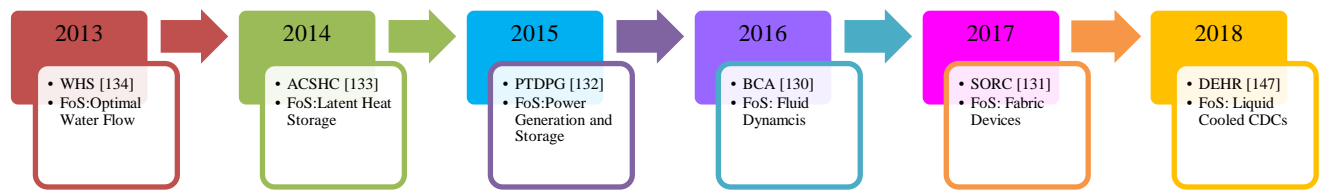

Figure 28: Evolution of Waste Heat Utilization Techniques

A summary of these related techniques (waste heat utilization) and their comparison based on objective, optimization parameter and metric along with open research challenges is given in Table 24.

**3.9.2 Waste Heat Utilization based Taxonomy**

Based on existing literature, waste heat utilization consists of the following components: i) focus of study, ii) location-aware, iii) heat transfer method and iv) cooling method as shown in Figure 29. Each of these taxonomy elements are discussed below along with relevant examples. The comparison of existing techniques based on our waste heat utilization taxonomy is given in Table 25.

Table 24: Comparison of Existing Techniques (Waste Heat Utilization) and Open Research Challenges

| Technique | Organization | Objective | Metric | Optimization Parameter | Citations | Open Research Challenges |
|---|---|---|---|---|---|---|
| BCA [130] | National Technical University of Athens, Greece | Improve Waste heat utilization | Coefficient of Performance (CoP) | Energy utilization | 60 | Power densities of servers are increasing by using stacked and multi-core server designs which further increases the cooling costs |
| SORC [131] | Imperial College London, UK | Improve power storage | Return Heat Index, CoP | Storage capacity | 4 | The energy efficiency of cloud datacenter can be improved by reducing the energy usage in cooling |
| PTDPG [132] | Deakin University, Australia | Reduce power consumption | Recirculation Ratio, CoP | Heat Transfer Rate | 59 | Implantation of CDC neared to free cooling resources reduce the cooling cost. |
| ACSHC [133] | Bavarian Center for Applied Energy Research (ZAE Bayern), Germany | Improve energy efficiency | Supply Heat Index, CoP | Energy saving | 17 | Due to consumption of large amount of energy, CDCs are acting as a heat generator |
| WHS [134] | Dublin Institute of Technology, Ireland | Reduce consumption of fuels | Recirculation Ratio | Temperature | 73 | Quality of air is degrading by shifting of air-based cooling systems to water-based. |
| DEHR [147] | Catalonia Institute for Energy Research, Spain | Improve heat reuse | Coefficient of Performance | Operational expenses | - | The waste heat reuse can be increased by reducing the use of fossil fuel, which further reduces the city $CO_2$ emissions. |

Waste heat utilization techniques convert the energy dissipated by network components into some useful work and provide advantages like: i) revenue generation due to selling of waste heat, ii) reduce operational cost, iii) heat can be used for vapor-based cooling systems, iv) reduce load on cooling equipment because it is utilized effectively to heat power plants and v) reduce carbon footprints. Waste heat utilization of cloud datacenters can be measured by a metric called Energy Reuse Effectiveness (ERE), which is a ratio of *reused energy* to *total energy* [132]. The quality of heat affects the amount of work, which can be done by utilizing heat.

*3.9.2.1 Focus of study:* Existing waste heat utilization techniques focusing on two different ways to utilize heat: i) vapor absorption based cooling systems and ii) give heat to co-located datacenter buildings [130] [131]. The first way is utilizing heat for *onsite cooling* using vapor absorption, in which heat is generated during the execution of workloads. The second way is to distribute the heat generated from cloud datacenters to the *heating model* using different modes of transfer. Heat modelling is an effective mechanism in thermal-aware scheduling to develop a relationship between eventual heat dissipation and energy consumed by computing devices.

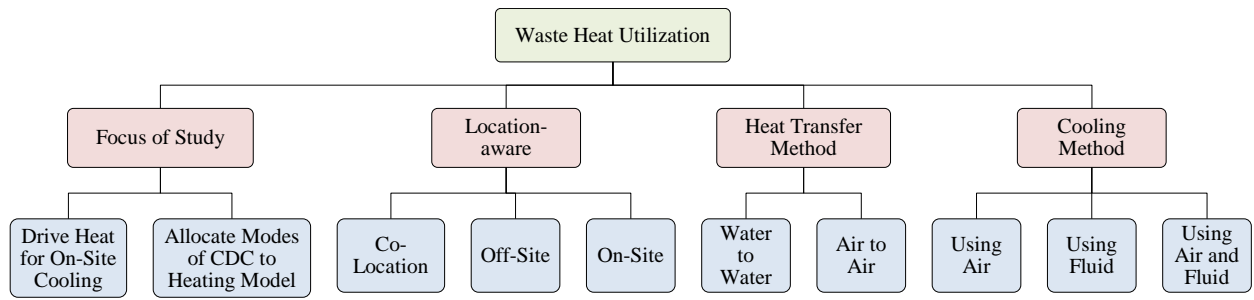

Figure 29: Taxonomy Based on Waste Heat Utilization

*3.9.2.2 Location-aware:* In waste heat utilization, heat can be recovered using three different localities: i) on-site, ii) off-site and iii) co-location as described in *Section 3.8.2.4*.

*3.9.2.3 Heat transfer method:* There are two different methods available for transferring heat: i) water to water and ii) air to air [133]. *Water-to-water* heat transfer method is based on a refrigerator mechanism, in which heat is transferred from the source side to the load side using conditioned fluid (hot or cold). Boiler or cooler can be used at both sides of the exchanger based on the purpose of the transfer. *Air to air* heat transfer method is working based on vapor compression refrigeration, which uses reverse-cycle air conditioners to transfer heat from one place to another.

*3.9.2.4 Cooling method:* As identified from literature, there are three types of cooling methods used in existing waste heat utilization techniques: i) using air, ii) using water and iii) using air and water [132] [134]. An evaporative cooler is a device that uses evaporation of water to cool *air* and it is working based on vapor-compression refrigeration cycles. On the other hand, the cooling effect is produced by consumption of *water* through evaporation. Both *water and air* based cooling mechanism are used by WHU techniques.

Table 25: Comparison of Existing Techniques Based on Taxonomy of Waste Heat Utilization

| Technique | Author | Focus of Study | Location-aware | Heat Transfer Method | Cooling Method |
|---|---|---|---|---|---|
| DEHR | Oro et al. [147] | Allocate Modes of CDC to Heating Model | On-site | Water to water | Using Air and Fluid |
| SORC | Freeman et al. [131] | Allocate modes of CDC to heating model | Off-site | Water to water | Using Fluid |
| BCA | Kilkovsky et al. [130] | Drive heat for onsite cooling | Co-location | Water to water | Using Air |
| PTDPG | Du et al. [132] | Drive heat for onsite cooling | On-site | Air to air | Using Fluid |
| ACSHC | Helm et al. [133] | Drive heat for onsite cooling | On-site | Air to air | Using Air and Fluid |
| WHS | Ayompe et al. [134] | Allocate modes of CDC to heating model | Co-location | Air to air | Using Air and Fluid |

## 4. Result Outcomes

The main aim of this systematic review is to discover the existing research related to sustainable cloud computing as per the research questions specified in Table 2. We have considered 142 research papers, 55 articles out of 142 are published in leading journals and the remaining are published in prominent workshops, symposiums and conferences on sustainable cloud computing. Table 26 lists the conferences, symposiums and journals publishing the research mostly related to sustainable cloud computing, which includes the number of research articles which report sustainable cloud computing as main research from each source.

Table 26: Journals/Conferences Reporting Most Sustainable Cloud Computing Related Research

| Publication Source | Publisher | Type | # | N |
|---|---|---|---|---|
| Future Generation Computer Systems | Elsevier | J | 7 | 16 |
| Concurrency and Computation: Practice and Experience | Wiley and Johns | J | 1 | 4 |
| ACM Computing Surveys | ACM | J | 5 | 9 |
| Ph.D Thesis | Arizona State University | PT | 1 | 1 |
| Renewable and Sustainable Energy Reviews | Elsevier | J | 4 | 11 |
| International Journal of Information Management | Elsevier | J | 2 | 5 |
| Transactions on Cloud Computing | IEEE | T | 5 | 14 |
| Parallel and Distributed Computing | Elsevier | J | 4 | 8 |
| Communications Magazine | IEEE | M | 2 | 6 |
| Cluster computing | Springer | J | 5 | 11 |
| Computers & Industrial Engineering | Elsevier | J | 1 | 3 |
| Computers in Industry | Elsevier | J | 1 | 4 |
| Journal of Cleaner Production | Elsevier | J | 2 | 4 |
| International Symposium on Sustainable Systems and Technology | IEEE | S | 2 | 5 |
| International Symposium on Systems Engineering | IEEE | S | 2 | 5 |
| International Symposium on Cloud Computing | ACM | S | 1 | 2 |

| Source | Publisher | Type | N | # |
|---|---|---|---|---|
| International Symposium on Advancement of Construction Management and Real Estate | Springer | S | 1 | 2 |
| Dependable Systems and Networks Workshops | IEEE | W | 1 | 3 |
| Recent Trends in Telecommunications Research Workshop | IEEE | W | 1 | 3 |
| White Paper | NoviFlow Inc | WP | 1 | 1 |
| Journal of manufacturing science and engineering | Elsevier | J | 1 | 2 |
| Mobile Networks and Applications | Elsevier | J | 1 | 2 |
| Journal of Software: Evolution and Process | Elsevier | J | 1 | 4 |
| Book Chapter of Large-Scale Distributed Systems and Energy Efficiency: A Holistic View | Wiley and Johns | BC | 1 | 5 |
| Transactions on Sustainable Computing | IEEE | T | 5 | 8 |
| Book Chapter of Intelligent distributed computing | Springer | BC | 1 | 2 |
| Journal of Network and Systems Management | Elsevier | J | 1 | 2 |
| Computer Network | Elsevier | J | 2 | 3 |
| Environmental modelling & software | Elsevier | J | 2 | 3 |
| Systems Journal | IEEE | J | 2 | 1 |
| International Conference on Network and Service Management | IEEE | C | 1 | 2 |
| International Conference on Cyber, Physical and Social Computing | IEEE | C | 1 | 4 |
| International Conference on Computing, Networking and Communications | IEEE | C | 1 | 5 |
| International Conference on Communications | IEEE | C | 1 | 5 |
| International Conference on Cloud Computing Technology and Science | IEEE | C | 1 | 2 |
| International Conference on Cyber-Physical Systems | IEEE | C | 1 | 2 |
| International Conference on Software Quality, Reliability and Security | IEEE | C | 1 | 3 |
| International Conference on Network Softwarization | IEEE | C | 1 | 4 |
| International Conference on Advance Computing | IEEE | C | 1 | 5 |
| International Conference on Design Science Research in Information Systems and Technology | Karlsruhe Institute of Technology | C | 1 | 5 |
| International Conference on Cloud Computing | IEEE | C | 1 | 2 |
| International Conference on Cloud Networking | IEEE | C | 1 | 2 |
| Journal of Intelligent & Fuzzy Systems | IoS Press | J | 1 | 3 |
| Ad Hoc Networks | Elsevier | J | 2 | 3 |
| Sustainable Computing: Informatics and Systems | Elsevier | J | 1 | 1 |
| Transactions on Parallel and Distributed Systems | IEEE | J | 3 | 5 |
| Ph.D Thesis | University of Trento | PT | 1 | 4 |
| Ph.D Thesis | University of Madrid | PT | 1 | 5 |
| Journal of Systems Architecture | Elsevier | J | 1 | 5 |
| Simulation Modelling Practice and Theory | Elsevier | J | 1 | 2 |
| IT Professional | IEEE | M | 1 | 2 |
| Transactions on Services Computing | IEEE | T | 1 | 3 |
| Transactions on Network and Service Management | Elsevier | T | 1 | 3 |
| Applied Energy | Elsevier | J | 3 | 6 |
| International journal of energy research | Elsevier | J | 1 | 2 |
| SIGMETRICS Performance Evaluation Review | ACM | M | 2 | 4 |
| Sustainability | MDPI | J | 1 | 5 |
| Energy | Elsevier | J | 1 | 5 |
| Energy Conversion and Management | Elsevier | J | 1 | 2 |
| Applied Thermal Engineering | Elsevier | J | 2 | 2 |
| Energy Procedia | Elsevier | J | 1 | 3 |
| Sustainable Cities and Society | Elsevier | J | 1 | 3 |
| Heat Transfer Engineering | Elsevier | J | 1 | 4 |
| Environmental Modelling & Software | Elsevier | J | 2 | 5 |
| IEEE Access | IEEE | J | 1 | 5 |
| Journal of Organizational and End User Computing | IGI Global | J | 1 | 2 |
| Computer Networks | Elsevier | J | 2 | 2 |
| Information and software technology | Elsevier | J | 1 | 3 |
| SIGOPS Operating Systems Review | ACM | J | 1 | 3 |
| Communications Surveys & Tutorials | IEEE | J | 1 | 1 |
| Computing | Springer | J | 1 | 2 |

J – Journal, C – Conference, W – Workshop, S – Symposium, T – Transactions, WP- White Paper, M- Magazine, BC -Book Chapter, PT- PhD Thesis and N – Number of studies reporting sustainable cloud computing as prime study, # – Total number of articles investigated.

It has been observed that conferences such as International Conference on Cloud Computing, International Conference on Advance Computing, International Conference on Cloud Computing Technology and Science contribute, International Symposium on Sustainable Systems and Technology and Workshop on Dependable Systems and Networks, contribute major part of research papers. Leading journals and IEEE Transactions like Future Generation of Computer Systems, Transactions on Sustainable Computing, Transactions on Cloud Computing, Concurrency and Computation: Practice and Experience, Renewable and Sustainable Energy Reviews, Journal of Parallel and Distributed Computing, Transactions on Parallel and Distributed Systems, ACM Computing Surveys, Environmental Modelling & Software and Applied Energy contributed expressively towards sustainable cloud computing, which is our review area.

Figure 30 depicts the number of research articles discussing various categories of sustainable cloud computing from year 2010 to 2018. The several drifts can be grasped for different categories of sustainable cloud computing. Research in the area of renewable energy is consistent since year 2012 and research on thermal-aware scheduling and cooling management increase abruptly in 2016 and becomes a hotspot area for sustainable cloud computing. The number of papers published in area of application design rose abruptly in 2014 and it was very important research area in year 2017. The maximum research work has been done on important areas of energy management. On the other hand, the number of papers published in the area of virtualization, capacity planning and waste heat utilization are stable throughout the years. Figure 31 shows the total number of publications of sustainable cloud computing and it is clearly shows that research on sustainable cloud computing is growing exponentially.

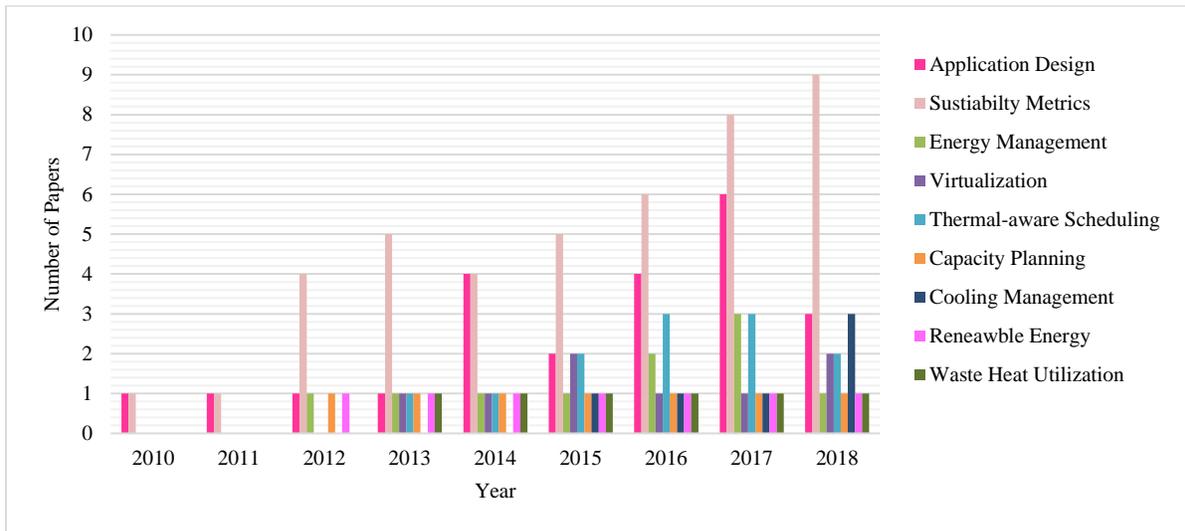

Figure 30: Different Categories of Sustainable Cloud Computing

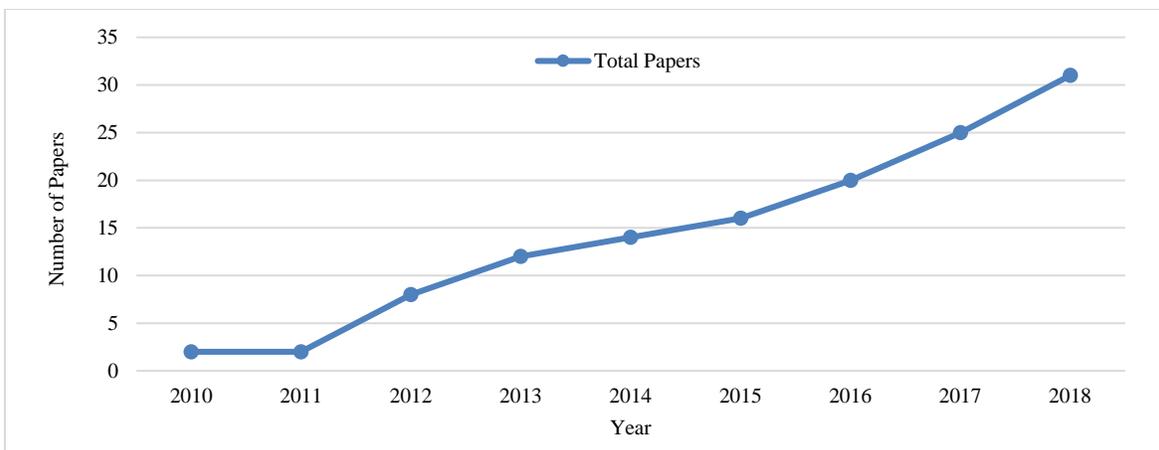

Figure 31: Year-wise Publications of Sustainable Cloud Computing

Figure 32 shows the number of citations discussing different categories of sustainable cloud computing from year 2010 to 2018. In year 2010-11, application design category was hotspot. In year 2017-18, energy management, cooling management, thermal-aware scheduling and renewable energy are important research areas. Figure 33 shows 28% of the research work were appeared in conferences, 32% of the literature published in journals, 21% of the studies were appeared in IEEE Transactions, 10% of the literature published in book chapters, 4% studies were appeared in workshops, 3% of the literature published in symposiums and 2% of the studies were appeared published in various conferences. The largest percentages of publications came from journals (91 papers) followed by conferences (50 papers). Literature reported that there are seven different types of study (introductory, review & survey, conceptual model, simulation, real testbed and Ph.D. thesis) in sustainable cloud computing. Figure 34 depicts that most of the research work (59%) has been on simulation based environments. Figure 35 and Figure 36 depicts the percentage of research articles which are focusing on various QoS parameters and we have identified different QoS parameters for the perspective of user as well as cloud provider from year 2010 to 2018 respectively.

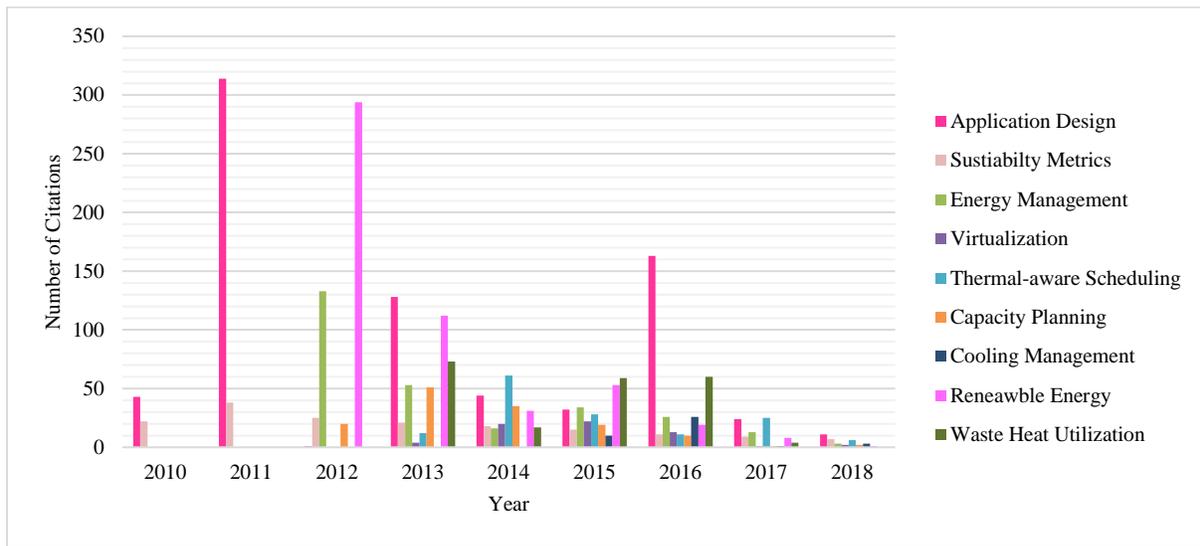

Figure 32: Number of Citations of Different Categories

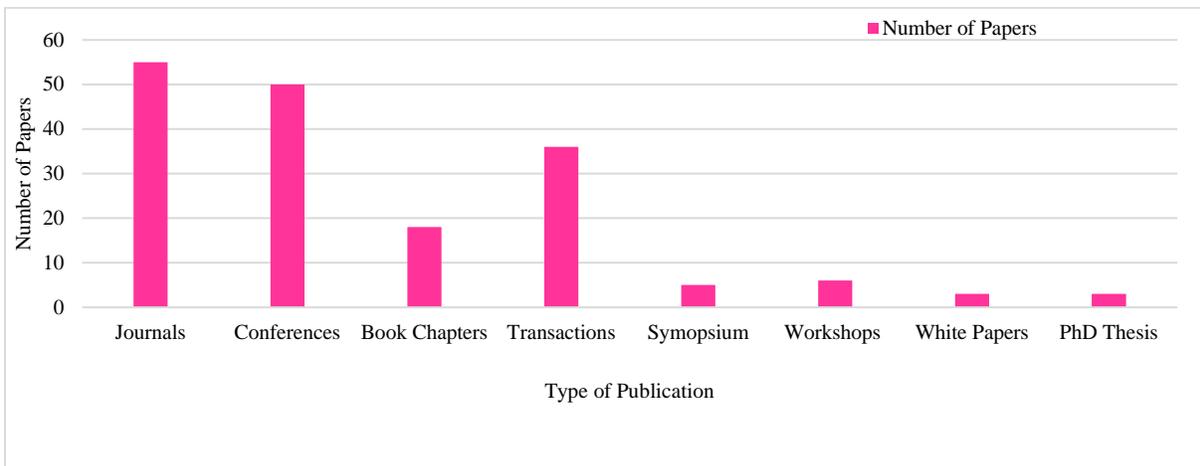

Figure 33: Type of Publications of Sustainable Cloud Computing

Response time (60%) is the most important QoS parameter for user while energy (40%) is the most important QoS parameter for cloud provider. We investigated a lack of research work in security and reliability as a QoS parameter. We described large number of research articles about cost, makespan, throughput and energy-efficiency as QoS parameters.

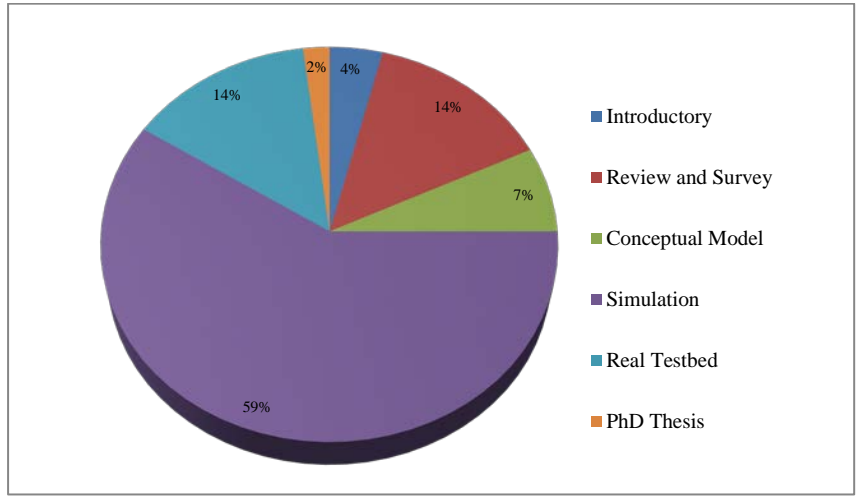

Figure 34: Type of Study of Sustainable Cloud Computing

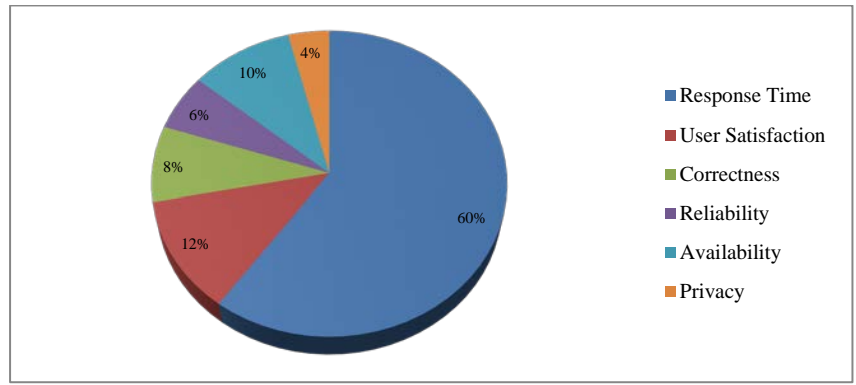

Figure 35: QoS Parameters for User's Perspective

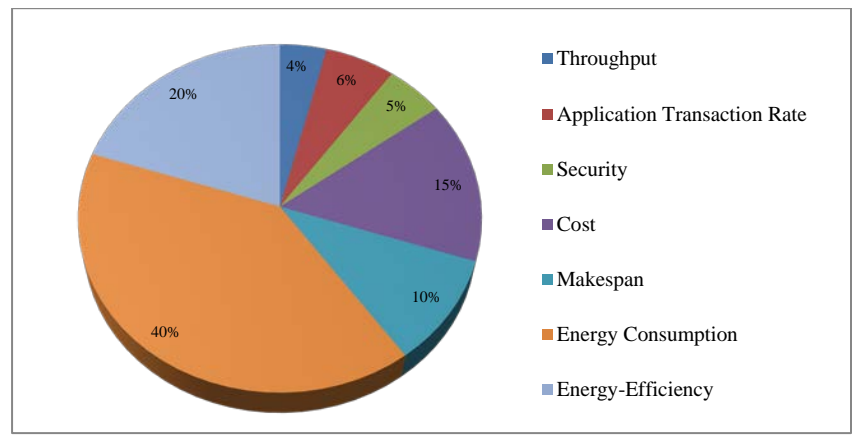

Figure 36: QoS Parameters for Cloud Provider's Perspective

## 5 OPEN CHALLENGES AND FUTURE DIRECTIONS: A SUMMARY

We surveyed 142 research papers in this systematic review and presented them in a categorized manner. The focus of our systematic review is broader than the existing surveys as discussed in *Table 1* (Section 2.1). This survey paper used methodical survey technique to conduct a systematic review and it comprises the most recent research work related to the sustainable cloud computing. In addition to the nine categories of sustainable cloud computing, we covered the other research issues related to the sustainability of emerging technologies such as Internet of Things and smart city. A systematic methodology has been used to develop an evolution of categories of sustainable cloud computing which identifies optimization parameters, metrics, open issues and Focus of Study (FoS). We explored and compared the existing techniques based on the proposed taxonomy. We documented the research issues addressed and open challenges that are still unresolved in sustainable cloud computing.

## 5.1 Open Challenges

Though a lot of progress has been achieved in sustainable cloud computing. Still there are many issues and challenges in this field that need to be addressed. Based on existing research, we have identified various open issues still pending in this area and described in Table 27.

Table 27: Open Challenges for different categories

| Category | Open Challenges |
|---|---|
| Application Design | 1. There is a need of effective communication among components of applications, so applications can scale easily in case of underloading or overloading of servers using a dynamic topology to add or remove resources automatically.<br>2. How to add an additional capacity to the application at runtime to scale individual services by avoiding the contention issues?<br>3. Which operations of the application could be handled asynchronously for effective load balancing at peak times?<br>4. What is the trade-off between scalability and availability during execution of an application?<br>5. How do we control access to application databases from other services running parallelly? |
| Sustainability Metrics | 1. How to measure the efficiency of the cloud datacenter by integrating different metrics for a particular context?<br>2. How to determine the age and location of cloud datacenter for overall comparison?<br>3. What are the acceptable levels of performance for every metric?<br>4. There is a need to define new metrics, which can measure security of CDC directly because existing security metrics dependent on compliance standards, SLA and governance?<br>5. How to measure the effect of IT load of the CDC on PUE significantly?<br>6. Existing metrics needs seasonal benchmarking to capture region and change of the season to measure performance accurately.<br>7. How to measure the energy consumption at sub-component or lower level of a CDC?<br>8. How to measure the cost of CDC based on space and energy used by CDC? |
| Capacity Planning | 1. What application parameters are considered to merge different applications?<br>2. What is the trade-off between capacity cost and resource utilization during merging of applications?<br>3. How migration of workloads or VMs can affect power infrastructure capacity?<br>4. How to plan an effective capacity to perform data recovery in case of disaster management?<br>5. What are the important user requirements, which effects the capacity of CDC during its technical design? |
| Energy Management | 1. What is the trade-off between availability of an application and energy-efficiency?<br>2. How power consumption affects the user satisfaction in terms of SLA?<br>3. How to reduce SLA violation rate while transferring resources from high scaling mode to low scaling mode?<br>4. How size of the cloud datacenter affects its energy efficiency?<br>5. How ICT is mainly responsible for large consumption of energy and carbon footprint generation?<br>6. How memory contention effects the energy consumption of CDC?<br>7. How to improve energy-efficiency based on traffic demands and QoS parameters?<br>8. How on-demand switching can affect energy consumption during starvation? |
| Virtualization | 1. What is the impact of VM consolidation and migration on preemption policy?<br>2. How to track dynamic load variations during VM load balancing?<br>3. What is the trade-off between energy utilization and network delay during VM migration.<br>4. Increasing the size of VM consumes more energy, which can increase service delay.<br>5. WAN based VM migration requires storage migration, which can be an overhead for cost-effective migration.<br>6. Security during VM migration is also an important issue because a VM state can be hijacked during its migration.<br>7. How to achieve VM migration and fault tolerance together in an efficient way? |
| Thermal-Aware Scheduling | 1. There is a need of effective thermal-aware scheduling techniques, which can execute workloads with minimum heat concentration and dissipation.<br>2. The complexity of scheduling and monitoring is increased due to variation of temperatures of the servers in CDC, which also causes vagueness in thermal profiling. To solve this problem, there is a need of dynamically updated thermal profiles instead of static, which will be updated automatically, and provide more accurate values of temperature.<br>3. What is the trade-off between TCO and PUE?<br>4. If scheduling is performed based on different thermal aspects, like inlet temperature and heat contribution, then admission control mechanisms at processor level and server level contradicts each other. |
| Cooling Management | 1. How to improve the datacenter cooling efficiency without effecting the temperature of CDC?<br>2. The consumption of large amount of energy can challenge the cooling management system of CDCs but reductions in energy consumption can also affect PUE of CDC.<br>3. What is the trade-off between cooling efficiency and PUE?<br>4. There is a need to change the location of cloud datacenters to reduce cooling costs and it can be done through placing the CDCs in areas that have availability of free cooling resources. |
| Renewable Energy | 1. The main challenges of renewable energy are unpredictability and capital cost of green resources.<br>2. Mostly, sites of commercial cloud datacenters are located away from abundant renewable energy resources. Consequently, moveable cloud datacenters are required to place nearer to the renewable energy sources to make cost effective.<br>3. Adoption of renewable energy in cloud datacenters is a research challenge of high capital cost.<br>4. The issue of unpredictability in supply of renewable energy and demand of CDCs is required to be addressed effectively.<br>5. The main reason of a large value of CUE is the total dependency on grid electricity which is generated using fossil fuels. Cooling measures and other computing devices also have an impact on the value of PUE. |
| Waste Heat Utilization | 1. The important issues of waste heat utilization techniques are: 1) high capital cost and 2) low heat quality.<br>2. The shifting of cooling systems from air based to water based creates new challenges such as:<br>   a. low quality of air due to mixing of exhaust air with cold air supply,<br>   b. smaller flow path in water based cooling systems and<br>   c. leakage of water into electronic equipment, which increases cost. |

**5.2 Implications for Research and Practice**

The systematic review has suggestions for perspective research scholars and practitioners who are already working in the area of sustainable cloud computing and looking for new ideas. A lot of research challenges are described for perspective researchers and professional experts. There is a need to integrate the different categories of sustainable cloud computing for better management of open issues related to cloud data center. Figure 37 shows the interactions among different categories of sustainable cloud computing. We have identified two different types of interaction among different categories: weak interaction (if any category depends the other category indirectly then it is considered as weak) and strong interaction (if one category depends the other category directly then it is considered as strong). Based on the existing research work and their citations, we have identified the importance of an induvial category along with their sub-categories. Further, future hotspot areas have been identified among different categories. The 360 Degree View (global and complete view) of the taxonomy of sustainable cloud computing is provided at the end of this research paper.

**5.3 Integrated: Sustainability vs. Reliability**

Sustainable cloud services are attracting more cloud customers and making it more profitable. Improving energy utilization, which reduces electricity bills and operational costs to enables sustainable cloud computing [14] [15]. On the other hand, to provide reliable cloud services, the business operations of different cloud providers such as Microsoft, Google, and Amazon are replicating services, which needs additional resources and increases energy consumption. To overcome this impact, a trade-off between energy consumption and reliability is required to provide cost-efficient cloud services. Existing energy efficient resource management techniques consume a huge amount of energy while executing workloads, which decreases resources leased from cloud datacenters [33].

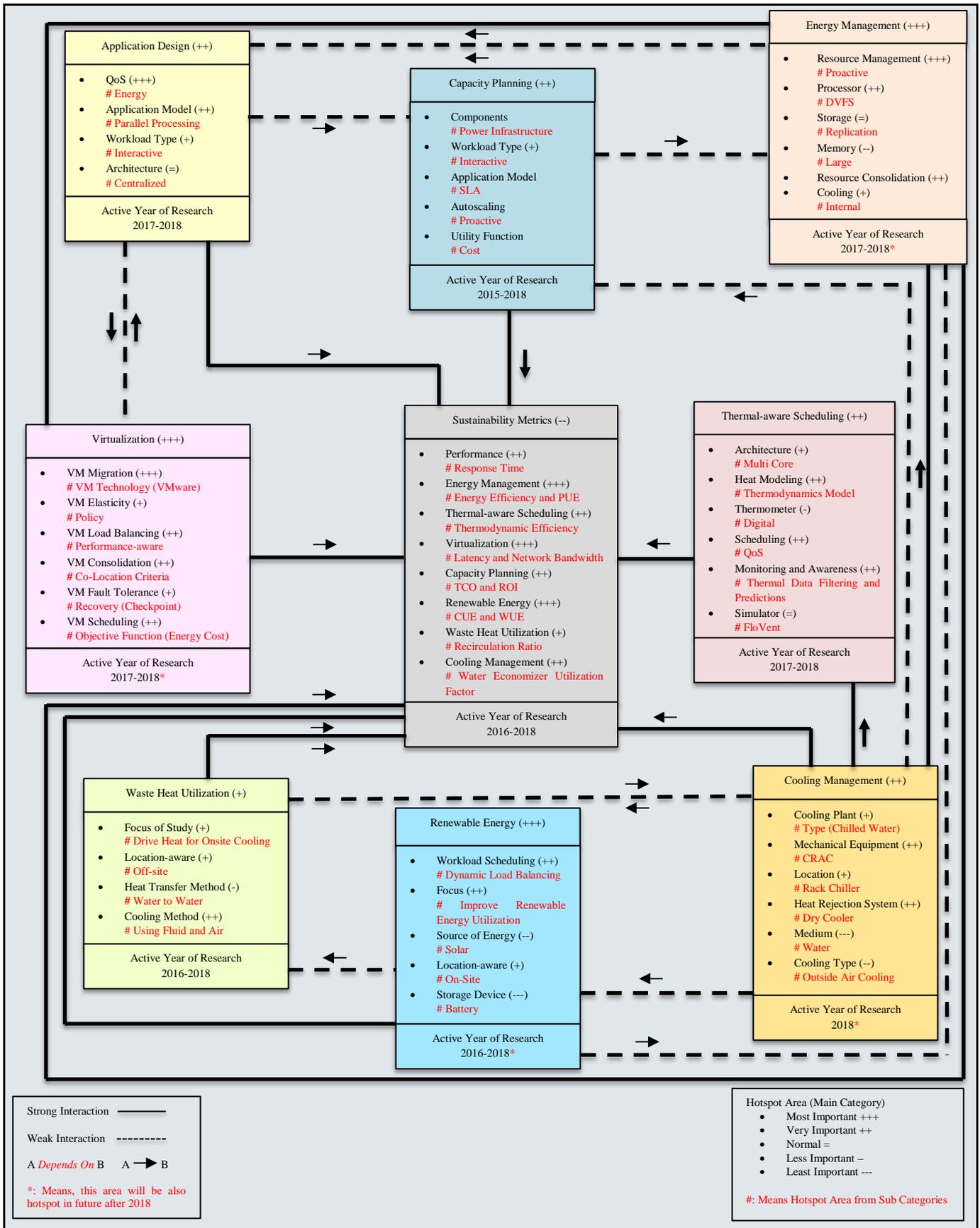

Figure 37: Interactions among Different Categories of Sustainable Cloud Computing (360 Degree view)

DVFS based energy management techniques reduced energy consumption, but response time and service delay are increased due to the switching of resources between high scaling and low scaling modes [1]. Further, reliability of the system component is also affected by excessive turning on/off servers. Power modulation decreases the reliability of server components like storage devices, memory etc. By reducing energy consumption of CDCs, we can improve the resource utilization, reliability and performance of the server. Therefore, there is a need of new energy-aware resource management techniques to reduce power consumption without affecting the reliability of cloud services.

**5.4 Emerging Trends and their Impact**

Emerging trends can be observed in three different contexts: (1) applications (e.g., Big Data), (2) technologies (e.g., Software-defined network), and (3) techniques (e.g., deep learning). They either present significant business opportunity for Cloud service providers (to offer new SaaS services using Clouds by supporting emerging applications) in many interesting ways or help them in efficient management and utilization of their Cloud infrastructures in reliable and energy-efficient manner.

With the enormous success of Cloud computing, there are many emerging applications such as big data and smart cities [166] are harnessing it. These applications can be created using existing application/programming models such as those discussed in *Section 3.1* (Application Design) or new ones to support the rapid creation of applications or ease of programming. When new application design/programming models are proposed, there is a need for new approaches for resource management and application scheduling to ensure the delivery of services of these Cloud applications in an energy efficient manner. All these factors will impact Cloud software platforms [167]. Furthermore, new technological trends [168] such as Software-Defined Networks [165], Internet of Things (IoT), and Containers are presenting new opportunities for better management of data centres to support execution of emerging applications in an energy efficient manner. For example, Internet of Things (IoT)-based sensors and actuators help better management of cooling systems of CDC infrastructures.

The success of Cloud computing is also leading to the creation of larger CDC infrastructures, which is contributing to growing complexities in the management of CDCs. To better manage their resources and monitor application execution and resource status, recently works as [167] are starting to use deep learning approaches for detecting anomalies. Such deep learning techniques need to be developed for better management of resources and workloads of emerging application models in a reliable and energy efficient manner without affecting the quality of service delivered to users.

**6. SUSTAINABLE CLOUD COMPUTING ARCHITECTURE: A CONCEPTUAL MODEL**

To resolve the above-mentioned challenges, there is a requirement of cloud computing architectures that can provide sustainable and reliable cloud services. Earlier models by Arlitt et al. [19] and Guitart [87] have been very innovative, but as the research has persistently grown in the field of sustainable cloud computing, there is a necessity for new conceptual model to cover other important aspects of sustainability. A model proposed by Arlitt et al. [19] is only focused on cooling management of cloud datacenters and energy storage management, while a model proposed by Guitart [87] is focused on management of cooling, infrastructure and workloads. Therefore, our model augments the previous aspects and covers all the important components such as i) layered architecture which comprises of software (application management), platform (workload and VM/Resource management) and infrastructure management, ii) cooling management, iii) energy management (renewable and grid energy) and iv) thermal-aware aspects of cloud datacenter and describes the interactions among them. Figure 38 shows the conceptual model for sustainable cloud computing in the form of layered architecture, which offers holistic management of cloud computing resources to make cloud services more energy-efficient, sustainable and reliable. The three main components of proposed architecture are as follows:

1. *Software as a Service (SaaS):* At this layer, application manager is deployed to handle the incoming user workloads, which can be interactive or batch style and transfer to workload manager for further action. At SaaS level, the QoS requirements of different applications can be defined in terms of SLA. QoS requirements for an application can be deadline, budget etc. For example, different encoding techniques can be used for video on-demand application based on the QoS requirements of cloud user.

2. *Platform as a Service (PaaS)*: At this layer, controller or middleware is deployed to control the important aspects of the system. Resource manager and scheduler follows different provisioning and scheduling policies for efficient management of cloud resources, which can improve energy efficiency and resource utilization to make CDCS more sustainable. There are four sub units of controller: workload manager, VM/resource manager, and manager and their functions are described below:

    a) *Workload manager* manages the incoming workloads from the application manager and identifies the QoS requirement for every workload for their successful execution and transfer the QoS information of workload to the VM/resource manager.
    b) *VM/resource manager* provisions and schedules the cloud resources for workload execution based on QoS requirements of workload using physical machines or virtual machines.
    c) *Manager* controls the two modules of infrastructure: i) *Green energy resource manager* and ii) *cooling manager*

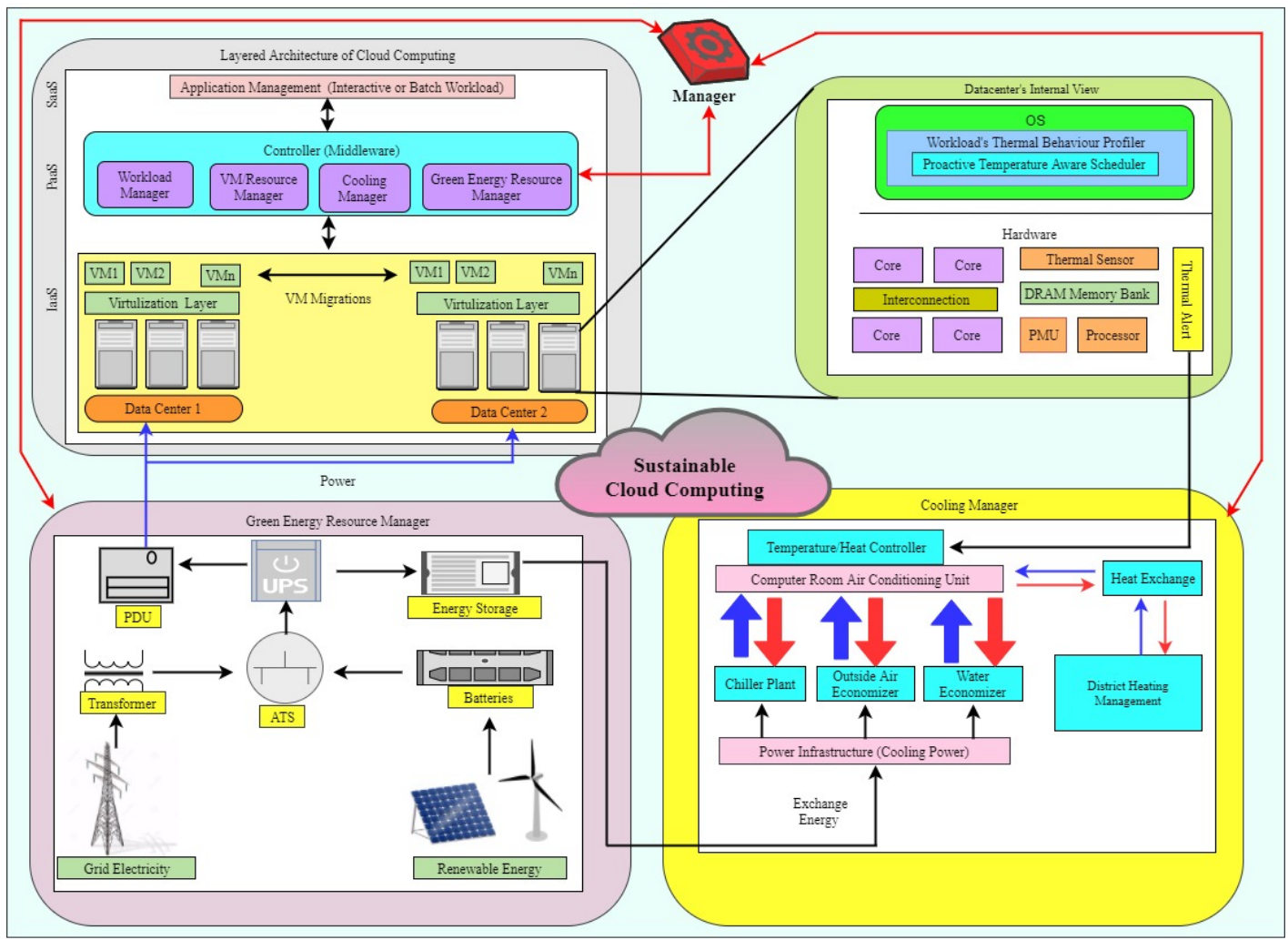

Figure 38: Conceptual Model for Sustainable Cloud Computing

3. *Infrastructure as a Service (IaaS):* This layer contains the information about cloud datacenters and virtual machines. VM migrations are performed to balance the load at virtualization layer for efficient execution of workloads. The proactive temperature aware scheduler is used to monitor the temperature variation of different VMs running at different cores. Power Management Unit (PMU) is integrated to power all the hardware executing the VMs. Dynamic Random-Access Memory (DRAM) stores the current states of VMs. Thermal profiling and monitoring techniques are used to analyze the temperature variations of CDCs based on the value of temperature as monitored by thermal sensors. Thermal alerts will be generated if temperature is higher than the threshold value and heat controller will take an action to control the temperature without affecting the performance of the CDC. The electricity coming from Uninterruptible Power Supply (UPS) is used to run the cooling devices to control the temperature. District heating management is integrated in which, the temperature is controlled by using chiller plant, outside air economizer and water economizer. *Green energy resource manager* controls the power generated from renewable energy resources and fossil fuels. To enable sustainable cloud environment, renewable energy is more preferred as compared to grid energy. If there is execution of deadline oriented workloads, then grid energy can be used to maintain the reliability of cloud services. The sources of renewable energy are solar and wind. Batteries are used to store the renewable energy. Automatic Transfer Switch (ATS) is used to manage the energy coming from both sources (renewable and non-renewable) and forward to UPS. Further, Power Distribution Unit (PDU) is used to transfer the electricity to all the CDCs and cooling devices.

## 7. SUMMARY AND CONCLUSIONS

The usage of large number of cloud datacenters results in a huge amount of energy consumption and produces significant amount of carbon footprints, which has become the greatest challenge of the 21st century. On the other hand, the utilization of a combination of grid and renewable energy to run cloud datacenter in smart cities can save energy to a large extent. Consequently, there is a need to manage both energy and QoS together to enable sustainable and energy-efficient cloud services. Existing energy-aware resource management techniques and policies mainly focus on VM consolidation to reduce energy consumption of servers only, but other resources such as networks, storage, memory and cooling consume a huge amount of energy. There is a need for an efficient scheduling of the traffic flows between servers in the cloud datacenter to save energy. Therefore, holistic management of all the resources (networks, memory, processors, cooling and storage) is required to enable sustainable cloud computing. Further, the effect of QoS on SLA is required to be addressed in holistic management techniques. Moreover, self-aware or autonomic management of cloud resources in a holistic manner can manage both energy consumption and QoS simultaneously, which can improve the sustainability of cloud computing systems. In addition, dynamically changing the variable clock rates of processor can also optimize the usage of energy. It has also been recommended that *follow the renewable* can motivate the cloud providers to locate their cloud datacenters nearer to green energy resources and load can be distributed geographically. However, geographical distribution of resources affects the QoS of networks, which is an open research challenge for the community. Unfortunately, the requirement for processing of a huge amount of data and high performance simultaneously can also consume large amounts of energy. To solve this problem, there is a need to manage energy consumption, SLAs and QoS at same time. Further, there is a need of self-aware management of cloud resources holistically to address these research issues. Currently, research community is working in this direction, but more advanced research is required to ensure the energy efficiency and sustainability of cloud services.

In this paper, we proposed a taxonomy of sustainable cloud computing to analyze existing techniques for sustainability such as application design, sustainability metrics, capacity planning, energy management, virtualization, thermal-aware scheduling, cooling management, renewable energy and waste heat utilization for CDCs. Further, the taxonomy mapping based comparison has been described. A conceptual model for sustainable cloud computing has been proposed. Through a detailed analysis of the related studies in the context of taxonomy, we are able to identify and propose various future research directions.

Following facts can be further concluded:

- VM consolidation techniques can minimize energy consumption of servers.
- Optimization scheduling of traffic flows between servers is required.
- There is a need of dynamic task scheduling for energy and QoS optimization.
- New system architectures and algorithms can geographically distribute the CDC.
- There is a need of interplay between IoT-enabled cooling systems and CDC manager.
- Maximum utilization of renewable energy powered resources is required for holistic management of resources and workloads.

We hope that this systematic review will be helpful for practitioners and researchers who want to pursue research in any area of sustainable cloud computing.

## ACKNOWLEDGEMENTS

We thank Patricia Arroba, Minxian Xu and Shashikant Ilager for their useful suggestions.

[137] Mehiar Dabbagh, Bechir Hamdaoui, Ammar Rayes, and Mohsen Guizani. 2017. Shaving Datacenter Power Demand Peaks Through Energy Storage and Workload Shifting Control, *IEEE Transactions on Cloud Computing*, 2017. DOI: https://doi.org/10.1109/TCC.2017.2744623

[138] Juarez, Fredy, Jorge Ejarque, and Rosa M. Badia. 2018. Dynamic energy-aware scheduling for parallel task-based application in cloud computing. Future Generation Computer Systems 78 (2018): 257-271.

[139] Anik Mukherjee, R. P. Sundarraj, and Kaushik Dutta. 2017. Users' time preference based stochastic resource allocation in cloud spot market: cloud provider's perspective. *In Designing the Digital Transformation: DESRIST 2017 Research in Progress Proceedings of the 12th International Conference on Design Science Research in Information Systems and Technology*. Karlsruhe, Germany. 30 May-1 Jun. Karlsruher Institut für Technologie (KIT), 2017.

[140] Suleiman Onimisi Aliyu, Feng Chen, Ying He, and Hongji Yang. 2017. A Game-Theoretic Based QoS-Aware Capacity Management for Real-Time EdgeIoT Applications. In *Proceedings of the IEEE International Conference on Software Quality, Reliability and Security (QRS)*, 2017: 386-397.

[141] D. Kanapram, R. Rapuzzi, G. Lamanna, and M. Repetto. 2017. A framework to correlate power consumption and resource usage in cloud infrastructures. In *Proceedings of the IEEE International Conference on Network Softwarization (NetSoft)*, 2017: 1-5.

[142] Chao Jin, Bronis R. de Supinski, David Abramson, Heidi Poxon, Luiz DeRose, Minh Ngoc Dinh, Mark Endrei, and Elizabeth R. Jessup. 2016. A survey on software methods to improve the energy efficiency of parallel computing. *The International Journal of High Performance Computing Applications* (2016).

[143] Sukhpal Singh, Inderveer Chana, 2013. Consistency Verification and Quality Assurance (CVQA) Traceability Framework for SaaS", In *Proceeding of 3rd IEEE International Advance Computing Conference (IACC-2013)*, February 22-23, 2013, India.

[144] Junaid Shuja, Kashif Bilal, Sajjad Ahmad Madani, and Samee U. Khan. 2014. Data center energy efficient resource scheduling. *Cluster Computing* 17, no. 4 (2014): 1265-1277.

[145] Junaid Shuja, Raja Wasim Ahmad, Abdullah Gani, Abdelmuttlib Ibrahim Abdalla Ahmed, Aisha Siddiqa, Kashif Nisar, Samee U. Khan, and Albert Y. Zomaya. 2017. Greening emerging IT technologies: techniques and practices. *Journal of Internet Services and Applications* 8, no. 1 (2017).

[146] Ignacio Aransay, Marina Zapater, Patricia Arroba, and José M. Moya. 2015. A Trust and Reputation system for energy optimization in Cloud data centers. In *Proceeding of the IEEE 8th International Conference on Cloud Computing (CLOUD)*, 2015: 138-145.

[147] Oró, Eduard, Ricard Allepuz, Ingrid Martorell, and Jaume Salom. 2018. Design and economic analysis of liquid cooled data centres for waste heat recovery: A case study for an indoor swimming pool. *Sustainable Cities and Society* 36 (2018): 185-203.

[148] Khosravi, Atefeh, and Rajkumar Buyya. 2018. Short-Term Prediction Model to Maximize Renewable Energy Usage in Cloud Data Centers. In *Sustainable Cloud and Energy Services*, pp. 203-218. Springer, Cham, 2018.

[149] Triantafyllidis, Charalampos P., Rembrandt HEM Koppelaar, Xiaonan Wang, Koen H. van Dam, and Nilay Shah. 2018. An integrated optimization platform for sustainable resource and infrastructure planning. *Environmental Modelling & Software* 101 (2018): 146-168.

[150] Ndukaife, Theodore A., and AG Agwu Nnanna. 2018. Optimization of Water Consumption in Hybrid Evaporative Cooling Air Conditioning Systems for Data Center Cooling Applications. *Heat Transfer Engineering* (2018): 1-15. DOI: https://doi.org/10.1080/01457632.2018.1436418

[151] Wu, Jiahong, Yuan Jin, and Jianguo Yao. 2018. EC 3: Cutting Cooling Energy Consumption Through Weather-Aware Geo-Scheduling Across Multiple Datacenters. *IEEE Access* 6 (2018): 2028-2038.

[152] Sahana, Sudipta, Rajesh Bose, and Debabrata Sarddar. 2018. Server Utilization-Based Smart Temperature Monitoring System for Cloud Data Center. In *Industry Interactive Innovations in Science, Engineering and Technology*, pp. 309-319. Springer, Singapore, 2018.

[153] Matsuoka, Morito, Kazuhiro Matsuda, and Hideo Kubo. 2017. Liquid immersion cooling technology with natural convection in data center. In *Proceedings of the IEEE 6th International Conference on Cloud Networking (CloudNet)*, 2017: 1-7.

[154] Liu, Qiang, Yujun Ma, Musaed Alhussein, Yin Zhang, and Limei Peng. 2016. Green data center with IoT sensing and cloud-assisted smart temperature control system. *Computer Networks* 101 (2016): 104-112.

[155] Manousakis, Ioannis, Íñigo Goiri, Sriram Sankar, Thu D. Nguyen, and Ricardo Bianchini. Coolprovision: Underprovisioning datacenter cooling. In *Proceedings of the Sixth ACM Symposium on Cloud Computing*, pp. 356-367. ACM, 2015.

[156] Gill, Sukhpal Singh, and Rajkumar Buyya. 2018. Resource Provisioning Based Scheduling Framework for Execution of Heterogeneous and Clustered Workloads in Clouds: from Fundamental to Autonomic Offering. *Journal of Grid Computing* (2018): 1-33. DOI: https://doi.org/10.1007/s10723-017-9424-0

[157] Chinnathambi, Sathya, Agilan Santhanam, Jeyarani Rajarathinam, and M. Senthilkumar. 2018. Scheduling and checkpointing optimization algorithm for Byzantine fault tolerance in cloud clusters. *Cluster Computing* (2018): 1-14. DOI: https://doi.org/10.1007/s10586-018-2375-9

[158] Sotiriadis, Stelios, Nik Bessis, and Rajkumar Buyya. 2018. Self-managed virtual machine scheduling in Cloud systems. *Information Sciences* Vol. 433–434, Pages 381-400, 2018.

[159] Ranjbari, Milad, and Javad Akbari Torkestani. 2018. A learning automata-based algorithm for energy and SLA efficient consolidation of virtual machines in cloud data centers. *Journal of Parallel and Distributed Computing* 113 (2018): 55-62.

[160] Ashraf, Adnan, and Ivan Porres. 2018. Multi-objective dynamic virtual machine consolidation in the cloud using ant colony system. *International Journal of Parallel, Emergent and Distributed Systems* 33, no. 1 (2018): 103-120.

[161] Beechu, Naresh Kumar Reddy, Vasantha Moodabettu Harishchandra, and Nithin Kumar Yernad Balachandra. 2017. High-performance and energy-efficient fault-tolerance core mapping in NoC. *Sustainable Computing: Informatics and Systems* 16 (2017): 1-10.

[162] Dastagiraiah, C., V. Krishna Reddy, and K. V. Pandurangarao. 2018. Dynamic Load Balancing Environment in Cloud Computing Based on VM Ware Off-Loading." In *Data Engineering and Intelligent Computing*, pp. 483-492. Springer, Singapore, 2018.

[163] Al-Dhuraibi, Yahya, Fawaz Paraiso, Nabil Djarallah, and Philippe Merle. 2017. Autonomic vertical elasticity of docker containers with elasticdocker." In *Proceedings of the IEEE 10th International Conference on Cloud Computing (CLOUD)*, 2017: 472-479.

[164] Al-Dhuraibi, Yahya, Faiez Zalila, Nabil Djarallah, and Philippe Merle. 2018. Coordinating Vertical Elasticity of both Containers and Virtual Machines. In *CLOSER 2018-8th International Conference on Cloud Computing and Services Science*, pp. 1-8. 2018.

[165] Jungmin Son and Rajkumar Buyya, A Taxonomy of Software-Defined Networking (SDN)-Enabled Cloud Computing, 2018. *ACM Computing Surveys*, Volume 51, No. 3, Article No. 59, Pages: 1-36, ISSN 0360-0300, ACM Press, New York, USA, June 2018.

[166] Santana, Eduardo Felipe Zambom, Ana Paula Chaves, Marco Aurelio Gerosa, Fabio Kon, and Dejan S. Milojicic. 2017. Software platforms for smart cities: Concepts, requirements, challenges, and a unified reference architecture. *ACM Computing Surveys* (CSUR) 50, no. 6 (2017): 78.

[167] Singh, Sukhpal, and Inderveer Chana. 2016. A survey on resource scheduling in cloud computing: Issues and challenges. *Journal of Grid Computing* 14, no. 2 (2016): 217-264.

[168] Rajkumar Buyya et. al., A Manifesto for Future Generation Cloud Computing: Research Directions for the Next Decade, *ACM Computing Surveys* (CSUR), Available at: https://arxiv.org/abs/1711.09123
66

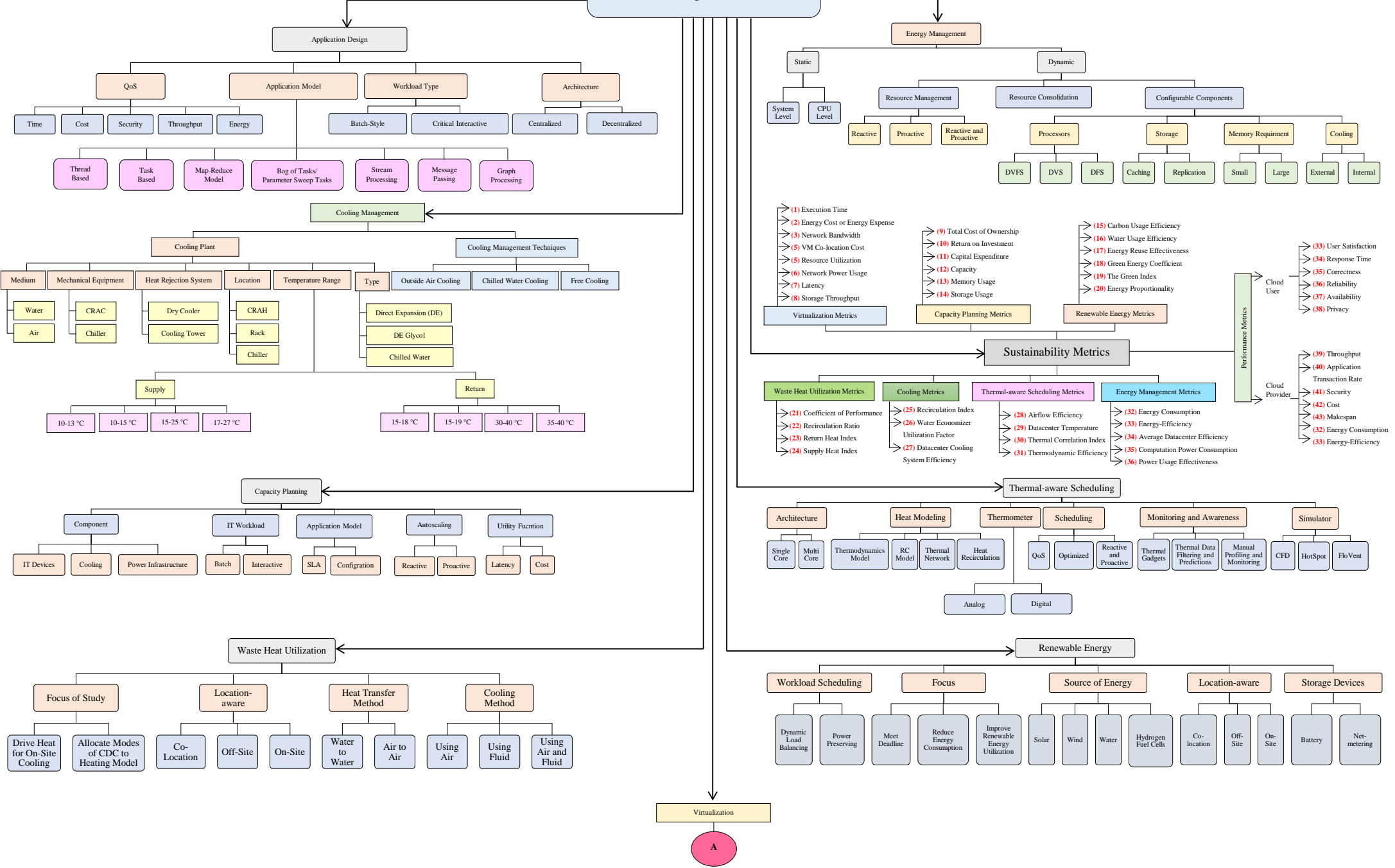

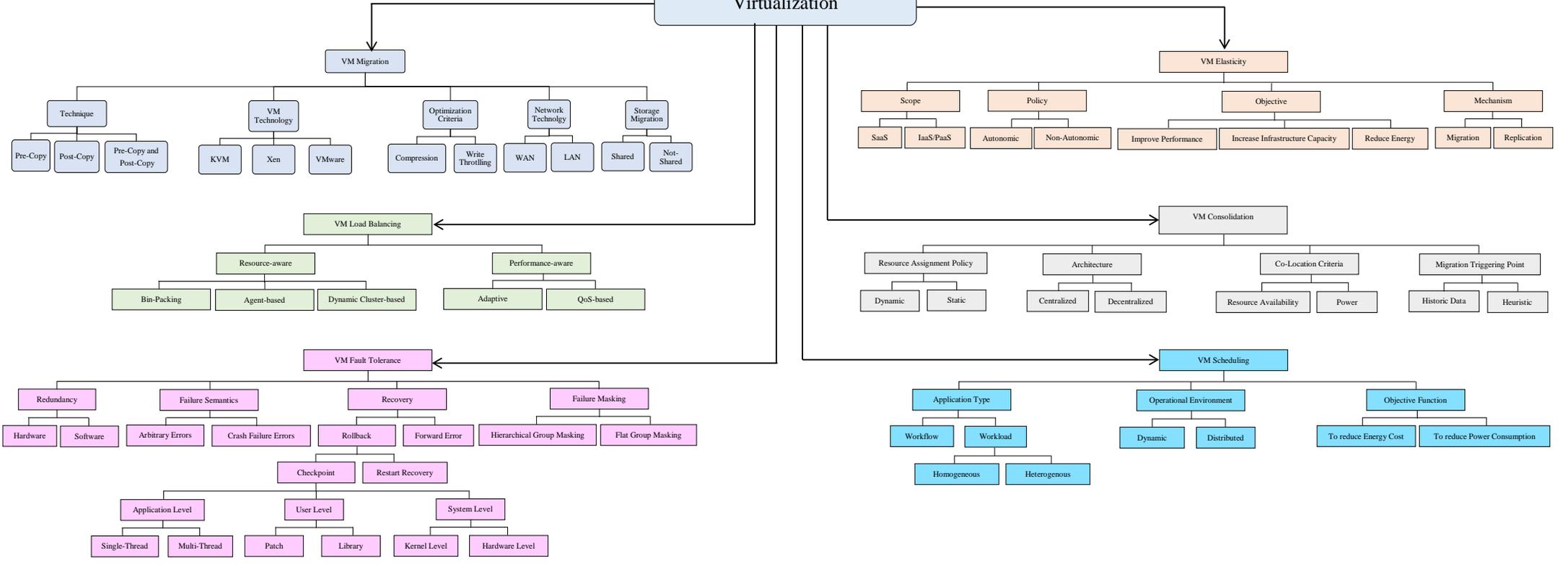